\documentclass[tightenlines,a4paper,altaffilletter,twoside,secnumarabic,
               showpacs,showkeys,twocolumn,floatfix,nofootinbib,prd]{revtex4}

\usepackage[english]{babel}
\usepackage{amsmath,amssymb}
\usepackage{slashed}
\usepackage{epsfig}
\usepackage{graphicx,color}
\usepackage[dvips,hyperfootnotes=false]{hyperref}
\usepackage{url}

\renewcommand{\url}{\nolinkurl}

\allowdisplaybreaks
\definecolor{gray}{rgb}{0.6,0.6,0.6}

\newcommand{\bra}[1]{\ensuremath{\langle #1 |}}   
\newcommand{\ket}[1]{\ensuremath{| #1 \rangle}}   
\newcommand{\sprod}[2]{\ensuremath{\langle #1 | #2 \rangle}}  

\newcommand{\sthchooz}{\ensuremath{\sin^2 2\theta_{13}}}

\newcommand{\ldm}{\ensuremath{{\Delta m_{31}^2}}}          
\newcommand{\sdm}{\ensuremath{{\Delta m_{21}^2}}}
\newcommand{\epssin}{\ensuremath{{\eps \sin 2\theta_{13}}}}

\newcommand{\eps}{\varepsilon}
\newcommand{\diag}{{\rm diag}}

\newcommand{\SuperK}{{\sf Super-Kamiokande}}

\newcommand{\TtoK}{{\sf T2K}}
\newcommand{\TtoHK}{{\sf T2HK}}
\newcommand{\NOvA}{{\sf NO$\nu$A}}
\newcommand{\Chooz}{{\sf Chooz}}
\newcommand{\DoubleChooz}{{\sf Double Chooz}}
\newcommand{\DCext}{{\sf DC-200}}
\newcommand{\DayaBay}{{\sf Daya Bay}}

\newcommand{\GLoBES}{{\sf GLoBES}}

\begin{document}

\title{Non-standard neutrino interactions in reactor and superbeam experiments}
\author{Joachim Kopp}    \email[Email: ]{jkopp@mpi-hd.mpg.de}
\author{Manfred Lindner} \email[Email: ]{lindner@mpi-hd.mpg.de}
\author{Toshihiko Ota}   \email[Email: ]{toshi@mpi-hd.mpg.de}
\affiliation{Max--Planck--Institut f\"ur Kernphysik, \\
             Postfach 10 39 80, 69029 Heidelberg, Germany}
\author{Joe Sato}        \email[Email: ]{joe@phy.saitama-u.ac.jp}
\affiliation{Department of Physics, Saitama University, Shimo-Okubo 255, \\
             Sakura-ku, Saitama, 338-8570, Japan}

\begin{abstract}
  The formalism of non-standard four-fermion interactions provides a convenient,
  model-independent way of parameterizing a wide class of ``new physics'' scenarios.
  In this article, we study the performance of reactor and superbeam neutrino
  experiments in the presence of such non-standard interactions (NSI). Due to
  interference between the standard and non-standard amplitudes, sizeable effects
  are to be expected if the NSI parameters are close to their current upper limits.
  We derive approximate formulas for the relevant oscillation probabilities including
  NSI, and show how the leading effects can be understood intuitively even without
  any calculations. We will present a classification of all possible NSI
  according to their impact on reactor and superbeam experiments, and it will turn
  out that these experiments are highly complementary in terms of their sensitivity
  to the non-standard parameters. The second part of the paper is devoted to
  detailed numerical simulations, which will demonstrate how a standard oscillation
  fit of the mixing angle $\theta_{13}$ may fail if experimental data is affected
  by NSI. We find that for some non-standard terms, reactor and superbeam
  experiments would yield seemingly conflicting results, while in other cases,
  they may agree well with each other, but the resulting value for $\theta_{13}$
  could be far from the true value. This offset may be so large that the true
  $\theta_{13}$ is even ruled out erroneously. In the last section of the paper,
  we demonstrate that reactor and superbeam data can actually establish the
  presence of non-standard interactions. Throughout our discussion, we
  pay special attention to the impact of the complex phases, and of the near
  detectors.
\end{abstract}

\ \pagebreak
\begin{flushright}
  STUPP-07-192
\end{flushright}
\maketitle

\section{Introduction}

At the dawn of the era beyond the standard model, a plethora of new theoretical
models has been devised to resolve many of the experimental and theoretical
shortcomings of our current picture of elementary particles. However, in the
context of future experiments, it is often desirable to describe new physics
in a more model-independent way. One possibility to achieve this is through
effective four-fermion operators, so-called non-standard interactions (NSI),
which arise naturally in the presence of heavy mediator fields. In this
article, we shall focus in particular on NSI in the neutrino sector, which
have been discussed on general phenomenological grounds
in~\cite{Wolfenstein:1977ue,Valle:1987gv,Guzzo:1991hi,
Roulet:1991sm,Bergmann:1999rz,Hattori:2002uw,Garbutt:2003ih,Blennow:2005qj},
and in the context of specific models in~\cite{DeGouvea:2001mz,Ota:2005et,
Honda:2007yd,Honda:2007wv}. The importance of NSI for neutrino oscillation physics
has been pointed out in a pioneering work by Grossman~\cite{Grossman:1995wx},
and many authors have studied their impact on solar
neutrinos~\cite{Bergmann:2000gp,Berezhiani:2001rt,Friedland:2004pp,Miranda:2004nb},
atmospheric neutrinos~\cite{Gonzalez-Garcia:1998hj,Bergmann:1999pk,Fornengo:2001pm,
Gonzalez-Garcia:2004wg,Friedland:2004ah,Friedland:2005vy}, conventional and
upgraded neutrino beams~\cite{Bergmann:1998ft,Ota:2001pw,Ota:2002na,Honda:2006gv,
Kitazawa:2006iq,Friedland:2006pi,Blennow:2007pu}, neutrino
factories~\cite{Ota:2001pw,Gonzalez-Garcia:2001mp,Huber:2001zw,Gago:2001xg,
Huber:2002bi,Campanelli:2002cc,Blennow:2005qj,Bueno:2000jy,Kopp:2007mi},
beta beams~\cite{Adhikari:2006uj}, supernova neutrinos~\cite{Fogli:2002xj,
Duan:2006jv,EstebanPretel:2007yu}, cosmological relic neutrinos~\cite{Mangano:2006ar},
$e^+ e^-$ colliders~\cite{Berezhiani:2001rs}, neutrino-electron
scattering~\cite{Barranco:2005ps}, and neutrino-nucleus
scattering~\cite{Barranco:2005yy,Barranco:2007tz}. Existing experimental bounds
are presented in~\cite{Davidson:2003ha}.

Our main interest in this work will be on non-standard interactions in
upcoming reactor and accelerator neutrino experiments. Although the main
design goal for these experiments is the precision measurement of the
standard oscillation parameters, the search for deviations from the
standard framework is an equally interesting part of their physics
program. Moreover, while in the race for the standard oscillation
parameters, reactor and beam experiments are competing, we will show
that their results will be highly complementary when one is interested
in non-standard physics.

In the numerical simulations which we are going to present, we will focus
on the experiments \TtoK~\cite{Itow:2001ee,T2KProposal}, \NOvA~\cite{Ayres:2004js},
\DoubleChooz~\cite{Ardellier:2004ui,Ardellier:2006mn}, and a hypothetical 200~t
reactor experiment~\cite{Huber:2006vr}. Of course, the analytical results
apply also to other experiments such as \TtoHK~\cite{Itow:2001ee} and
\DayaBay~\cite{Guo:2007ug}.

We will first introduce our formalism in Sec.~\ref{sec:formalism}, and give a
detailed discussion of the different possible Lorentz structures and their
relevance for reactor and superbeam experiments. Although this discussion may
seem rather technical, it will ultimately allow us to greatly simplify the problem
and considerably reduce the number of parameters. In Sec.~\ref{sec:theory}, we
will then present approximate expressions for the oscillation probabilities
including NSI for the $\bar{\nu}_e \rightarrow \bar{\nu}_e$, $\nu_\mu \rightarrow \nu_e$,
and $\nu_\mu \rightarrow \nu_\mu$ channels. We will also show in an intuitive way why
certain NSI terms appear in these expressions, and others do not. Sec.~\ref{sec:simulation}
is devoted to a discussion of numerical simulation techniques, and of the
specific experiments which we have simulated. In Sec.~\ref{sec:th13fits},
we show how the data from these experiments may be misinterpreted, if NSI
are not taken into account in the fits. We will finally demonstrate in
Sec.~\ref{sec:dis-reach}, that a combined analysis of reactor and superbeam
data may allow for the actual discovery of a wide variety of non-standard
interactions by goodness-of-fit arguments. Our conclusions will be presented
in Sec.~\ref{sec:conclusions}.

\section{The formalism of non-standard interactions}
\label{sec:formalism}

\subsection{The NSI Lagrangian}

It is well known that in the low energy regime, weak neutrino interactions can be
described by effective four-fermion operators like
\begin{align}
  \mathcal{L}_{\nu} &= \frac{G_{F}}{\sqrt{2}}
        \left[ \bar{\nu}_\alpha \gamma^{\rho} (1 - \gamma^{5}) \ell_\alpha \right]
               \left[ \bar{f}^\prime \gamma_{\rho} (1 - \gamma^{5}) f \right],
  \label{eq:SM-Lagrangian-CC}
\intertext{and}
  \mathcal{L}_{\rm MSW} &= \frac{G_{F}}{\sqrt{2}}
        \left[ \bar{\nu}_\alpha \gamma^{\rho} (1 - \gamma^{5}) \nu_\alpha \right]
               \left[ \bar{f} \gamma_{\rho} (1 - \gamma^{5}) f \right],
  \label{eq:SM-Lagrangian-MSW}
\end{align}
where $\nu_\alpha$ is the neutrino field of flavor $\alpha$, $\ell_\alpha$
is the corresponding charged lepton field, and $f$, $f^\prime$ are the
components of an arbitrary weak doublet.

The low-energy fingerprint of many ``new physics'' scenarios has
a structure similar to Eqs.~\eqref{eq:SM-Lagrangian-CC} and~\eqref{eq:SM-Lagrangian-MSW},
and the corresponding operators are called non-standard interactions. If we
consider only lepton number conserving operators, the most general NSI Lagrangian
reads
\begin{align}
  \mathcal{L}_{\rm NSI} &= \mathcal{L}_{V \pm A} + \mathcal{L}_{S \pm P}
                            + \mathcal{L}_T,
\end{align}
where the different terms are classified according to their Lorentz structure
in the following way:
\begin{align}
  \mathcal{L}_{V \pm A} &= \nonumber\\
  &\hspace{-0.8cm}
  \frac{G_{F}}{\sqrt{2}} \sum_{f, f^\prime} \tilde{\eps}^{s,f,f^\prime, V \pm A}_{\alpha\beta}
      \left[ \bar{\nu}_\beta \gamma^{\rho} (1 - \gamma^{5}) \ell_\alpha \right] \!
      \left[ \bar{f}^\prime \gamma_{\rho} (1 \pm \gamma^{5}) f \right] + \nonumber\\
  &\hspace{-0.8cm}
  \frac{G_{F}}{\sqrt{2}} \sum_{f} \! \tilde{\eps}^{m,f, V \pm A}_{\alpha\beta} \!
      \left[ \bar{\nu}_\alpha \gamma^{\rho} (1 - \gamma^{5}) \nu_\beta \right] \!\!
      \left[ \bar{f} \gamma_{\rho} (1 \pm \gamma^{5}) f \right] + {\rm h.c.},
  \label{eq:NSI-Lagrangian-VpmA} \\
  \mathcal{L}_{S \pm P} &=
  \frac{G_{F}}{\sqrt{2}} \sum_{f, f^\prime} \tilde{\eps}^{s,f,f^\prime, S \pm P}_{\alpha\beta}
      \left[ \bar{\nu}_\beta (1 + \gamma^{5}) \ell_\alpha \right] \!
      \left[ \bar{f}^\prime (1 \pm \gamma^{5}) f \right],
  \label{eq:NSI-Lagrangian-SpmP} \\
  \mathcal{L}_{T} &=
  \frac{G_{F}}{\sqrt{2}} \sum_{f, f^\prime} \tilde{\eps}^{s,f,f^\prime,T}_{\alpha\beta}
      \left[ \bar{\nu}_\beta \sigma^{\rho\tau} \ell_\alpha \right] \!
      \left[ \bar{f}^\prime \sigma_{\rho\tau} f \right].
  \label{eq:NSI-Lagrangian-T}
\end{align}
Here, $G_F$, is the Fermi constant, $\nu$ and $\ell$ are the neutrino and charged
lepton fields, and the $f$'s represent the interaction partners of the neutrinos. The
dimensionless parameters $\tilde{\eps}$ give the strength of the non-standard interactions
relative to $G_F$, where an upper index $s$ stands for NSI in the neutrino source or
detector, while $m$ denotes non-standard matter effects, i.e.\ NSI affecting the
propagation. In general, the $\tilde{\eps}^s$ can be arbitrary complex matrices,
while the $\tilde{\eps}^m$ have to be hermitian.

Note that we have required the neutrino fields to be purely left-handed, since
processes involving right handed neutrinos would require either a neutrino helicity
flip, or their amplitudes would have to contain at least two NSI terms (e.g.\ one
to create the right-handed neutrino and one to absorb it), and would therefore be
strongly suppressed. This constraint on the neutrino chirality in particular
forbids $\nu\nu f f$ terms in $\mathcal{L}_{S \pm P}$ and $\mathcal{L}_{T}$.

Before proceeding, let us give a simple estimate which relates the magnitude
of the $\tilde{\eps}$ parameters to the corresponding new physics scale
$M_{\rm NSI}$~\cite{Gonzalez-Garcia:2001mp}: If
we assume the non-standard interactions to be mediated by some intermediate
particles with a mass of order $M_{\rm NSI}$, the effective vertices in
Eqs.~\eqref{eq:NSI-Lagrangian-VpmA} -- \eqref{eq:NSI-Lagrangian-T} will be suppressed
by $1 / M_{\rm NSI}^2$ in the same way as the standard weak interactions are suppressed
by $1 / M_{\rm W}^2$. Therefore we expect
\begin{align}
  |\tilde{\eps}| \sim \frac{M_{\rm W}^2}{M_{\rm NSI}^2}.
  \label{eq:epsilon-estimate}
\end{align}

\subsection{Relevance of the different NSI terms to reactor and superbeam experiments}

We see from Eqs.~\eqref{eq:NSI-Lagrangian-VpmA} -- \eqref{eq:NSI-Lagrangian-T}
that the number of possible NSI terms is very large. However, the number of
parameters for our discussion of reactor and superbeam experiments can be greatly
reduced by a few simple, but rather technical arguments. Many of these arguments
are based on constraints coming from the requirement of interference between the
standard and non-standard amplitudes. Of course, the total interaction rate will
also contain pure NSI terms, for which these constraints do not apply; but they
are suppressed by $\tilde{\eps}^2$, and can therefore be assumed to be negligible
compared to the interference terms, which are linear in $\tilde{\eps}$. The
following arguments are also summarized in Tab.~\ref{tab:Lorentz-structure}.

\begin{table*}
  \centering
  {\bf Reactor source and detector ($f=u$, $f^\prime=d$)} \\ \vspace{0.2 cm}
  \begin{ruledtabular}
  \begin{tabular}{l|c|c|c|c|c|c}
          & \multicolumn{3}{c|}{Source} & \multicolumn{3}{c}{Detector} \\
          & \parbox{2.7cm}{$\ell_\alpha=e$} & \parbox{2.7cm}{$\ell_\alpha=\mu$} & \parbox{2.7cm}{$\ell_\alpha=\tau$}
          & \parbox{2.7cm}{$\ell_\alpha=e$} & \parbox{2.7cm}{$\ell_\alpha=\mu$} & \parbox{2.7cm}{$\ell_\alpha=\tau$} \\\hline
    $V-A$ & \checkmark         & no $\mu$ production & no $\tau$ production & \checkmark         & no $\mu$ production & no $\tau$ production \\
    $V+A$ & \checkmark         & no $\mu$ production & no $\tau$ production & \checkmark         & no $\mu$ production & no $\tau$ production \\
    $S-P$ & strong constraints & no $\mu$ production & no $\tau$ production & strong constraints & no $\mu$ production & no $\tau$ production \\
    $S+P$ & strong constraints & no $\mu$ production & no $\tau$ production & strong constraints & no $\mu$ production & no $\tau$ production \\
    $T$   & strong constraints & no $\mu$ production & no $\tau$ production & strong constraints & no $\mu$ production & no $\tau$ production 
  \end{tabular}
  \end{ruledtabular}

  \vspace{0.5 cm}
  {\bf Superbeam source and detector ($f=u$, $f^\prime=d$)} \\ \vspace{0.2 cm}
  \begin{ruledtabular}
  \begin{tabular}{l|c|c|c|c|c|c}
          & \multicolumn{3}{c|}{Source} & \multicolumn{3}{c}{Detector} \\
          & \parbox{2.7cm}{$\ell_\alpha=e$} & \parbox{2.7cm}{$\ell_\alpha=\mu$} & \parbox{2.7cm}{$\ell_\alpha=\tau$}
          & \parbox{2.7cm}{$\ell_\alpha=e$} & \parbox{2.7cm}{$\ell_\alpha=\mu$} & \parbox{2.7cm}{$\ell_\alpha=\tau$} \\\hline
    $V-A$ & no $e$ production & \checkmark            & no $\tau$ production & \checkmark         & \checkmark   & no $\tau$ detection  \\
    $V+A$ & no $e$ production & \checkmark            & no $\tau$ production & \checkmark (mild supp.) & \checkmark (mild supp.) & no $\tau$ detection  \\
    $S-P$ & no $e$ production & \checkmark            & no $\tau$ production & strong constraints & chiral supp. & no $\tau$ detection  \\
    $S+P$ & no $e$ production & \checkmark            & no $\tau$ production & strong constraints & chiral supp. & no $\tau$ detection  \\
    $T$   & no $e$ production & no $P$-odd part       & no $\tau$ production & strong constraints & chiral supp. & no $\tau$ detection  
  \end{tabular}
  \end{ruledtabular}

  \vspace{0.5 cm}
  {\bf Propagation ($f = e, u, d$)} \\ \vspace{0.2 cm}
  \parbox{4 cm}{
  \begin{ruledtabular}
  \begin{tabular}{l|c@{\hspace{1.2cm}}}
    $V-A$ & \checkmark \\
    $V+A$ & \checkmark
  \end{tabular}
  \end{ruledtabular}
  }
  \caption{Classification of the vertices from Eqs.~\eqref{eq:NSI-Lagrangian-VpmA}
    -- \eqref{eq:NSI-Lagrangian-T} according to their impact on reactor and
    superbeam experiments. Terms marked with \checkmark\ can give a sizeable
    contribution; for all other terms, the reason for their suppression is given
    (see text for details).}
  \label{tab:Lorentz-structure}
\end{table*}

\begin{enumerate}
  \item The standard production and detection processes of reactor and superbeam
    neutrinos are, on the fundamental level, decays of $u$ quarks into $d$ quarks,
    or vice-versa. Since interference of standard and non-standard amplitudes
    requires the external particles to be identical, only the $\tilde{\eps}^{s,f,f^\prime}$
    terms with $f = u$, $f^\prime = d$ will be relevant, and we will henceforth
    simply omit the indices $f$ and $f^\prime$.

  \item For the non-standard matter effects, only coupling to electrons, up quarks, and
    down quarks is important.

  \item Non-standard couplings involving $\tau$ leptons are irrelevant since
    $\tau$ production is impossible in reactor and beam sources, and is not considered
    as a detection process here, although it might in principle be possible
    for high energy superbeam neutrinos. Hence we take
    \begin{align}
      \tilde{\eps}^{s,V \pm A}_{\tau\beta} = \tilde{\eps}^{s,S \pm P}_{\tau\beta}
                                           = \tilde{\eps}^{s,T}_{\tau\beta} = 0
    \end{align}
    For the same reason, processes involving muons can be neglected in reactor
    experiments, and processes involving electrons can be neglected in the superbeam
    source, since they constitute subdominant backgrounds even in the standard
    framework.

  \item In the couplings to muons, there is still room for non-$(V-A)(V-A)$ contributions.
    For neutrino production in pion decay, the effect of \mbox{$(S+P)(S \pm P)$} type
    NSI is even enhanced by a factor of~\cite{Herczeg:1995kd,Vainshtein:1975sv}
    \begin{align}
      \omega = \frac{m_{\pi}}{m_{\mu}} \frac{m_{\pi}}{m_{u} + m_{d}} \sim 20,
      \label{eq:chiral-enhancement}
    \end{align}
    and the importance of this enhancement for accelerator neutrino experiments has
    been pointed out in~\cite{Ota:2002na}. However, there exist limits on the muon
    helicity in pion decay~\cite{Fetscher:1984da,Yao:2006px}, which ensure that, in
    spite of the enhancement, \mbox{$(S+P)(S \pm P)$} type NSI cannot affect the
    neutrino oscillation amplitude by more than a few per cent.

  \item Tensor interactions are impossible in pion decay since the decay operator
    must have a parity-odd component.

  \item In the detection processes involving muons, the $(S+P)(S \pm P)$ and $TT$ terms
    are chirally suppressed by the smallness of $m_\mu$ compared
    to the typical superbeam energies of $\mathcal{O}(1\ {\rm GeV})$. As mentioned
    above, the leading effect in the total event rate is given by the interference
    of the non-standard amplitude and the standard $(V-A)(V-A)$ amplitude. This interference
    can only occur if the initial and final state particles have identical helicities,
    so for $(S+P)(S \pm P)$ and $TT$ type non-standard interactions, a mass-suppressed
    helicity flip of the muon is required. We can also see the emergence of the suppression
    factor explicitly by considering the Dirac traces, which have to be evaluated
    when calculating the cross section. For example, in the case of $(S+P)(S+P)$ NSI, the
    spin sum in the interference term of standard and non-standard amplitudes is
    \begin{align}
      \hspace{1.3cm} 
            &\sum_{\rm spins} {\rm Tr} \big[ \gamma^\rho (1 - \gamma^5) \mu \bar{\mu}
                    (1 - \gamma^5) \nu \bar{\nu} \big] \nonumber\\
            &\hspace{0.2cm} \cdot {\rm Tr}\big[ \gamma_\rho (1 - \gamma^5) u \bar{u}
                    (1 - \gamma^5) d \bar{d} \big] \nonumber\\
        = \ &\sum_{\rm spins} {\rm Tr} \big[ \gamma^\rho (1 - \gamma^5) (\slashed{p}_\mu + m_\mu)
                    (1 - \gamma^5) (\slashed{p}_\nu + m_\nu) \big] \nonumber\\
            &\hspace{0.2cm} \cdot {\rm Tr}\big[ \gamma_\rho (1 - \gamma^5) (\slashed{p}_u + m_u)
                    (1 - \gamma^5) (\slashed{p}_d + m_d) \big].
      \label{eq:chiral-supp}
    \end{align}
    Similar equations can be derived for $(S+P)(S-P)$ and $TT$ interactions.
    Due to the orthogonality property of the chirality projection operators,
    a contribution proportional to $m_\mu$ remains of the first trace in
    Eq.~\eqref{eq:chiral-supp}, and a contribution proportional to $m_u$ from
    the second. This leads to a suppression factor of $\mathcal{O}( m_\mu m_u / E^2)$.
    Low energy neutrinos ($E \lesssim 1$~GeV) interact with whole
    nucleons, rather than single quarks, therefore $m_u$ should be replaced by
    the much larger nucleon mass $m_n$, so that in this case, the overall
    chiral suppression is only of $\mathcal{O}(m_\mu/E)$. At typical superbeam
    energies around 1~GeV, we are in the transition regime between neutrino-nucleon
    interactions (quasielastic scattering and resonance scattering) and
    neutrino-quark interactions (deep-inelastic scattering)~\cite{Paschos:2000be,
    Paschos:2001np}.
    
  \item For $(V-A)(V+A)$ interactions involving muons in the detector, chiral
    suppression occurs only for the hadronic interaction partners, and
    according to our above discussion, it is not very pronounced for them.
    Therefore, $(V-A)(V+A)$ type interactions may in general be important
    for the cross sections, and modify their overall magnitude as well as
    their energy dependence.
    
  \item From measurements of the electron angular distribution in nuclear $\beta$
    decays, $(S+P)(S \pm P)$ and $TT$ couplings to electrons are strongly
    constrained~\cite{Lee:1956qn,Wu:1957my,Feynman:1958ty,Hardy:2004dm,
    Severijns:2006dr}. Consequently, we take
    \begin{align}
      \tilde{\eps}^{s,S \pm P}_{e\beta} = \tilde{\eps}^{s,T}_{e\beta} = 0
    \end{align}
    for $\beta = e,\mu,\tau$.
    
  \item There is still room for $(V-A)(V+A)$ type terms involving electrons,
    because these terms differ from the standard model term only in the quark current,
    which cannot be directly measured. Limits exist only for the effective vector and
    axial-vector couplings of protons and neutrons~\cite{Severijns:2006dr}, but due
    to the non-perturbative nature of the strong interactions, these cannot be easily
    related to the couplings of the fundamental quarks.
    
    If $(V-A)(V+A)$ couplings to electrons exist, the processes in which they appear will
    in general have an energy dependence different from that of the corresponding
    standard processes. For anti-neutrino production in nuclear reactors, however, this
    difference is completely negligible because the neutrino spectrum from nuclear $\beta$
    decay is governed by kinematical effects and by the Fermi function, which describes
    final state Coulomb interactions.
    
    The cross section for the inverse $\beta$ decay process, which is used to detect reactor
    anti-neutrinos, is derived from empirical values for the effective vector and axial-vector
    couplings, so any possible $(V-A)(V+A)$ contribution is automatically taken into
    account properly.

    Finally, $(V-A)(V+A)$ interactions involving electrons in the beam detector, are
    midly chirally suppressed, in analogy to $(V-A)(V+A)$ interactions involving muons.

  \item As we have seen in Eqs.~\eqref{eq:NSI-Lagrangian-VpmA} --
    \eqref{eq:NSI-Lagrangian-T}, non-standard matter effects can only have a
    $(V-A)(V-A)$ or $(V-A)(V+A)$ Lorentz structure, as long as we restrict
    the discussion to left-handed neutrinos. For the computation of the
    coherent forward scattering amplitude, the factor
    $\left[ \bar{f} \gamma_{\rho} (1 \pm \gamma^{5}) f \right]$ has to be
    averaged over the neutrino trajectory, and for unpolarized matter at
    rest, the only contribution is $N_f = \bar{f} \gamma^0 f$, the
    fermion density appearing in the matter potential. Since $N_f$ is
    independent of the axial current, we conclude that both possible
    Lorentz structures would have the same impact on the non-standard
    matter effects.
\end{enumerate}
To conclude this discussion, we would like to emphasize again that non-$(V-A)(V-A)$
Lorentz structures can play an important role in reactor and superbeam experiments.
However, these experiments do not have the capability to distinguish different
Lorentz structures, unless the spectral distortion caused by $(V-A)(V+A)$ terms
in the superbeam detector is taken into account. In the following, we will
neglect this spectral distortion for simplicity, assuming that it is anyway
hidden by the systematical uncertainties in the neutrino cross sections.

\subsection{Hamiltonian approach to non-standard interactions in neutrino oscillations}
\label{sec:Hamiltonian}

The obsevrations from the previous section can be exploited to further reduce
the number of free parameters in our problem. To this end, we define effective
couplings $\eps^s_{\alpha\beta}$, $\eps^d_{\alpha\beta}$, and $\eps^m_{\alpha\beta}$,
corresponding to non-standard interactions in the production, detection,
and propagation processes. $\eps^s_{\alpha\beta}$ describes a non-standard admixture
of flavor $\beta$ to the neutrino state which is produced in association with a
charged lepton of flavor $\alpha$. This means, that the neutrino source does not
produce a pure flavor neutrino eigenstate $\ket{\nu_\alpha}$, but rather a state
\begin{align}
  \ket{\nu^s_\alpha} = \ket{\nu_\alpha}
    + \sum_{\beta=e,\mu,\tau} \eps^s_{\alpha\beta} \ket{\nu_\beta}.
\end{align}
Similarly, the detector is sensitive not to the normal weak eigenstates,
but to the combination
\begin{align}
  \bra{\nu^d_\beta} = \bra{\nu_\beta}
    + \sum_{\alpha=e,\mu,\tau} \eps^d_{\alpha\beta} \bra{\nu_\alpha}.
\end{align}
Note that in $\eps^s_{\alpha\beta}$, the first index corresponds to the flavor
of the charged lepton, and the second to that of the neutrino, while in
$\eps^d_{\alpha\beta}$, the order is reversed. We have chosen this convention
to be consistent with the literature.

In general, the matrices $(1+\eps^s)$ and $(1+\eps^d)$ are non-unitary, i.e.\
the source and detection states are not required to form complete orthonormal
sets of basis vectors in the Hilbert space:
\begin{align}
  \sum_{\alpha = e,\mu,\tau} \ket{\nu^{s}_{\alpha}} \bra{\nu^{s}_{\alpha}}
          &\neq 1, &
  \sum_{\beta = e,\mu,\tau} \ket{\nu^{d}_{\beta}} \bra{\nu^{d}_{\beta}}
          &\neq 1, \\
  \sprod{\nu^{s}_{\alpha}}{\nu^{s}_{\beta}}
          &\neq \delta_{\alpha\beta} &
  \sprod{\nu^{d}_{\alpha}}{\nu^{d}_{\beta}}
          &\neq \delta_{\alpha\beta}.
\end{align}

We can read off from Tab.~\ref{tab:Lorentz-structure} that
the $3\times3$ coupling matrix $\eps^s_{\alpha\beta}$ receives contributions
from $\tilde{\eps}^{s,u,d,V \pm A}_{\alpha\beta}$ and
$\tilde{\eps}^{s,u,d,S \pm P}_{\alpha\beta}$, while $\eps^d_{\alpha\beta}$
and $\eps^m_{\alpha\beta}$ are built up only from $(V-A)(V \pm A)$
contributions.

Since the coefficients $\eps^s_{e\alpha}$ and $\eps^d_{\alpha e}$ (for
$\alpha = e,\mu,\tau$) both originate from $\tilde{\eps}^{s,u,d,V \pm A}$, the
$(V - A)(V \pm A)$ coupling to up and down quarks, we have the constraint
\begin{align}
  \eps^s_{e\alpha} = \eps^{d*}_{\alpha e},
  \label{eq:eps-e-constraint}
\end{align}
which again reduces the number of independent parameters by 3. The aforementioned
spectral distortion and mild chiral suppression in the superbeam detector could
invalidate Eq.~\eqref{eq:eps-e-constraint}, but we will neglect it in the following.

Similarly, the $(V - A)(V \pm A)$ part of $\eps^s_{\mu\alpha}$ and
$\eps^{d}_{\alpha\mu}$ are the same, and since the $(S+P)(S \pm P)$ Lorentz
structures have less impact in $\eps^d_{\alpha\mu}$ than in
$\eps^{s}_{\mu\alpha}$, we will typically have
\begin{align}
  |\eps^s_{\mu\alpha}| \gtrsim |\eps^d_{\alpha\mu}|
  \label{eq:eps-mu-constraint}
\end{align}
(barring fine-tuned cancellation effects). If we assume all non-standard
interactions to be of the $(V-A)(V-A)$ type, as is sometimes done in the
literature, the constraints from Eqs.~\eqref{eq:eps-e-constraint} and
\eqref{eq:eps-mu-constraint} are tightened to $\eps^s = (\eps^d)^\dag$.

It is clear from Tab.~\ref{tab:Lorentz-structure} that coupling to $\tau$
leptons is irrelevant in our case, so we can also take
\begin{align}
  \eps^s_{\tau\beta} = \eps^d_{\alpha\tau} = 0
\end{align}
for all $\alpha$, $\beta$.

$\eps^m$ is an additive contribution to the Mikheyev-Smirnov-Wolfenstein (MSW)
potential in the flavor basis, $V_{\rm MSW} = a_{\rm CC} \, \diag(1, 0, 0)$,
which now becomes
\begin{align}
  \tilde{V}_{\rm MSW} = a_{\rm CC}
  \begin{pmatrix}
    1 + \eps^m_{ee}       & \eps^m_{e\mu}       & \eps^m_{e\tau}  \\
        \eps^{m*}_{e\mu}  & \eps^m_{\mu\mu}     & \eps^m_{\mu\tau} \\
        \eps^{m*}_{e\tau} & \eps^{m*}_{\mu\tau} & \eps^m_{\tau\tau}
  \end{pmatrix},
  \label{eq:V-MSW-NSI}
\end{align}
with $a_{\rm CC} = 2 \sqrt{2} G_F N_e E$.
Recall from Eq.~\eqref{eq:NSI-Lagrangian-VpmA} that the diagonal entries in
this matrix have to be real, so that the Hamiltonian will remain hermitian,
and can be diagonalized by a unitary mixing matrix.

Since we are interested in a combined analysis of reactor and superbeam experiments,
it is important to keep in mind that the effective $\eps$ matrices are \emph{the
same} for both types of experiments, because, under the assumptions and approximations
discussed above, those entries which may be relevant in both of them ($\eps^m_{\alpha\beta}$
and $\eps^d_{\alpha e}$) are identical in both cases.

The oscillation probability is obtained as
\begin{align}
  P_{\nu^s_\alpha \rightarrow \nu^d_\beta}
    &= |\bra{\nu^d_\beta} e^{-i H L} \ket{\nu^s_\alpha}|^2 \nonumber\\
    &= \big| (1 + \eps^d)_{\gamma\beta} \, \big( e^{-i H L} \big)_{\gamma\delta}
             (1 + \eps^s)_{\alpha\delta} \big|^2           \nonumber\\
    &= \Big| \Big[ \big( 1 + \eps^d \big)^T \,\, e^{-i H L} \,\,
                   \big( 1 + \eps^s \big)^T \Big]_{\beta\alpha} \Big|^2,
  \label{eq:P-ansatz}
\end{align}
where
\begin{widetext}
\begin{align}
  H_{\alpha\beta} &= \frac{1}{2E} \left[
           U_{\alpha j} \begin{pmatrix}
                          0 &      & \\
                            & \sdm & \\
                            &      & \ldm
                        \end{pmatrix}_{jk}
          (U^\dag)_{k \beta} + (\tilde{V}_{\rm MSW})_{\alpha\beta}
       \right].
  \label{eq:H-ansatz}
\end{align}
The Pontecorvo-Maki-Nakagawa-Sakata (PMNS) matrix $U$ is parameterized as
\begin{align}
  U = \begin{pmatrix}
        c_{12} c_{13}  &  s_{12} c_{13}  &  s_{13} e^{-i\delta_{\rm CP}}  \\
        -s_{12} c_{23} - c_{12} s_{13} s_{23} e^{i\delta_{\rm CP}} &
           c_{12} c_{23} - s_{12} s_{13} s_{23} e^{i\delta_{\rm CP}} & c_{13} s_{23} \\
        s_{12} s_{23} - c_{12} s_{13} c_{23} e^{i\delta_{\rm CP}} &
           -c_{12} s_{23} - s_{12} s_{13} c_{23} e^{i\delta_{\rm CP}} & c_{13} c_{23}
      \end{pmatrix}.
  \label{eq:UPMNS}
\end{align}
\end{widetext}
As usual, $s_{ij}$ and $c_{ij}$ denote the sine and cosine of the mixing angle
$\theta_{ij}$, and $\delta_{\rm CP}$ is the (Dirac) CP phase.

For anti-neutrinos, we have to replace $\eps^s$, $\eps^d$, and $\eps^m$ by their
complex conjugates in the above equations, and reverse the signs of $a_{\rm CC}$
and $\delta_{\rm CP}$.

\subsection{Perturbative calculation of oscillation probabilities}
\label{sec:perturbative}

In practice, it is very convenient to expand the oscillation probabilities in a
perturbative series with respect to the small quantities $\theta_{13}$, $\sdm/\ldm$,
and $|\eps^{s,m,d}_{\alpha\beta}|$ instead of attempting to evaluate
Eq.~\eqref{eq:P-ansatz} exactly. Following a procedure similar to the one explained
in the appendix of~\cite{Akhmedov:2004ny}, the first order expansion reads
\begin{align}
  P_{\nu^s_\alpha \rightarrow \nu^d_\beta}
    &=   P_{\nu^s_\alpha \rightarrow \nu^d_\beta}^{(0)}
       + P_{\nu^s_\alpha \rightarrow \nu^d_\beta}^{(1)}
       + \ldots,
\intertext{where}
  P_{\nu^s_\alpha \rightarrow \nu^d_\beta}^{(0)}
    &=   \Big| \Big[ e^{-i H^{(0)} L} \Big]_{\beta\alpha} \Big|^2,        \\
  P_{\nu^s_\alpha \rightarrow \nu^d_\beta}^{(1)}
    &=   \Big[ e^{-i H^{(0)} L} \Big]^*_{\beta\alpha}
         \Big[ e^{-i H^{(0)} L} \big( 1 + \eps^s \big)^T \Big]_{\beta\alpha}
                                                                 \nonumber\\
    &\hspace{-0.8cm}
       + \Big[ e^{-i H^{(0)} L} \Big]^*_{\beta\alpha}
         \Big[ \big( 1 + \eps^d \big)^T e^{-i H^{(0)} L} \Big]_{\beta\alpha}
                                                                 \nonumber\\
    &\hspace{-0.8cm}
       -i \Big[ e^{-i H^{(0)} L} \Big]^*_{\beta\alpha}
          \bigg[ \int_0^L dx \, e^{-i H^{(0)} (L-x)}
                 H^{(1)} e^{-i H^{(0)} x} \bigg]_{\beta\alpha}   \nonumber\\
    &\hspace{-0.8cm}
       + {\rm h.c.},
\intertext{and}
  H_{\alpha\beta}^{(0)}
    &= \frac{1}{2E} \left[
       U^{(0)} \begin{pmatrix}
                  0 &   & \\
                    & 0 & \\
                    &   & \ldm
               \end{pmatrix} U^{(0)\dag} \right.                 \nonumber\\
  &\hspace{2cm} \left.
        + a_{\rm CC} \begin{pmatrix}
                       1 &   & \\
                         & 0 & \\
                         &   & 0
                     \end{pmatrix}
        \right],                                                          \\
  H_{\alpha\beta}^{(1)}
    &= \frac{1}{2E} \left[
       U^{(0)} \begin{pmatrix}
                  0 &      & \\
                    & \sdm & \\
                    &      & 0
                \end{pmatrix} U^{(0)\dag} \right.                \nonumber\\
  &\hspace{2.0cm}
        + U^{(1)} \begin{pmatrix}
                  0 &   & \\
                    & 0 & \\
                    &   & \ldm
                \end{pmatrix} U^{(0)\dag}                        \nonumber\\
  &\hspace{2.0cm}
        + U^{(0)} \begin{pmatrix}
                  0 &   & \\
                    & 0 & \\
                    &   & \ldm
                \end{pmatrix} U^{(1)\dag}                        \nonumber\\
  &\hspace{2.0cm} \left.
        + a_{\rm CC} \begin{pmatrix}
                       \eps^m_{ee}       & \eps^m_{e\mu}       & \eps^m_{e\tau}   \\
                       \eps^{m*}_{e\mu}  & \eps^m_{\mu\mu}     & \eps^m_{\mu\tau} \\
                       \eps^{m*}_{e\tau} & \eps^{m*}_{\mu\tau} & \eps^m_{\tau\tau}
                     \end{pmatrix}
        \right].
\end{align}
By $U^{(0)}$, we denote the PMNS matrix for $\theta_{13} = 0$, and $U^{(1)}$
contains the first order terms in $\theta_{13}$. The unperturbed Hamiltonian,
$H_{\alpha\beta}^{(0)}$, can be easily diagonalized exactly, so that the matrix
exponentials in the above equations can be evaluated. It is straightforward to
extend the expansion to higher orders.

\section{Modified neutrino oscillation probabilities for reactor and superbeam
         experiments}
\label{sec:theory}

To study the impact of non-standard interactions on reactor and superbeam
experiments, we need in particular the oscillation probabilities
$P_{\bar{\nu}^s_e \rightarrow \bar{\nu}^d_e}$, $P_{\nu^s_\mu \rightarrow \nu^d_e}$,
and $P_{\nu^s_\mu \rightarrow \nu^d_\mu}$, to which these experiments are
sensitive. We will first present approximate analytic formulas for these
quantities in Secs.~\ref{sec:ee} to~\ref{sec:mumu}, and then discuss
them in Sec.~\ref{sec:interpretation}. All approximations were carried out
with the perturbative method described in the previous section.
We have checked that the expressions presented in this section
reduce to the well-known standard oscillation results if NSI are absent
by comparing them to the expressions derived in~\cite{Arafune:1997hd,Ota:2000hf,
Akhmedov:2004ny}. Moreover, we have verified all formulas numerically,
term by term, using Mathematica.

To simplify the notation, let us make the abbreviations $s_{ij} =  \sin\theta_{ij}$,
$c_{ij} = \cos\theta_{ij}$, $s_{2\times ij} = \sin 2\theta_{ij}$, and
$c_{2\times ij} = \cos 2\theta_{ij}$. Moreover, it will be convenient
to split the $\eps$ parameters into their real and imaginary parts by writing
$\eps^{s,m,d}_{\alpha\beta} = |\eps^{s,m,d}_{\alpha\beta}|
\exp(i \phi^{s,m,d}_{\alpha\beta})$. To keep our results as general as
possible, we will \emph{not} impose the constraint from
Eq.~\eqref{eq:eps-e-constraint}, but treat $\eps^s$ and $\eps^d$ as
completely independent matrices. Thus, our formulas will be also applicable
to experiments with fundamentally different production and detection processes,
e.g.\ to a neutrino factory, where the production occurs through a purely leptonic
$\nu\nu e \mu$ vertex, while the detection process $\nu\ell u d$ involves
coupling to quarks. For reactors and superbeams, it is, of course,
straightforward to impose Eq.~\eqref{eq:eps-e-constraint} a posteriori.

\subsection{The $\bar{\nu}_e \rightarrow \bar{\nu}_e$ channel}
\label{sec:ee}

In a reactor experiment, the ratio $L/E$ is chosen close to the first atmospheric
oscillation maximum, so we can safely neglect terms proportional to $\sdm/\ldm$.
Moreover, matter effects are irrelevant, i.e.\ we can take $a_{\rm CC} = 0$.
Finally, we will neglect terms suppressed by $s_{13}^3$,
$\eps s_{13}^2$, or $\eps^2$. The last approximation implies
that we only consider the interference terms between standard and non-standard
contributions, but not the pure, incoherent, NSI effect. This is justified
for most realistic extensions of the Standard Model, where $\eps \ll 1$, but
it has been pointed out in~\cite{Kitazawa:2006iq} that, from current model-independent
experimental limits, the NSI might even dominate over the standard oscillations in some
situations. We find for the oscillation probability
\begin{align}
  P_{\bar{\nu}^s_e \rightarrow \bar{\nu}^d_e} &=
        1 - 4 s_{13}^2 \sin^2 \frac{\ldm L}{4E}                         \nonumber\\
    &\hspace{0.5cm}
      + 2 |\eps^s_{ee}| \cos\phi^s_{ee} + 2 |\eps^d_{ee}|\cos\phi^d_{ee}\nonumber\\
    &\hspace{0.5cm}
      - 4 |\eps^s_{e\mu}| s_{13} s_{23}
        \cos(\delta_{\rm CP} - \phi^s_{e\mu})  \sin^2 \frac{\ldm L}{4E} \nonumber\\
    &\hspace{0.5cm}
      + 2 |\eps^s_{e\mu}| s_{13} s_{23}
        \sin(\delta_{\rm CP} - \phi^s_{e\mu}) \sin \frac{\ldm L}{2E}    \nonumber\\
    &\hspace{0.5cm}
      - 4 |\eps^s_{e\tau}| s_{13} c_{23}
        \cos(\delta_{\rm CP} - \phi^s_{e\tau}) \sin^2 \frac{\ldm L}{4E} \nonumber\\
    &\hspace{0.5cm}
      + 2 |\eps^s_{e\tau}| s_{13} c_{23} 
        \sin(\delta_{\rm CP} - \phi^s_{e\tau}) \sin \frac{\ldm L}{2E}   \nonumber\\
    &\hspace{0.5cm}
      - 4 |\eps^d_{\mu e}| s_{13} s_{23}
        \cos(\delta_{\rm CP} + \phi^d_{\mu e}) \sin^2 \frac{\ldm L}{4E} \nonumber\\
    &\hspace{0.5cm}
      - 2 |\eps^d_{\mu e}| s_{13} s_{23}
        \sin(\delta_{\rm CP} + \phi^d_{\mu e}) \sin \frac{\ldm L}{2E}   \nonumber\\
    &\hspace{0.5cm}
      - 4 |\eps^d_{\tau e}| s_{13} c_{23}
        \cos(\delta_{\rm CP} + \phi^d_{\tau e}) \sin^2 \frac{\ldm L}{4E}\nonumber\\
    &\hspace{0.5cm}
      - 2 |\eps^d_{\tau e}| s_{13} c_{23}
        \sin(\delta_{\rm CP} + \phi^d_{\tau e}) \sin \frac{\ldm L}{2E}  \nonumber\\
    &\hspace{0.5cm}
      + \mathcal{O} \Big( \frac{\sdm}{\ldm} \Big)
      + \mathcal{O} ( \eps s_{13}^2 )
      + \mathcal{O} ( s_{13}^3 )
      + \mathcal{O} ( \eps^2 ).
  \label{eq:Pee}
\end{align}
It is interesting to remark that, due to the $\eps^s_{ee}$ and $\eps^d_{ee}$
terms, this expression can be different from unity even for $\ldm L/4E \ll 1$,
i.e.\ at the near detector (ND) site. Indeed, we obtain in this case
\begin{align}
  P_{\bar{\nu}^s_e \rightarrow \bar{\nu}^d_e}^{\rm ND} &=
     1 + 2 |\epsilon^{s}_{ee}| \cos \phi^{s}_{ee}
       + 2 |\epsilon^{d}_{ee}| \cos \phi^{d}_{ee}
       + | \epsilon^{s}_{ee} |^{2} + | \epsilon^{d}_{ee} |^{2} \nonumber\\
    &\hspace{0.5cm}
       + 2 |\epsilon^{s}_{ee}| |\epsilon^{d}_{ee}|
         \big[
             \cos(\phi^{s}_{ee} + \phi^{d}_{ee}) 
           + \cos(\phi^{s}_{ee} - \phi^{d}_{ee})
         \big]                                                 \nonumber\\
    &\hspace{0.5cm}
       + 2 |\epsilon^{s}_{e\mu}| |\epsilon^{d}_{\mu e}|
         \cos(\phi^{s}_{\mu e} + \phi^{d}_{e\mu})              \nonumber\\
    &\hspace{0.5cm}
       + 2 |\epsilon^{s}_{e\tau}| |\epsilon^{d}_{\tau e}|
         \cos(\phi^{s}_{\tau e} + \phi^{d}_{e\tau})            \nonumber\\
    &\hspace{0.5cm}
       + \mathcal{O}\Big( \frac{\ldm L}{4 E} \Big)
       + \mathcal{O} ( \eps^3 ),
  \label{eq:Pee-ND}
\end{align}
where we have taken into account also second order terms in $\eps$, which may be
important in the near detector due to the large event rates. Eq.~\eqref{eq:Pee-ND}
corresponds to an overall rescaling of the neutrino flux, which, however, will be
hard to detect in a realistic experiment due to the systematical flux uncertainty.

\subsection{The $\nu_\mu \rightarrow \nu_e$ channel}
\label{sec:mue}

In the derivation of $P_{\nu^s_\mu \rightarrow \nu^d_e}$, we will relax
our approximations from the previous section, and take into account also terms
of $\mathcal{O}(s_{13} \sdm/\ldm)$, $\mathcal{O}([\sdm/\ldm]^2)$, and
$\mathcal{O}(\eps \sdm/\ldm)$, to reproduce the correct $\delta_{\rm CP}$
dependence. For experiments with a relatively short baseline, such as \TtoK,
it is justified to assume vacuum oscillations, if the $\eps^m$ parameters are
$\lesssim \mathcal{O}(0.1)$ (in Sec.~\ref{sec:th13fits}, we will discuss cases
where this is not true, and we will see that non-standard matter effects can
then be large in \TtoK). The vacuum oscillation probability reads
\begin{align}
  P_{\nu^s_\mu \rightarrow \nu^d_e}^{\rm vac} &=
      4 s_{13}^{2} s_{23}^{2} \sin^{2} \frac{\Delta m_{31}^{2} L}{4E}       \nonumber\\
  &\hspace{-0.7cm}
    + \Big( \frac{\sdm}{\ldm} \Big)^2 c_{23}^2 s_{2 \times 12}^2
      \Big( \frac{\Delta m_{31}^{2} L}{4E} \Big)^2                          \nonumber\\
  &\hspace{-0.7cm}
    + \frac{\sdm}{\ldm} s_{13} s_{2 \times 12} s_{2 \times 23} \cos\delta_{\rm CP}
      \frac{\Delta m_{31}^{2} L}{4E} \sin \frac{\Delta m_{31}^{2} L}{2E}
                                                                            \nonumber\\
  &\hspace{-0.7cm}
    - 2 \frac{\sdm}{\ldm} s_{13} s_{2 \times 12} s_{2 \times 23} \sin\delta_{\rm CP}
      \frac{\Delta m_{31}^{2} L}{4E} \sin^{2} \frac{\Delta m_{31}^{2} L}{4E}\nonumber\\
  &\hspace{-0.7cm}
    - 4  |\epsilon^{s}_{\mu e}| s_{13} s_{23}
      \cos(\phi^{s}_{\mu e} + \delta_{\rm CP}) \sin^{2} \frac{\ldm L}{4E}   \nonumber\\
  &\hspace{-0.7cm}
    - 2 |\epsilon^{s}_{\mu e}| s_{13} s_{23}
      \sin(\phi^{s}_{\mu e} + \delta_{\rm CP}) \sin \frac{\ldm L}{2E}       \nonumber\\
  &\hspace{-0.7cm}
    - 4 |\epsilon^{d}_{\mu e}| s_{13} c_{2 \times 23} s_{23}
      \cos(\phi^{d}_{\mu e} + \delta_{\rm CP}) \sin^{2} \frac{\ldm L}{4E}   \nonumber\\
  &\hspace{-0.7cm}
    - 2 |\epsilon^{d}_{\mu e}| s_{13} s_{23}
      \sin(\phi^{d}_{\mu e} + \delta_{\rm CP}) \sin \frac{\ldm L}{2E}       \nonumber\\
  &\hspace{-0.7cm}
    + 4 |\epsilon^{d}_{\tau e}| s_{13} s_{2 \times 23} s_{23} 
      \cos(\phi^{d}_{\tau e} + \delta_{\rm CP}) \sin^{2} \frac{\ldm L}{4E}  \nonumber\\
  &\hspace{-0.7cm}
    - |\epsilon^{s}_{\mu e}| \frac{\sdm}{\ldm} s_{2 \times 12} c_{23}
      \sin\phi^{s}_{\mu e} \frac{\ldm L}{2E}                                \nonumber\\
  &\hspace{-0.7cm}
    + 2 |\epsilon^{d}_{\mu e}| \frac{\sdm}{\ldm} s_{2 \times 12} s_{23}^2 c_{23} 
      \cos \phi^{d}_{\mu e} \frac{\ldm L}{4E} \sin \frac{\ldm L}{2E}        \nonumber\\
  &\hspace{-0.7cm}
    - |\epsilon^{d}_{\mu e}| \frac{\sdm}{\ldm} s_{2 \times 12} c_{23}
      \sin \phi^{d}_{\mu e} \frac{\ldm L}{2E}                               \nonumber\\
  &\hspace{4.0cm} \cdot
      \left[1 - 2 s_{23}^2 \sin^2\frac{\ldm L}{2E} \right]                  \nonumber\\
  &\hspace{-0.7cm}
    + 2 |\epsilon^{d}_{\tau e}| \frac{\sdm}{\ldm} s_{2 \times 12} s_{23} c_{23}^2
      \cos \phi^{d}_{\mu e} \frac{\ldm L}{4E} \sin \frac{\ldm L}{2E}        \nonumber\\
  &\hspace{-0.7cm}
    + 2 |\epsilon^{d}_{\tau e}| \frac{\sdm}{\ldm} s_{2 \times 12} s_{23} c_{23}^2
      \sin \phi^{d}_{\mu e} \frac{\ldm L}{2E} \sin^2\frac{\ldm L}{4E}       \nonumber\\
  &\hspace{-0.7cm}
    + \mathcal{O}\Big( \Big[ \frac{\sdm}{\ldm} \Big]^3 \Big)
    + \mathcal{O}\Big( \Big[ \frac{\sdm}{\ldm} \Big]^2 s_{13} \Big)
    + \mathcal{O}\Big( \frac{\sdm}{\ldm} s_{13}^2 \Big)                     \nonumber\\
  &\hspace{-0.7cm}
    + \mathcal{O} ( s_{13}^3 )
    + \mathcal{O}\Big( \eps \Big[ \frac{\sdm}{\ldm} \Big]^2 \Big)
    + \mathcal{O}\Big( \eps \tilde{s}_{13} \frac{\sdm}{\ldm} \Big)          \nonumber\\
  &\hspace{-0.7cm}
    + \mathcal{O} ( \eps s_{13}^2 )
    + \mathcal{O} ( \eps^2 ).
  \label{eq:Pmue-vac}
\end{align}
The corresponding expression for the near detector is
\begin{align}
  P_{\nu^s_\mu \rightarrow \nu^d_e}^{\rm vac, ND} &=
    |\eps^s_{\mu e}|^2 + |\eps^d_{\mu e}|^2
    + 2 |\eps^s_{\mu e}| |\eps^d_{\mu e}| \cos(\phi^s_{\mu e} - \phi^d_{\mu e})
                                                                 \nonumber\\
  &\hspace{0.5cm}
    + \mathcal{O}\Big( \frac{\ldm L}{4 E} \Big)
    + \mathcal{O} ( \eps^3 ).
\end{align}
If the baseline is longer, as is the case e.g.\ in \NOvA, matter effects
are important. To keep the notation concise in this case, we define the
effective 13-mixing angle in matter, which is given to lowest order by
\begin{align}
  \tilde{s}_{13} \equiv \frac{\ldm}{\ldm - a_{\rm CC}} s_{13} + \mathcal{O}(s_{13}^2).
\end{align}
The oscillation probability is then
\begin{widetext}
\begin{align}
  P_{\nu^s_\mu \rightarrow \nu^d_e}^{\rm mat} &=
      4 \tilde{s}_{13}^{2} s_{23}^{2} \sin^{2} \frac{(\ldm - a_{\rm CC})L}{4E}
                                                         \nonumber\\
  &\hspace{0.5cm}
    + \Big( \frac{\sdm}{\ldm} \Big)^2 c_{23}^2 s_{2 \times 12}^2
      \Big( \frac{\ldm}{a_{\rm CC}} \Big)^2 \sin^{2} \frac{a_{\rm CC} L}{4E}
                                                         \nonumber\\
  &\hspace{0.5cm}
    - \frac{\sdm}{\ldm} \tilde{s}_{13} s_{2 \times 12} s_{2 \times 23} \cos\delta_{\rm CP}
      \frac{\ldm}{a_{\rm CC}}
      \left[   \sin^{2} \frac{a_{\rm CC} L}{4E}
             - \sin^{2} \frac{\ldm L}{4E}
             + \sin^{2} \frac{(\ldm - a_{\rm CC})L}{4E}
      \right]                                           \nonumber\\
  &\hspace{0.5cm}
  - \frac{1}{2} \frac{\sdm}{\ldm} \tilde{s}_{13} s_{2 \times 12} s_{2 \times 23} \sin\delta_{\rm CP}
      \frac{\ldm}{a_{\rm CC}}
      \left[   \sin \frac{a_{\rm CC} L}{2E} 
             - \sin \frac{\ldm L}{2E}
             + \sin \frac{(\ldm - a_{\rm CC})L}{2E}
      \right]                                           \nonumber\\
  &\hspace{0.5 cm}
    - 4 |\epsilon^{s}_{\mu e}| \tilde{s}_{13} s_{23} \cos(\phi^{s}_{\mu e} + \delta_{\rm CP})
      \sin^{2} \frac{(\ldm - a_{\rm CC})L}{4E}  \nonumber\\
  &\hspace{0.5 cm}
    - 2 |\epsilon^{s}_{\mu e}| \tilde{s}_{13} s_{23} \sin(\phi^{s}_{\mu e} + \delta_{\rm CP})
      \sin \frac{(\ldm - a_{\rm CC})L}{2E}      \nonumber\\
  &\hspace{0.5 cm}
    + 4 |\epsilon^{d}_{\mu e}| \tilde{s}_{13} s_{23} \cos(\phi^{d}_{\mu e} + \delta_{\rm CP})
      \left[   c_{23}^{2} \sin^{2} \frac{a_{\rm CC}L}{4E}
             - c_{23}^{2} \sin^{2} \frac{\ldm L}{4E}
             + s_{23}^{2} \sin^{2} \frac{(\ldm - a_{\rm CC})L}{4E}
      \right]                                           \nonumber\\
  &\hspace{0.5 cm}
    + 2 |\epsilon^{d}_{\mu e}| \tilde{s}_{13} s_{23} \sin(\phi^{d}_{\mu e} + \delta_{\rm CP})
      \left[   c_{23}^{2} \sin \frac{a_{\rm CC}L}{2E}
             - c_{23}^{2} \sin \frac{\ldm L}{2E}
             - s_{23}^{2} \sin \frac{(\ldm - a_{\rm CC})L}{2E}
      \right]                                           \nonumber\\
  &\hspace{0.5 cm}
    - 4 |\epsilon^{d}_{\tau e}| \tilde{s}_{13} s_{23}^{2} c_{23}
      \cos(\phi^{d}_{\tau e} + \delta_{\rm CP})
      \left[   \sin^{2} \frac{a_{\rm CC}L}{4E}
             - \sin^{2} \frac{\ldm L}{4E}
             - \sin^{2} \frac{(\ldm - a_{\rm CC})L}{4E}
      \right]                                           \nonumber\\
  &\hspace{0.5 cm}
    - 2 |\epsilon^{d}_{\tau e}| \tilde{s}_{13} s_{23}^{2} c_{23}
      \sin(\phi^{d}_{\tau e} + \delta_{\rm CP})
      \left[   \sin \frac{a_{\rm CC}L}{2E}
             - \sin \frac{\ldm L}{2E}
             + \sin \frac{(\ldm - a_{\rm CC})L}{2E}
      \right]                                           \nonumber\\
  &\hspace{0.5 cm}
    - 4 |\epsilon^{m}_{e\mu}| \tilde{s}_{13} s_{23} c_{23}^{2}
      \cos(\phi^{m}_{e\mu} + \delta_{\rm CP})
      \left[   \sin^{2} \frac{a_{\rm CC} L}{4E}
             - \sin^{2} \frac{\ldm L}{4E}
             + \sin^{2} \frac{(\ldm - a_{\rm CC})L}{4E}
      \right]                                           \nonumber\\
  &\hspace{0.5 cm}
    - 2 |\epsilon^{m}_{e\mu}| \tilde{s}_{13} s_{23} c_{23}^{2}
      \sin(\phi^{m}_{e\mu} + \delta_{\rm CP})
      \left[   \sin \frac{a_{\rm CC} L}{2E} 
             - \sin \frac{\ldm L}{2E}
             + \sin \frac{(\ldm - a_{\rm CC})L}{2E}
      \right]                                           \nonumber\\
  &\hspace{0.5 cm}
    + 8 |\epsilon^{m}_{e\mu}| \tilde{s}_{13} s_{23}^{3}
      \cos(\phi^{m}_{e\mu} + \delta_{\rm CP}) \frac{a_{\rm CC}}{\ldm - a_{\rm CC}}
      \sin^{2} \frac{(\ldm - a_{\rm CC})L}{4E} \nonumber \\
  &\hspace{0.5 cm}
    + 4 |\epsilon^{m}_{e\tau}| \tilde{s}_{13} s_{23}^{2} c_{23}
      \cos(\phi^{m}_{e\tau} + \delta_{\rm CP})
      \left[   \sin^{2} \frac{a_{\rm CC} L}{4E}
             - \sin^{2} \frac{\ldm L}{4E}
             + \sin^{2} \frac{(\ldm - a_{\rm CC})L}{4E}
      \right]                                           \nonumber\\
  &\hspace{0.5 cm}
    + 2 |\epsilon^{m}_{e\tau}| \tilde{s}_{13} s_{23}^{2} c_{23}
      \sin(\phi^{m}_{e\tau} + \delta_{\rm CP})
      \left[   \sin \frac{a_{\rm CC} L}{2E} 
             - \sin \frac{\ldm L}{2E}
             + \sin \frac{(\ldm - a_{\rm CC})L}{2E}
      \right]                                           \nonumber\\
  &\hspace{0.5 cm}
    + 8 |\epsilon^{m}_{e\tau}| \tilde{s}_{13} s_{23}^{2} c_{23}
      \cos(\phi^{m}_{e\tau} + \delta_{\rm CP}) \frac{a_{\rm CC}}{\ldm - a_{\rm CC}}
      \sin^{2} \frac{(\ldm - a_{\rm CC})L}{4E}  \nonumber\\
  &\hspace{0.5 cm}
    + 2 |\epsilon^{s}_{\mu e}| \frac{\sdm}{\ldm} s_{2 \times 12} c_{23}
      \cos \phi^{s}_{\mu e} \frac{\ldm}{a_{\rm CC}}
      \sin^{2} \frac{a_{\rm CC} L}{4E}                 \nonumber\\
  &\hspace{0.5 cm}
    - |\epsilon^{s}_{\mu e}| \frac{\sdm}{\ldm} s_{2 \times 12} c_{23}
      \sin \phi^{s}_{\mu e} \frac{\ldm}{a_{\rm CC}}
      \sin \frac{a_{\rm CC} L}{2E}                     \nonumber\\
  &\hspace{0.5 cm}
    - 2 |\epsilon^{d}_{\mu e}| \frac{\sdm}{\ldm} s_{2 \times 12} c_{23} 
      \cos \phi^{d}_{\mu e} \frac{\ldm}{a_{\rm CC}}
      \left[   c_{23}^{2} \sin^{2} \frac{a_{\rm CC} L}{4E}
             - s_{23}^{2} \sin^{2} \frac{\ldm L}{4E}
             + s_{23}^{2} \sin^{2} \frac{(\ldm - a_{\rm CC}) L}{4E}
      \right]                                           \nonumber \\
  &\hspace{0.5 cm}
    - |\epsilon^{d}_{\mu e}| \frac{\sdm}{\ldm} s_{2 \times 12} c_{23} 
      \sin \phi^{d}_{\mu e} \frac{\ldm}{a_{\rm CC}}
      \left[   c_{23}^{2} \sin \frac{a_{\rm CC} L}{2E}
             + s_{23}^{2} \sin \frac{\ldm L}{2E}
             - s_{23}^{2} \sin \frac{(\ldm - a_{\rm CC}) L}{2E}
      \right]                                           \nonumber\\
  &\hspace{0.5 cm}
    + 2 |\epsilon^{d}_{\tau e}| \frac{\sdm}{\ldm} s_{2 \times 12} s_{23} c_{23}^2
      \cos \phi^{d}_{\tau e} \frac{\ldm}{a_{\rm CC}}
      \left[   \sin^{2} \frac{a_{\rm CC} L}{4E}
             + \sin^{2} \frac{\ldm L}{4E}
             - \sin^{2} \frac{(\ldm - a_{\rm CC}) L}{4E}
      \right]                                           \nonumber\\
  &\hspace{0.5 cm}
    + |\epsilon^{d}_{\tau e}| \frac{\sdm}{\ldm} s_{2 \times 12} s_{23} c_{23}^2
      \sin \phi^{d}_{\tau e} \frac{\ldm}{a_{\rm CC}}
      \left[   \sin \frac{a_{\rm CC} L}{2E}
             - \sin \frac{\ldm L}{2E}
             + \sin \frac{(\ldm - a_{\rm CC}) L}{2E}
      \right]                                           \nonumber\\
  &\hspace{0.5 cm}
    + 4 |\epsilon^{m}_{e\mu}| \frac{\sdm}{\ldm} s_{2 \times 12} c_{23}^3
      \cos \phi^{m}_{e\mu} \frac{\ldm}{a_{\rm CC}}
      \sin^{2} \frac{a_{\rm CC} L}{4E}                 \nonumber\\
  &\hspace{0.5 cm}
    - 2 |\epsilon^{m}_{e\mu}| \frac{\sdm}{\ldm} s_{2 \times 12} s_{23}^2 c_{23}
      \cos \phi^{m}_{e\mu} \frac{\ldm}{\ldm - a_{\rm CC}}
      \left[   \sin^{2} \frac{a_{\rm CC} L}{4E}
             - \sin^{2} \frac{\ldm L}{4E}
             + \sin^{2} \frac{(\ldm - a_{\rm CC}) L}{4E}
      \right]                                           \nonumber\\
  &\hspace{0.5 cm}
    + |\epsilon^{m}_{e\mu}| \frac{\sdm}{\ldm} s_{2 \times 12} s_{23}^2 c_{23}
      \sin \phi^{m}_{e\mu} \frac{\ldm}{\ldm - a_{\rm CC}}
      \left[   \sin \frac{a_{\rm CC} L}{2E}
             - \sin \frac{\ldm L}{2E}
             + \sin \frac{(\ldm - a_{\rm CC}) L}{2E}
      \right]                                           \nonumber\\
  &\hspace{0.5 cm}
    - 4 |\epsilon^{m}_{e\tau}| \frac{\sdm}{\ldm} s_{2 \times 12} s_{23} c_{23}^2
      \cos \phi^{m}_{e\tau} \frac{\ldm}{a_{\rm CC}}
      \sin^{2} \frac{a_{\rm CC} L}{4E}                 \nonumber\\
  &\hspace{0.5 cm}
    - 2 |\epsilon^{m}_{e\tau}| \frac{\sdm}{\ldm} s_{2 \times 12} s_{23} c_{23}^2
      \cos \phi^{m}_{e\tau} \frac{\ldm}{\ldm -a_{\rm CC}}
      \left[   \sin^{2} \frac{a_{\rm CC} L}{4E}
             - \sin^{2} \frac{\ldm L}{4E}
             + \sin^{2} \frac{(\ldm - a_{\rm CC}) L}{4E}
      \right]                                           \nonumber\\
  &\hspace{0.5 cm}
    + |\epsilon^{m}_{e\tau}| \frac{\sdm}{\ldm} s_{2 \times 12} s_{23} c_{23}^2
      \sin \phi^{m}_{e\tau} \frac{\ldm}{\ldm - a_{\rm CC}}
      \left[   \sin \frac{a_{\rm CC} L}{2E}
             - \sin \frac{\ldm L}{2E}
             + \sin \frac{(\ldm - a_{\rm CC}) L}{2E}
      \right]                                           \nonumber\\
  &\hspace{0.5 cm}
      + \mathcal{O}\Big( \Big[ \frac{\sdm}{\ldm} \Big]^3 \Big)
      + \mathcal{O}\Big( \Big[ \frac{\sdm}{\ldm} \Big]^2 s_{13} \Big)
      + \mathcal{O}\Big( \frac{\sdm}{\ldm} s_{13}^2 \Big)
      + \mathcal{O} ( s_{13}^3 )                        \nonumber\\
  &\hspace{0.5 cm}
      + \mathcal{O}\Big( \eps \Big[ \frac{\sdm}{\ldm} \Big]^2 \Big)
      + \mathcal{O}\Big( \eps s_{13} \frac{\sdm}{\ldm} \Big)
      + \mathcal{O} ( \eps s_{13}^2 )
      + \mathcal{O} ( \eps^2 ).
  \label{eq:Pmue-mat}
\end{align}
Most of the $\mathcal{O}(\sdm/\ldm)$ terms contain factors of $\ldm / a_{\rm CC}$,
which can be large at low matter densities, and might therefore seem to spoil the
accuracy of the expansion in this case. However, the oscillatory terms in square
brackets become small as $\ldm/a_{\rm CC}$ becomes large, so that overall, the
$\mathcal{O}(\sdm/\ldm)$ terms remain subdominant even if the vaccum limit is
approached.

\subsection{The $\nu_\mu \rightarrow \nu_\mu$ channel}
\label{sec:mumu}

For $P_{\nu^s_\mu \rightarrow \nu^d_\mu}$, we obtain
\begin{align}
  P_{\nu^s_\mu \rightarrow \nu^d_\mu}^{\rm vac} &=
    1 - s^2_{2 \times 23} \sin^{2} \frac{\Delta m_{31}^{2} L}{4E}  \nonumber\\
  &\hspace{0.5 cm}
    + 2 |\epsilon^{s}_{\mu \mu}| \cos\phi^{s}_{\mu\mu} 
    + 2 |\epsilon^{d}_{\mu \mu}| \cos\phi^{d}_{\mu\mu}           \nonumber\\
  &\hspace{0.5 cm}
    - \big[   2 |\epsilon^{s}_{\mu \mu}| \cos\phi^{s}_{\mu\mu} 
            + 2 |\epsilon^{d}_{\mu \mu}| \cos\phi^{d}_{\mu\mu} \big]
      s_{2\times 23}^{2} \sin^{2} \frac{\Delta m_{31}^{2} L}{4E} \nonumber\\
  &\hspace{0.5 cm}
    - 2 \left(   |\epsilon^{s}_{\mu\tau}| \cos\phi^{s}_{\mu\tau}
               + |\epsilon^{d}_{\tau\mu}| \cos\phi^{d}_{\tau\mu} \right)
      c_{2 \times 23} s_{2 \times 23}
      \sin^{2} \frac{\Delta m_{31}^{2} L}{4E}                    \nonumber\\
  &\hspace{0.5 cm}
    + \left(   |\epsilon^{s}_{\mu\tau}| \sin\phi^{s}_{\mu\tau}
             + |\epsilon^{d}_{\tau\mu}| \sin\phi^{d}_{\tau\mu} \right)
      s_{2 \times 23} \sin \frac{\Delta m_{31}^{2} L}{2E}        \nonumber\\
  &\hspace{0.5cm}
      + \mathcal{O}\Big( \frac{\sdm}{\ldm} \Big)
      + \mathcal{O} ( s_{13} )
      + \mathcal{O} ( \eps^2 ).
  \label{eq:Pmumu-vac}
  \intertext{in vacuum, and}
  P_{\nu^s_\mu \rightarrow \nu^d_\mu}^{\rm mat} &=
    1 - s^2_{2 \times 23} \sin^{2} \frac{\Delta m_{31}^{2} L}{4E}  \nonumber\\
  &\hspace{0.5 cm}
    + 2 |\epsilon^{s}_{\mu \mu}| \cos\phi^{s}_{\mu\mu} 
    + 2 |\epsilon^{d}_{\mu \mu}| \cos\phi^{d}_{\mu\mu}   \nonumber\\
  &\hspace{0.5 cm}
    - \big[   2 |\epsilon^{s}_{\mu \mu}| \cos\phi^{s}_{\mu\mu} 
            + 2 |\epsilon^{d}_{\mu \mu}| \cos\phi^{d}_{\mu\mu} \big]
      s_{2\times 23}^{2} \sin^{2} \frac{\ldm L}{4E}   \nonumber\\
  &\hspace{0.5 cm}
    - 2 \left(   |\epsilon^{s}_{\mu\tau}| \cos\phi^{s}_{\mu\tau}
               + |\epsilon^{d}_{\tau\mu}| \cos\phi^{d}_{\tau\mu} \right)
      c_{2 \times 23} s_{2 \times 23}
      \sin^{2} \frac{\ldm L}{4E}                      \nonumber\\
  &\hspace{0.5 cm}
    + \left(   |\epsilon^{s}_{\mu\tau}| \sin\phi^{s}_{\mu\tau}
             + |\epsilon^{d}_{\tau\mu}| \sin\phi^{d}_{\tau\mu} \right)
      s_{2 \times 23} \sin \frac{\ldm L}{2E}          \nonumber\\
  &\hspace{0.5 cm}
    - |\epsilon^{m}_{\mu\tau}| \left[
         s_{2 \times 23}^{3} \cos\phi^{m}_{\mu\tau}
         \frac{a_{\rm CC} L}{2E} \sin \frac{\ldm L}{2E}
       + 4 s_{2 \times 23} c_{2 \times 23}^{2} \cos\phi^{m}_{\mu\tau}
         \frac{a_{\rm CC}}{\ldm} \sin^{2} \frac{\ldm L}{4E}
     \right]                                                 \nonumber\\
  &\hspace{0.5 cm}
    + \frac{1}{2} |\epsilon^{m}_{\mu\mu}| \left[
         s_{2 \times 23}^{2} c_{2 \times 23}
         \frac{a_{\rm CC} L }{2E} \sin \frac{\ldm L}{2E}
       - 4  s_{2 \times 23}^{2} c_{2 \times 23}
         \frac{a_{\rm CC}}{\ldm} \sin^{2} \frac{\ldm L}{4E}
     \right]                                                \nonumber\\
  &\hspace{0.5 cm}
    - \frac{1}{2} |\epsilon^{m}_{\tau\tau}| \left[
         s_{2 \times 23}^{2} c_{2 \times 23}
         \frac{a_{\rm CC} L}{2E} \sin \frac{\ldm L}{2E}
       - 4 s_{2 \times 23}^{2} c_{2 \times 23}
         \frac{a_{\rm CC}}{\ldm} \sin^{2} \frac{\ldm L}{4E}
     \right]                                                \nonumber\\
  &\hspace{0.5cm}
      + \mathcal{O}\Big( \frac{\sdm}{\ldm} \Big)
      + \mathcal{O} ( s_{13} )
      + \mathcal{O} ( \eps^2 ).
  \label{eq:Pmumu-mat}
\end{align}
in matter.
\end{widetext}

\subsection{Interpretation}
\label{sec:interpretation}

\begin{figure}
  \begin{center}
    \includegraphics[width=8.5cm]{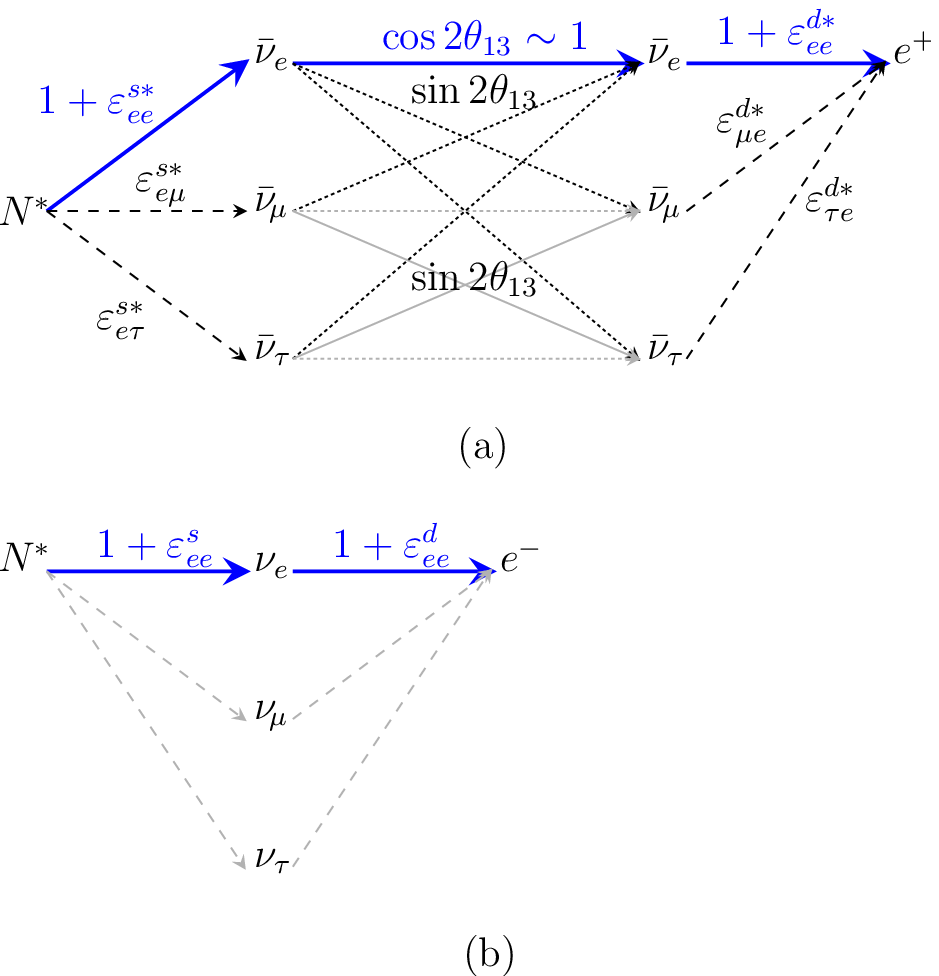}
  \end{center}
  \caption{Possible contributions of $\eps^s$ and $\eps^d$ to the event
    rate in the far detector (a), and the near detector (b) of a reactor
    $\bar{\nu}_e$ disappearance experiment. Thick blue lines indicate the
    reaction chain for standard oscillations. Dotted lines indicate processes
    that are suppressed by standard three-flavor effects proportional to
    $\theta_{13}$ or $\sdm/\ldm$, while dashed lines represent transitions that
    are suppressed due to non-standard interactions. Paths which would only
    be accessible in the presence of two different non-standard effects, are
    shown in light gray since they are usually subdominant.}
  \label{fig:diagrams-sd-reactor}
\end{figure}

\begin{figure}
  \begin{center}
    \includegraphics[width=8.5cm]{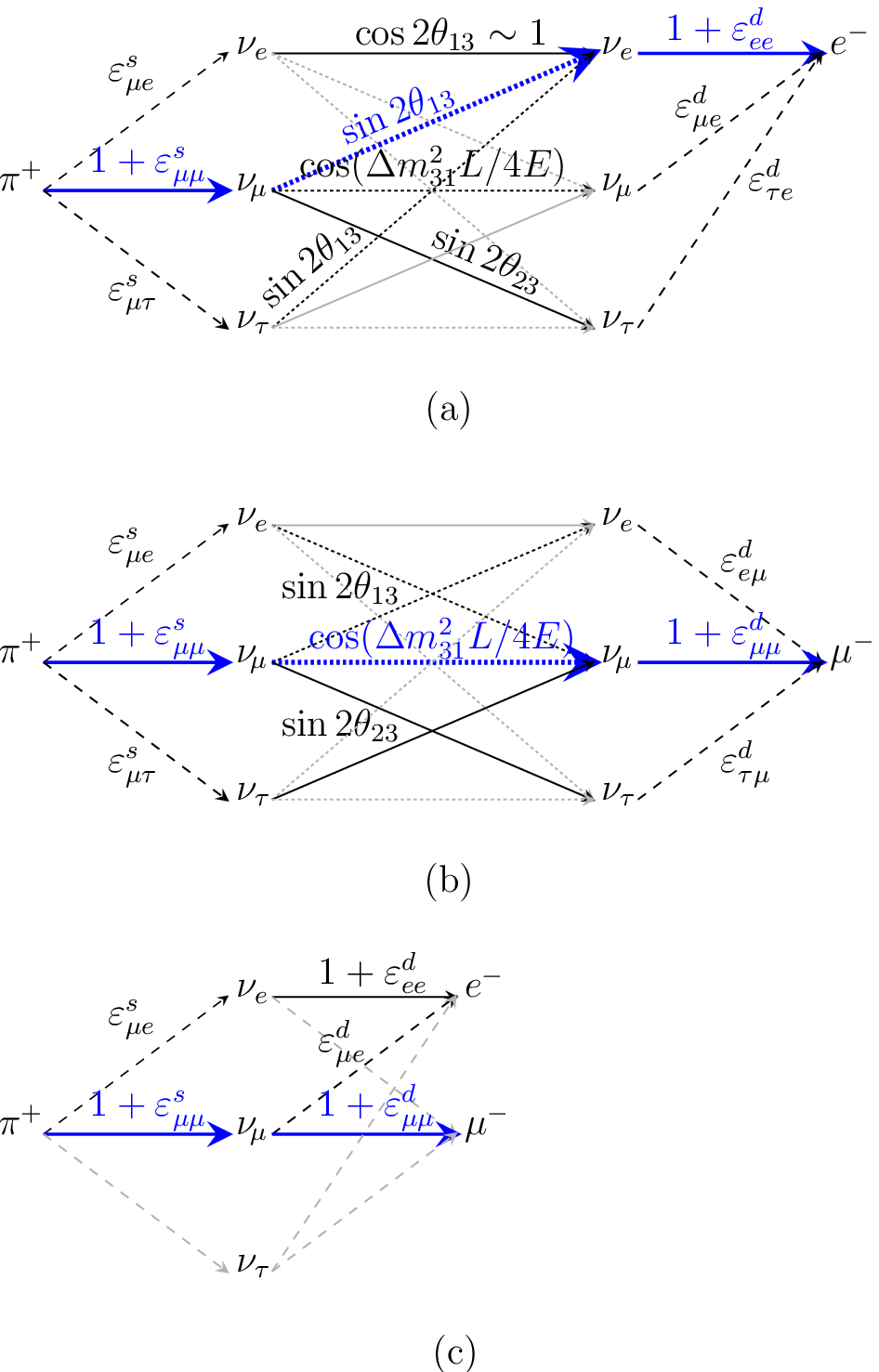}
  \end{center}
  \caption{Possible contributions of $\eps^s$ and $\eps^d$ to the event
    rate in a superbeam experiment for the appearance channel (a), the
    disappearance channel (b), and in the near detector (c). The meaning of
    the colors and line styles is the same as in Fig.~\ref{fig:diagrams-sd-reactor}.}
  \label{fig:diagrams-sd-beam}
\end{figure}

\begin{figure}
  \begin{center}
    \includegraphics[width=8.5cm]{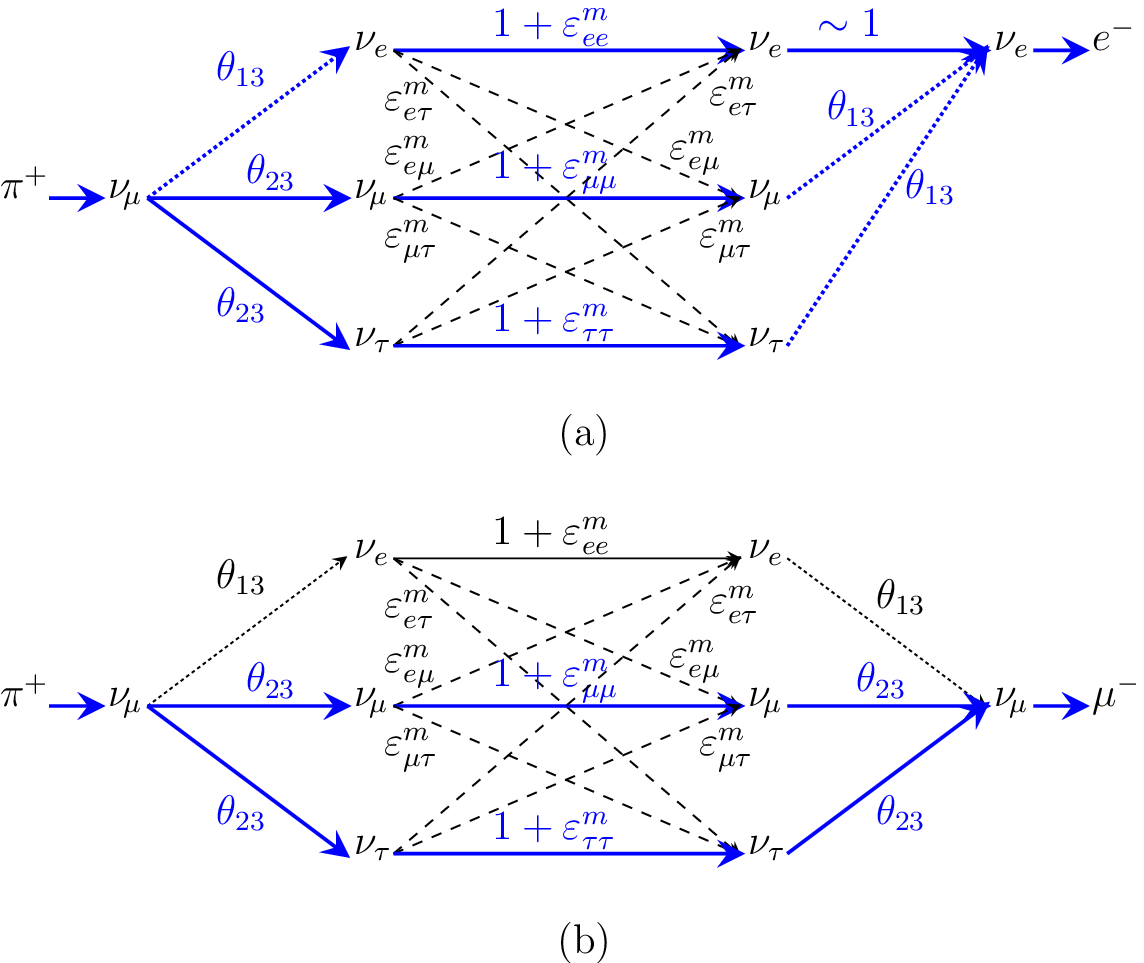}
  \end{center}
  \caption{Possible contributions of $\eps^m$ to the event rate in the
    superbeam appearance channel (a), and in the corresponding disappearance
    channel (b). The meaning of the colors and line styles is the same as in
    Fig.~\ref{fig:diagrams-sd-reactor}.}
  \label{fig:diagrams-m}
\end{figure}

To understand the physical origin of the formulas derived in the previous sections,
let us consider Figs.~\ref{fig:diagrams-sd-reactor} --~\ref{fig:diagrams-m}, where
we show schematically the possible reaction chains that a neutrino can follow before
its detection. We also indicate the respective suppression factors of the transition
amplitude, but to simplify the discussion and to improve the clarity of the figures, we
do not explicitly show contributions proportional to $\sdm/\ldm$, which would appear
in concurrence to the $\theta_{13}$-suppressed processes if $\theta_{13}$ is very small.
Dotted lines indicate suppression due to standard effects, while dashed lines represent
transitions that are suppressed by the non-standard parameters. The thick blue (black)
paths are those followed by the standard oscillation channels, while light gray lines
indicate paths that are suppressed by more than one $\eps$ parameter, and are therefore
mostly negligible. In Tab.~\ref{tab:suppression-factors} we summarize the same
considerations in tabular form. In the first part of the discussion, we will assume
$\eps^s$ and $\eps^d$ to be completely independent in order to keep the discussion
as general as possible. The constraints from Eqs.~\eqref{eq:eps-e-constraint}
and \eqref{eq:eps-mu-constraint} will only be implemented afterwards.

On the reactor side, we can read off from Fig.~\ref{fig:diagrams-sd-reactor} and
Tab.~\ref{tab:suppression-factors}, that, in the presence of just one type
of non-standard interaction, only $\eps^s_{ee}$, $\eps^s_{e\mu}$, $\eps^s_{e\tau}$,
$\eps^d_{ee}$, $\eps^d_{\mu e}$, and $\eps^d_{\tau e}$ are relevant. Of these,
$\eps^s_{ee}$ and $\eps^d_{ee}$ have an $\mathcal{O}(\eps)$ effect which could
even exceed standard oscillations if it were not for the near detector,
where these parameters would induce a similar effect as in the far detector.
Since the absolute reactor neutrino flux is not known precisely, the
measurement relies on the \emph{relative} counting rates in both
detectors, so that the impact of $\eps^s_{ee}$ and $\eps^d_{ee}$
is canceled. The remaining NSI, $\eps^s_{e\mu}$, $\eps^s_{e\tau}$,
$\eps^d_{\mu e}$, and $\eps^d_{\tau e}$, contribute to the oscillation
probability at $\mathcal{O}(\eps \sin 2\theta_{13})$ (or at $\mathcal{O}
(\eps \sin 2\theta_{13} + \eps^2)$, if the constraint $\eps^s_{e\alpha} =
\eps^{d*}_{\alpha e}$ from Eq.~\eqref{eq:eps-e-constraint} is implemented).
This can be comparable to the standard term, so that these NSI are
expected to have a large impact on the far detector. They do not affect
the near detector as long as only one of them is present, but if
Eq.~\eqref{eq:eps-e-constraint} is taken into account, the near detector will
receive an $\mathcal{O}(\eps^2)$ contribution that can be important in some
situations. All these considerations are nicely confirmed by Eq.~\eqref{eq:Pee}.

\begin{table*}
  \centering
  \begin{ruledtabular}
  \begin{tabular}{l|c|c|c|c|c}
    \bf NSI              & \multicolumn{2}{c|}{\bf Reactor}                      & \multicolumn{3}{c}{\bf Superbeam}                                                        \\
                         & \bf$\bar{\nu}_e$ disappearance & \bf Effect in ND     & \bf $\nu_e$ appearance& \bf $\nu_\mu$ disappearance & \bf Effect in ND   \\\hline
    None                 & 1                       & ---                         & $\sin 2\theta_{13}$   & $\cos(\ldm L/4E)$      & ---                     \\ \hline
                                                                 
    $\eps^s_{ee}$        & $\eps$                  & modified $\bar{\nu}_e$ flux & ---                   & ---                    & ---                     \\
    $\eps^s_{e\mu}$      & \fbox{$\epssin$}        & ---                         & ---                   & ---                    & ---                     \\
    $\eps^s_{e\tau}$     & \fbox{$\epssin$}        & ---                         & ---                   & ---                    & ---                     \\
    $\eps^s_{\mu e}$     & ---                     & ---                         & \fbox{$\eps$}         & $\epssin$              & modified $\nu_e$ flux   \\
    $\eps^s_{\mu\mu}$    & ---                     & ---                         & $\epssin$             & $\eps \cos(\ldm L/4E)$ & modified $\nu_\mu$ flux \\
    $\eps^s_{\mu\tau}$   & ---                     & ---                         & $\epssin$             & $\eps$                 & ---                     \\
    $\eps^s_{\tau e}$    & ---                     & ---                         & ---                   & ---                    & ---                     \\
    $\eps^s_{\tau\mu}$   & ---                     & ---                         & ---                   & ---                    & ---                     \\
    $\eps^s_{\tau\tau}$  & ---                     & ---                         & ---                   & ---                    & ---                     \\ \hline

    $\eps^d_{ee}$        & $\eps$                  & modified $\bar{\nu}_e$ flux & $\epssin$             & ---                    & ---                     \\
    $\eps^d_{e\mu}$      & ---                     & ---                         & ---                   & $\epssin$              & ---                     \\
    $\eps^d_{e\tau}$     & ---                     & ---                         & ---                   & ---                    & ---                     \\
    $\eps^d_{\mu e}$     & \fbox{$\epssin$}        & ---                         & $\eps\cos(\ldm L/4E)$ & ---                    & modified $\nu_e$ flux   \\
    $\eps^d_{\mu\mu}$    & ---                     & ---                         & ---                   & $\eps \cos(\ldm L/4E)$ & modified $\nu_\mu$ flux \\
    $\eps^d_{\mu\tau}$   & ---                     & ---                         & ---                   & ---                    & ---                     \\
    $\eps^d_{\tau e}$    & \fbox{$\epssin$}        & ---                         & \fbox{$\eps$}         & ---                    & ---                     \\
    $\eps^d_{\tau\mu}$   & ---                     & ---                         & ---                   & $\eps$                 & ---                     \\
    $\eps^d_{\tau\tau}$  & ---                     & ---                         & ---                   & ---                    & ---                     \\ \hline

    $\eps^s_{ee} =
     \eps^{d*}_{ee}$     & $\eps$                  & modified $\bar{\nu}_e$ flux & $\epssin$             & ---                    & ---                     \\
    $\eps^s_{e\mu} =
     \eps^{d*}_{\mu e}$  & \fbox{$\epssin+\eps^2$} & modified $\bar{\nu}_e$ flux & $\eps\cos(\ldm L/4E)$ & ---                    & modified $\nu_e$ flux   \\
    $\eps^s_{e\tau} =
     \eps^{d*}_{\tau e}$ & \fbox{$\epssin+\eps^2$} & modified $\bar{\nu}_e$ flux & \fbox{$\eps$}         & ---                    & ---                     \\
    $\eps^s_{\mu e} =
     \eps^{d*}_{e\mu}$   & ---                     & ---                         & \fbox{$\eps$}         & $\epssin+\eps^2$       & modified $\nu_e$ flux   \\
    $\eps^s_{\mu\mu} =
     \eps^{d*}_{\mu\mu}$ & ---                     & ---                         & $\epssin$             & $\eps\cos(\ldm L/4E)$  & modified $\nu_\mu$ flux \\
    $\eps^s_{\mu\tau} =
     \eps^{d*}_{\tau\mu}$& ---                     & ---                         & $\epssin$             & $\eps$                 & ---                     \\ \hline
                                                                                                       
    $\eps^m_{ee}$        & ---                     & ---                         & $\epssin$             & $\eps\sthchooz$        & ---                     \\
    $\eps^m_{e\mu}$      & ---                     & ---                         & \fbox{$\eps$}         & $\epssin$              & ---                     \\
    $\eps^m_{e\tau}$     & ---                     & ---                         & \fbox{$\eps$}         & $\epssin$              & ---                     \\
    $\eps^m_{\mu\mu}$    & ---                     & ---                         & $\epssin$             & $\eps\cos 2\theta_{23}$ \footnotemark[1] & ---  \\
    $\eps^m_{\mu\tau}$   & ---                     & ---                         & $\epssin$             & $\eps$                 & ---                     \\
    $\eps^m_{\tau\tau}$  & ---                     & ---                         & $\epssin$             & $\eps\cos 2\theta_{23}$ \footnotemark[1] & ---  \\
  \end{tabular}
  \end{ruledtabular}
  \footnotetext[1]{\mbox{The factor $\cos 2\theta_{23}$ cannot be derived
    from Fig.~\ref{fig:diagrams-m}, but only from Eq.~\eqref{eq:Pmumu-mat}.}}
  \caption{Classification of the reparameterized non-standard interactions $\eps^s$,
    $\eps^d$, and $\eps^m$ according to their impact on the transition amplitudes
    for reactor and superbeam experiments. For each NSI coupling, only the leading
    order effect is shown. The framed entries highlight those terms that are most
    relevant to the determination of $\theta_{13}$ (see text for details).}
  \label{tab:suppression-factors}
\end{table*}

For a superbeam experiment, Fig.~\ref{fig:diagrams-sd-beam} and
Tab.~\ref{tab:suppression-factors} show that, as long as only one type of NSI is taken
into account, $\eps^s_{\mu e}$, $\eps^s_{\mu\mu}$, $\eps^s_{\mu\tau}$ can affect the
production process, while $\eps^d_{ee}$, $\eps^d_{e\mu}$, $\eps^d_{\mu e}$,
$\eps^d_{\mu\mu}$, $\eps^d_{\tau e}$, and $\eps^d_{\tau\mu}$ may be important in the
detector. The propagation can be affected by all entries of $\eps^m$. As for the reactor
case, the suppression factors associated with these different processes can be understood
from simple physical arguments, which are confirmed in a more rigorous way by
Eqs.~\eqref{eq:Pmue-vac} and~\eqref{eq:Pmumu-vac}. Let us discuss the different types
of NSI in more detail:
\begin{itemize}
  \item 
  $\eps^d_{e\mu}$ and $\eps^d_{\tau\mu}$ affect only the disappearance channel
  and are therefore irrelevant for the measurement of $\theta_{13}$ (they may,
  however, lead to wrong results for the leading atmospheric parameters).

  \item
  $\eps^d_{\mu\mu}$ also affects the disappearance channel, but it can also lead
  to a modified $\nu_\mu$ rate in the near detector, and therefore to wrong assumptions
  on the initial neutrino flux. This, in turn, could lead to a misinterpretation
  of the far detector appearance measurement, so that the $\theta_{13}$ measurement
  is influenced indirectly. The effect in the near detector is suppressed by
  $\eps$, so it will affect the far detector analysis only at the subleading
  level of $\eps \sin^2 2\theta_{13}$.

  \item
  $\eps^s_{\mu\mu}$, $\eps^s_{\mu\tau}$ and $\eps^d_{ee}$ are relevant for the
  appearance channel, but the corresponding \emph{amplitude} is suppressed by
  $\eps \sin 2\theta_{13}$, so that in the oscillation \emph{probability}, we
  would obtain a subdominant contribution of $\mathcal{O}(\eps \sin^2 2\theta_{13})$
  from the interference of the standard and non-standard terms.

  \item
  $\eps^s_{\mu e}$, $\eps^d_{\mu e}$, and $\eps^d_{\tau e}$ have
  \emph{amplitudes} of $\mathcal{O}(\eps)$, i.e.\ they contribute to the
  appearance \emph{probability} on the level of $\eps \sin 2\theta_{13}$,
  which can be comparable to the leading contribution $\sim \sin^2 2\theta_{13}$.
  $\eps^d_{\mu e}$, however, is suppressed by a factor of $\cos(\ldm L/4E)$,
  which is small at the first atmospheric maximum around which the beam is
  centered. Note also that the modified $\nu_e$ flux in the near detector that
  is expected in the presence of $\eps^s_{\mu e}$ or $\eps^d_{\mu e}$ can help
  to actually detect the NSI, although part of it may be misinterpreted as a
  systematical error on the intrinsic beam background.

  \item
  Of the non-standard matter effects, only $\eps^m_{e\mu}$ and $\eps^m_{e\tau}$
  contribute at leading order to the appearance probability
  $P_{\nu^s_\mu \rightarrow \nu^d_e}^{\rm mat}$. Of these, $\eps^m_{e\mu}$
  is already strongly constrained experimentally~\cite{Davidson:2003ha},
  and so is not expected to have a large impact on reactor and superbeam experiments.
  $\eps^m_{e\tau}$, on the other hand, could contribute significantly
  to the superbeam appearance channel, in accordance with~\cite{Blennow:2005qj,
  Blennow:2007pu}. All other non-standard matter effects are suppressed by an additional
  power of $s_{13}$, (or, more correctly, $\tilde{s}_{13}$, which is, however, still small
  since we are far from the MSW resonance).
  
  It is interesting to observe that the sensitivity to $\eps^m_{ee}$ is very
  weak, although this type of interaction corresponds to a simple rescaling
  of the standard MSW potential. However, it is not a leading order effect,
  and therefore does not appear in our approximate formula, Eq.~\eqref{eq:Pmue-mat}.
  
  \item
  In $P_{\nu^s_\mu \rightarrow \nu^d_\mu}^{\rm mat}$, the dominant matter
  effect is $\eps^m_{\mu\tau}$, and since there is no $\theta_{13}$
  suppression from the interference with the standard amplitude, this effect
  is even stronger than those in $P_{\nu^s_\mu \rightarrow \nu^d_e}^{\rm mat}$.
  Note that, from Fig.~\ref{fig:diagrams-m}, one might expect
  $\eps^m_{\mu\mu}$ and $\eps^m_{\tau\tau}$ to be of similar strength
  as $\eps^m_{\mu\tau}$, but when one performs the calculation, it
  turns out that an additional suppression factor $c_{2 \times 23}$ appears
  (cf.\ Eq.~\eqref{eq:Pmumu-mat}).

  \item
  The implementation of the constraints $\eps^s_{e\alpha} = \eps^{d*}_{\alpha e}$
  (Eq.~\eqref{eq:eps-e-constraint}) and $|\eps^s_{\mu\alpha}| \gtrsim |\eps^d_{\alpha\mu}|$
  (Eq.~\eqref{eq:eps-mu-constraint}) does not lead to any new effects, except
  for the appearance of an additional $\eps^2$ term in the disappearance
  channel for $\eps^s_{\mu e} = \eps^{d*}_{e \mu}$.
\end{itemize}
Let us finally emphasize the crucial importance of the standard and
non-standard phases in the oscillation probabilities: The formulas from
Secs.~\ref{sec:ee} -- \ref{sec:mumu} reveal that unfavorable phase combinations
may suppress non-standard effects, even if the modulus of the corresponding
$\eps$ parameter is large.

\section{Simulation of reactor and superbeam experiments}
\label{sec:simulation}

To fully assess the high-level consequences of non-standard interactions for
realistic reactor and superbeam experiments, we have performed numerical
simulations using the \GLoBES\ software~\cite{Huber:2004ka,Huber:2007ji}.
We have considered the following scenarios:
\begin{itemize}
  \item \TtoK\ + \DoubleChooz.
    Our simulation of the \TtoK\ far detector, \SuperK, is based
    on~\cite{Huber:2002mx}. Most parameters are taken from the \TtoK\ letter of
    intent~\cite{Itow:2001ee}, and the systematical uncertainties are based
    on~\cite{Ishitsuka:2005qi}. We include a separate 1.0~kt water
    \v{C}erenkov near detector with similar properties as the far detector,
    and similar systematical uncertainties. To model the interplay
    of the two detectors, we introduce a common 10\% uncertainty on the neutrino
    flux, and a common 20\% error on the number of background events in the
    $\nu_e$ appearance channel. In the absence of non-standard
    interactions, these correlated errors would cancel completely, since the
    total neutrino flux and the background contribution are effectively
    calibrated by the near detector, but if $\eps^{s,d} \neq 0$, this calibration
    can be wrong, and there may be an observable effect. The neutrino interaction
    cross sections are taken from~\cite{Messier:1999kj,Paschos:2001np}. We assume
    3~years of neutrino running and 3~years of anti-neutrino running, each
    with a beam power of 0.77~MW. The fiducial far detector mass is 22.5~kt,
    and the baseline is 295~km. We consider $\nu_e$ appearance events as well
    as the $\nu_\mu$ disappearance signal. The background for the disappearance
    channel is made up of neutral current events, while for the appearance
    measurement, neutral current events, misidentified muons, and the intrinsic
    beam backgrounds can contribute.

    For the simulation of \DoubleChooz, we use the same parameters as
    in~\cite{Huber:2006vr}, and the cross sections for inverse beta decay
    are taken from~\cite{Vogel:1999zy}. As for \TtoK, we simulate the
    near and far detectors separately, but take into account the appropriate
    correlations between systematical errors. In particular, we introduce
    a 2.8\% flux normalization error, which is correlated between the
    near and far detectors, uncorrelated 0.6\% fiducial mass errors for
    both detectors, uncorrelated 0.5\% energy calibration uncertainties,
    and an 0.5\% bin-to-bin uncorrelated error. 
    
  \item \NOvA\ + \DCext, where \DCext\ refers to a reactor experiment similar
    to \DoubleChooz, but with a 200~t far detector~\cite{Huber:2006vr}. Such a
    large reactor experiment has a considerable sensitivity not only to the
    total event rate, but also to distortions of the energy spectrum.

    The simulation of the $\nu_e$ appearance signal in \NOvA\ is based
    on~\cite{Ayres:2004js}, while for the $\nu_\mu$ disappearance channel, we
    follow~\cite{Yang:2004}. We assume 3~years of neutrino running and
    3~years of anti-neutrino running, with a beam power of 1.12~MW.
    The far detector mass is 25~kt, and the baseline is 812~km, with an average
    matter density of 2.8~g/cm$^3$ along the trajectory, while the near detector
    has a mass of 0.0204~kt, and is located at 1~km from the target. Again,
    we introduce, in addition to the uncorrelated systematical errors
    from~\cite{Ayres:2004js,Yang:2004}, a correlated 10\% uncertainty on the total
    neutrino flux, and a correlated 20\% error on the $\nu_e$ background.
    
    The parameters and systematical errors of the  \DCext\ scenario are
    identical to those of \DoubleChooz.
\end{itemize}
Unless indicated otherwise, we calculate the respective event rates using the
following ``true'' values for the oscillation parameters~\cite{Maltoni:2004ei}:
\begin{align}
  \begin{split}
    \sin^{2} 2 \theta_{12}^{\text{true}} &= 0.84, \\ 
    \sin^{2} 2 \theta_{23}^{\text{true}} &= 1.0,  \\
    \sin^{2} 2 \theta_{13}^{\text{true}} &= 0.05, \\
    \delta_{\rm CP}^\text{true}          &= 0.0,  \\
    (\Delta m_{21}^{2})^{\text{true}}
      &= 7.9 \times 10^{-5} \text{ eV$^{2}$},     \\
    (\Delta m_{31}^{2})^{\text{true}}
      &= 2.6 \times 10^{-3} \text{ eV$^{2}$},
  \end{split}
\end{align}
and assume a normal mass hierarchy. To analyze the simulated data, we follow
the statistical procedure described in the appendix of~\cite{Huber:2002mx},
and define the following $\chi^2$ function
\footnote{In the implementation of superbeam experiments, we assume the events
to follow the Poisson distribution. However, for illustrative purposes, it is
sufficient to consider the more compact approximative Gaussian expression.}
\begin{multline}
  \chi^{2} = \min_{\lambda} \sum^{\text{channel}}_{j} \sum_{i}^{\text{bin}}
    \frac{\left| N_{ij} \left( \lambda^{\text{true}}, \eps^{\text{true}} \right)
      - N_{ij} \left(\lambda, \eps = 0\right) \right|^{2}}
        {N_{ij} ( \lambda^{\text{true}}, \eps^{\text{true}})} \\
      + {\rm Priors},
\label{eq:chiSq}
\end{multline}
where $N_{ij}$ denotes the number of events in the $i$-th energy bin for
oscillation channel $j$, the vector $\lambda = ( \theta_{12}, \theta_{13}, \theta_{23},
\delta_{\rm CP}, \sdm, \ldm, \vec{b} )$ contains the standard oscillation parameters
and the systematical biases $\vec{b}$, and $\eps$ represents the non-standard
parameters. In the fit, we marginalize $\chi^2$ over all standard oscillation
parameters and over the systematical biases, but since we want to study how a
standard-oscillation fit gets modified if there are non-standard interactions,
we keep the NSI parameters fixed at 0.%
\footnote{Of course, when computing confidence intervals for certain parameters,
we have to keep these parameters fixed as well.}
The prior terms implement external input from other experiments and have the form
$(x - x^{\rm true})^2 / \sigma_x^2$, where $x$ stands for any oscillation parameter
or systematical bias, and $\sigma_x$ is the corresponding externally given uncertainty.
We assume $\theta_{12}$ to be known to within 10\%, and $\sdm$ to within 5\% from
solar and reactor experiments~\cite{Maltoni:2004ei}. When analyzing the reactor
experiment alone, we additionally assume a 15\% uncertainty on $\theta_{23}$
and a 5\% error on $\ldm$. Beam experiments are themselves sensitive to
$\theta_{23}$ and $\ldm$, so we omit these priors for them.

\section{NSI-induced offsets and discrepancies in $\theta_{13}$ fits}
\label{sec:th13fits}

\begin{figure*}
  \begin{center}
    \begin{tabular}{cc}
      \includegraphics[width=8cm]{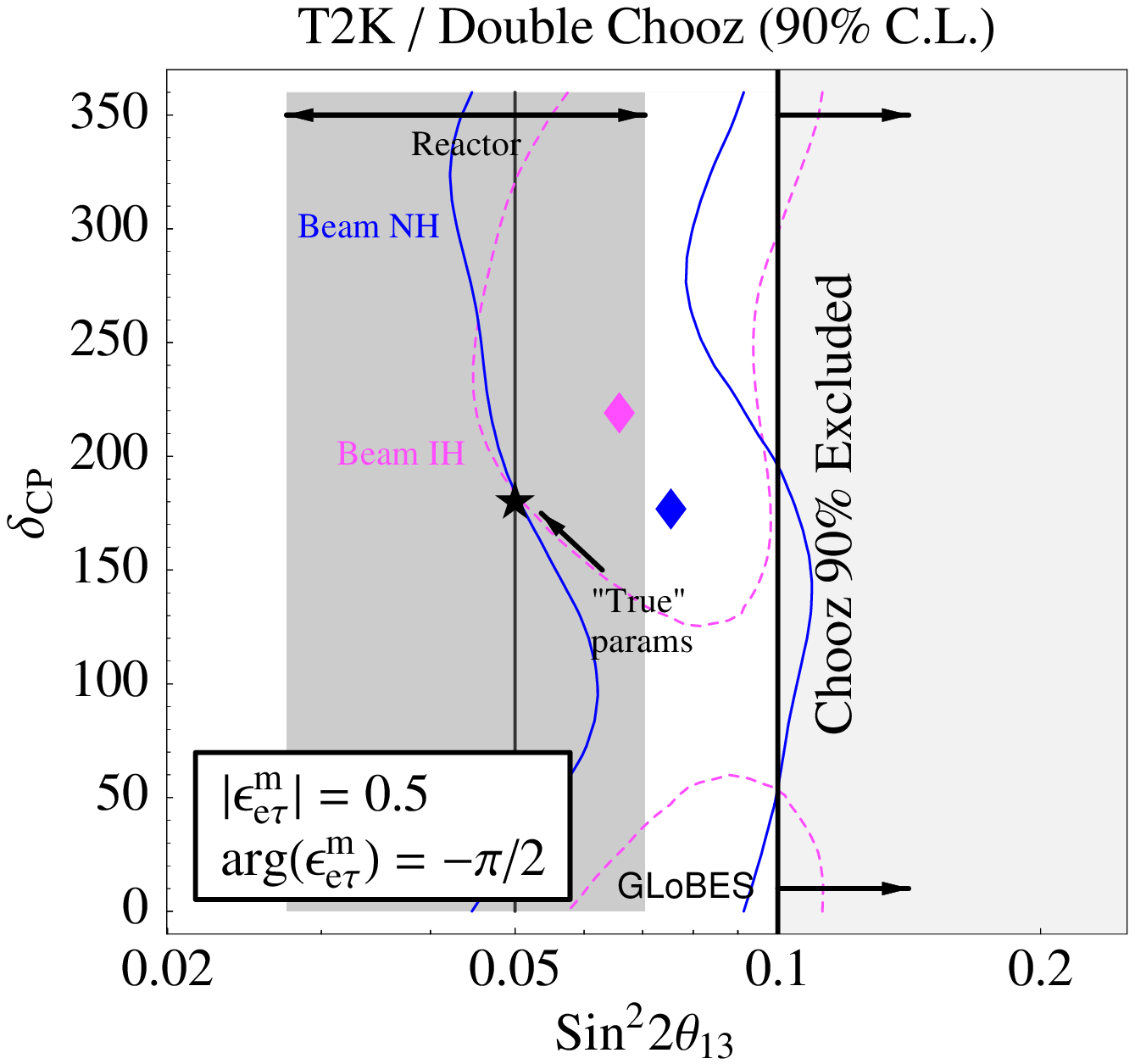} &
      \includegraphics[width=8cm]{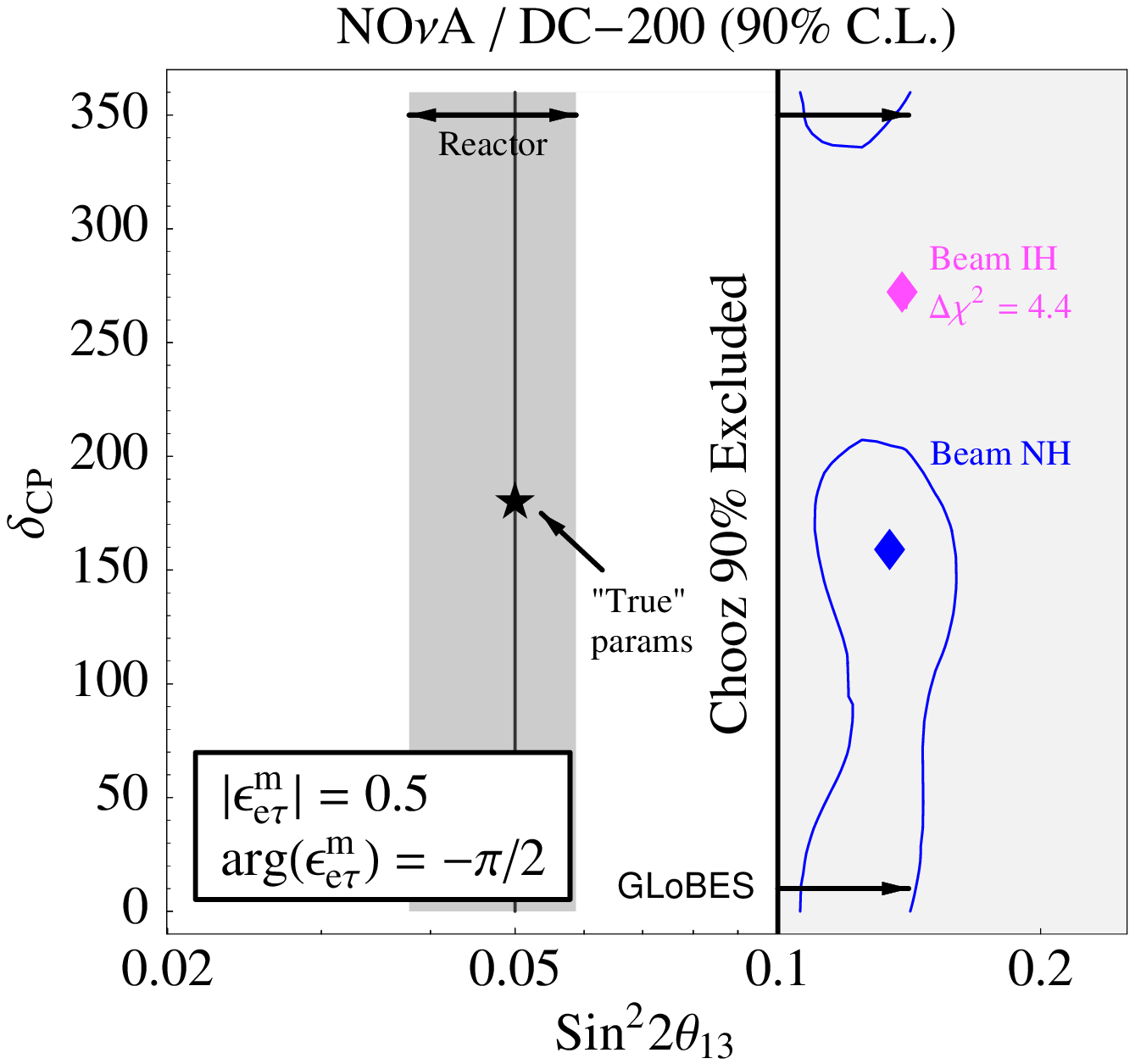} \\[0.3cm]
      \includegraphics[width=8cm]{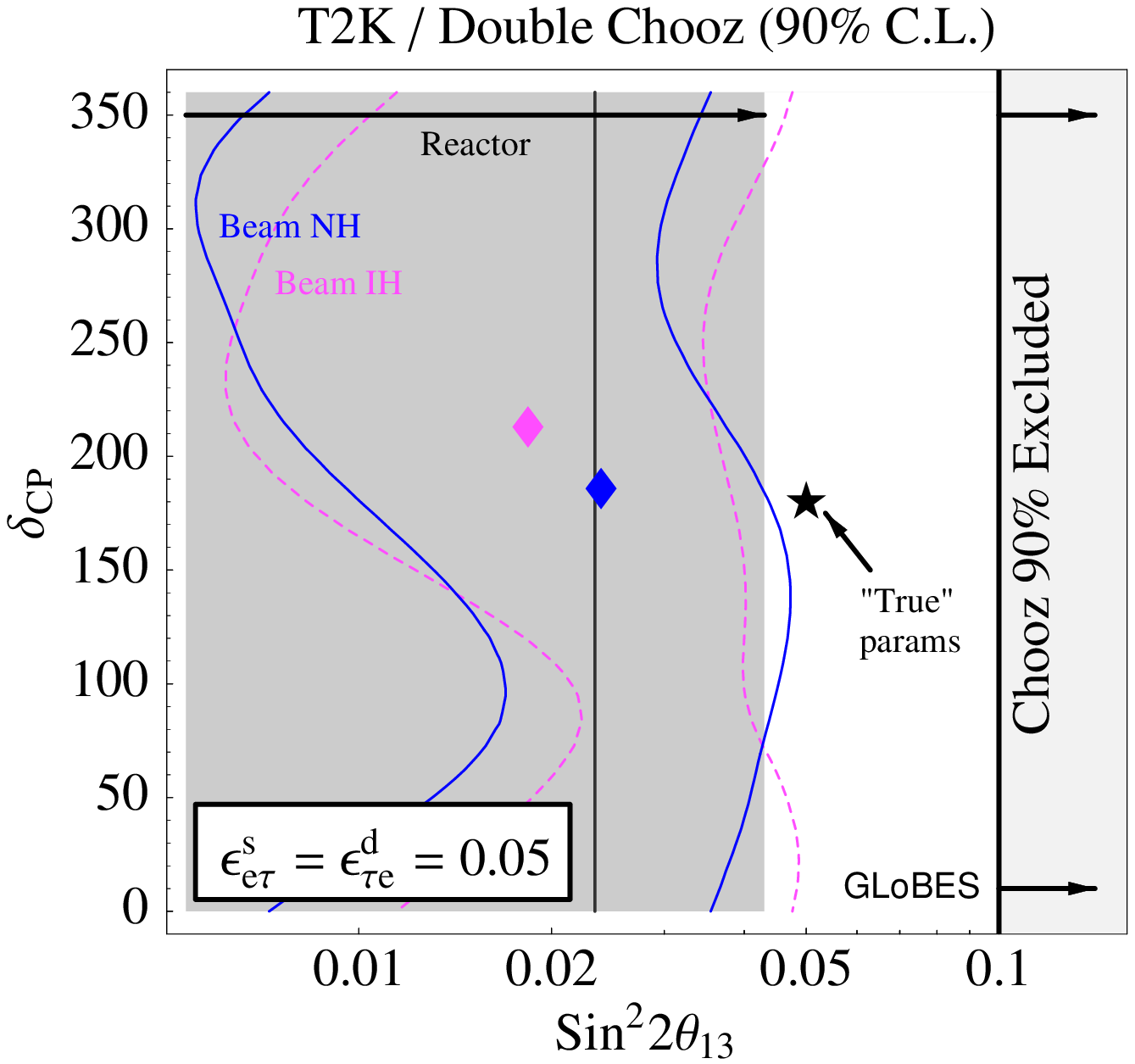} &
      \includegraphics[width=8cm]{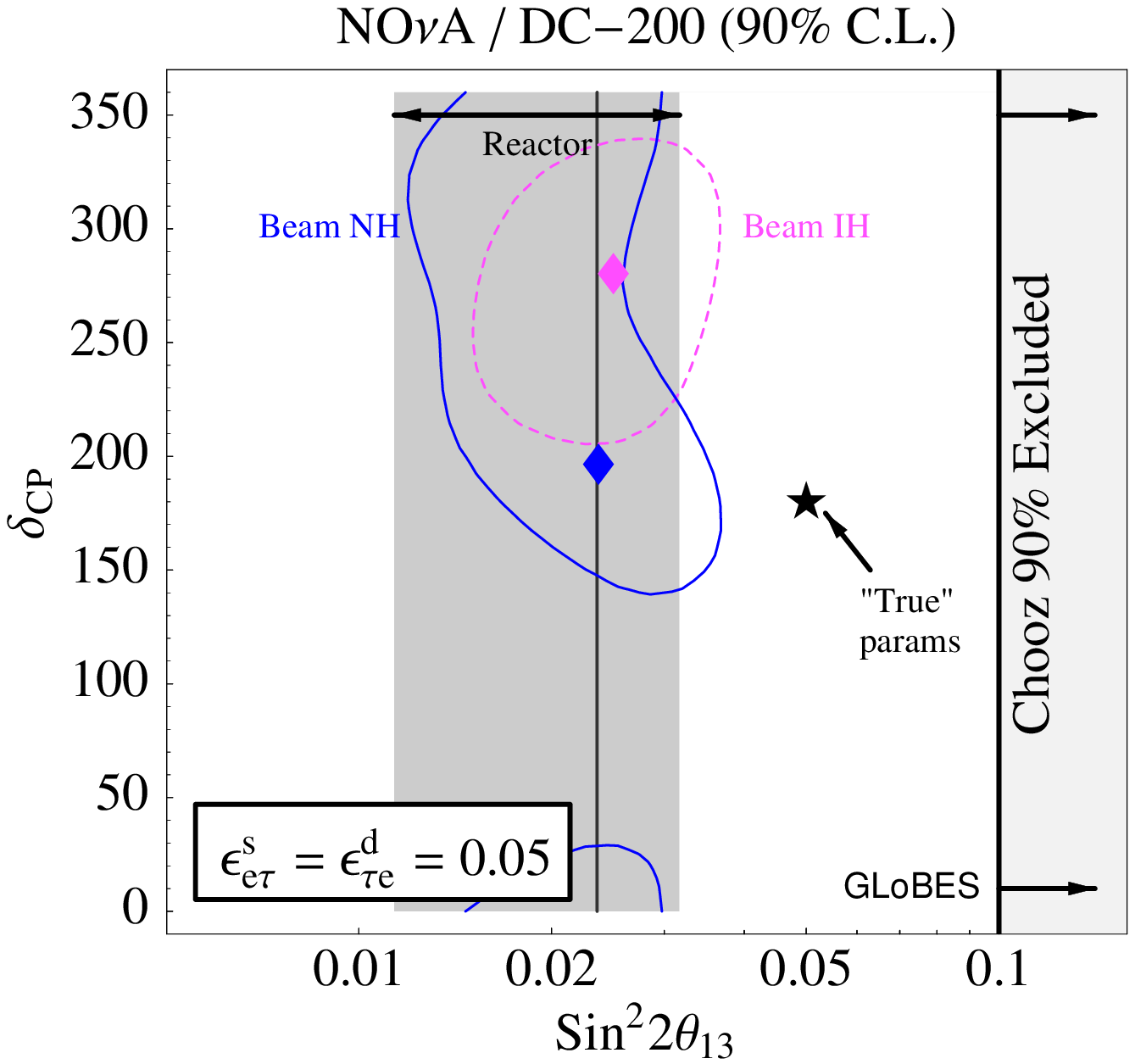}    
    \end{tabular}
  \end{center}
  \vspace{-0.5cm}
  \caption{Two examples for the errors that are introduced if non-standard
    interactions are neglected when fitting $\theta_{13}$ and $\delta_{\rm CP}$
    to the data of reactor and superbeam experiment. In the upper plots,
    a discrepancy arises between the two experiments (the \NOvA\ fit is even
    above the \Chooz\ bound), while in the lower plots, there is a common offset,
    leading to consistent results, but erroneously ``ruling out'' the true
    $\theta_{13}$ (indicated by the black star) at a high confidence level.
    The left hand plots are for \TtoK\ and \DoubleChooz, and the right hand
    ones are for \NOvA\ and \DCext. The gray shading represents the 90\%
    confidence region from the reactor experiment, and the vertical black
    line shows the corresponding best fit value for $\theta_{13}$. The
    90\% contours from the superbeam are shown as solid blue (dark gray) lines for
    a normal hierarchy fit, and as dashed pink (light gray) lines for an inverted
    hierarchy fit. The colored diamonds represent the corresponding best fit
    values. In interpreting the computed $\chi^2$ values, we have assumed
    2~degrees of freedom for the beam experiments, and one degree of freedom
    for the reactor setups.}
  \label{fig:th13delta}
\end{figure*}

Using the simulation techniques discussed in the previous section, we can now
determine the errors that are introduced when non-standard interactions are present
in reactor and superbeam experiments, but are not properly taken into account 
in the respective fits. Possible outcomes of such fits are shown in
Fig.~\ref{fig:th13delta} for our two scenarios. As ``true'' parameter values,
we have taken $\sthchooz = 0.05$ and $\delta_{\rm CP} = \pi$, and the NSI
contribution was assumed to be $\eps^m_{e \tau} = 0.5 e^{-i \pi/2 }$ in the
upper panel, and $\eps^s_{e \tau} = \eps^d_{\tau e} = 0.05$ (fulfilling
Eq.~\eqref{eq:eps-e-constraint}) in the lower panel. These NSI parameters
are rather large, but still consistent with current bounds~\cite{Davidson:2003ha,
Gonzalez-Garcia:2001mp}. According to the discussion in Sec.~\ref{sec:theory},
only the superbeam experiment should be affected in the first case, while in the
second case, there should be an impact on both experiments. The shaded areas
show the 90\% confidence regions for the reactor experiment, while the colored
contours are for the superbeam. The data has been calculated under the assumption
of a normal mass hierarchy, but the fit has been performed for both the normal
mass ordering (solid blue/dark gray contours) and for the inverted ordering
(dashed pink/light gray contours). The vertical black line and the colored
diamonds represent the respective best fit values, while the black star stands
for the assumed ``true'' parameter values. We have assumed two degrees of freedom
for the superbeam experiments, and one degree of freedom for the reactor setups,
which are insensitive to $\delta_{\rm CP}$.

\begin{figure*}
  \vspace{-0.6cm}
  \begin{center}
    \begin{tabular}{c@{\hspace{-0.5cm}}c@{\hspace{-0.5cm}}c}
      \includegraphics[width=6.0cm]{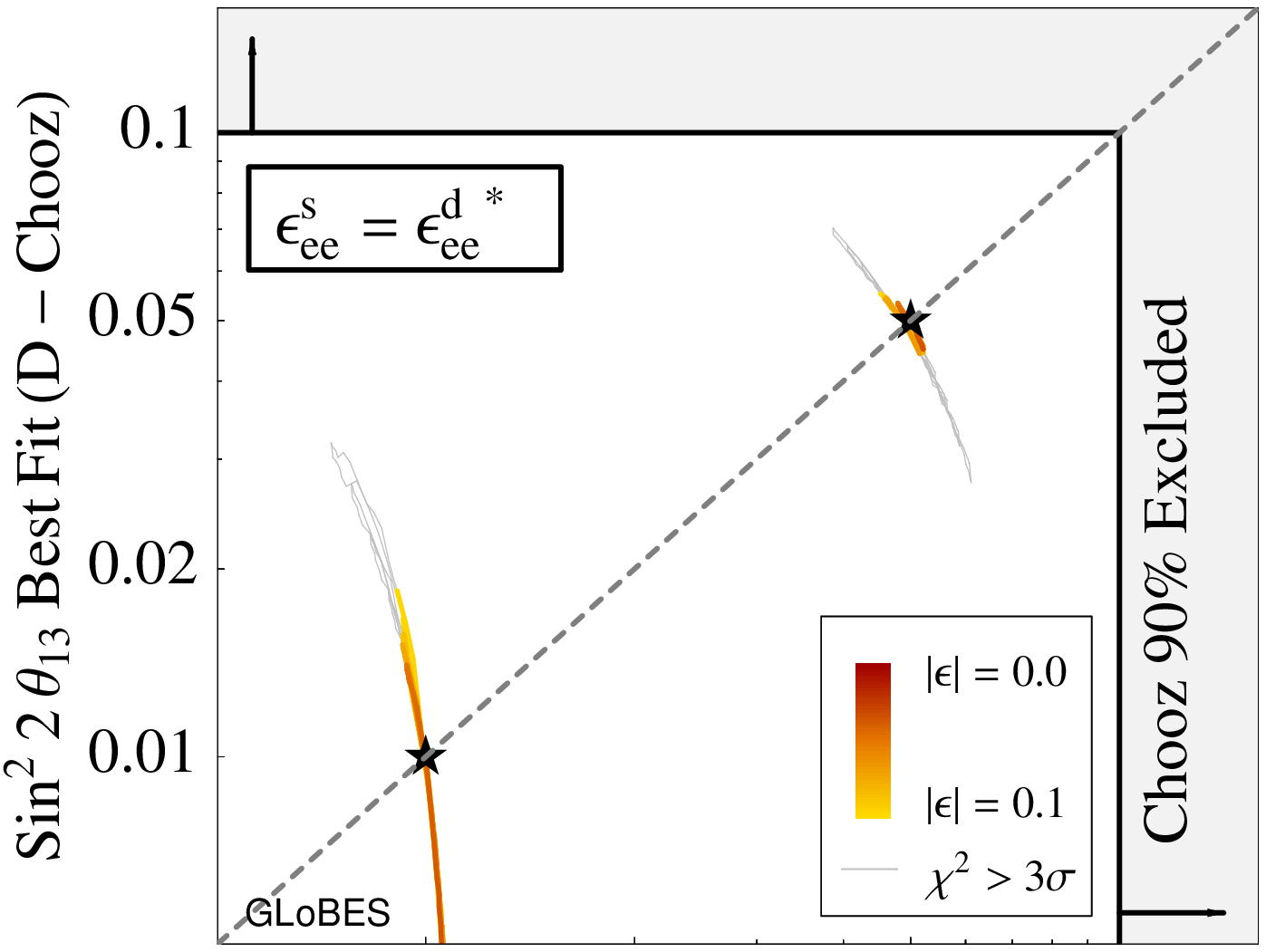}       &
      \includegraphics[width=6.0cm]{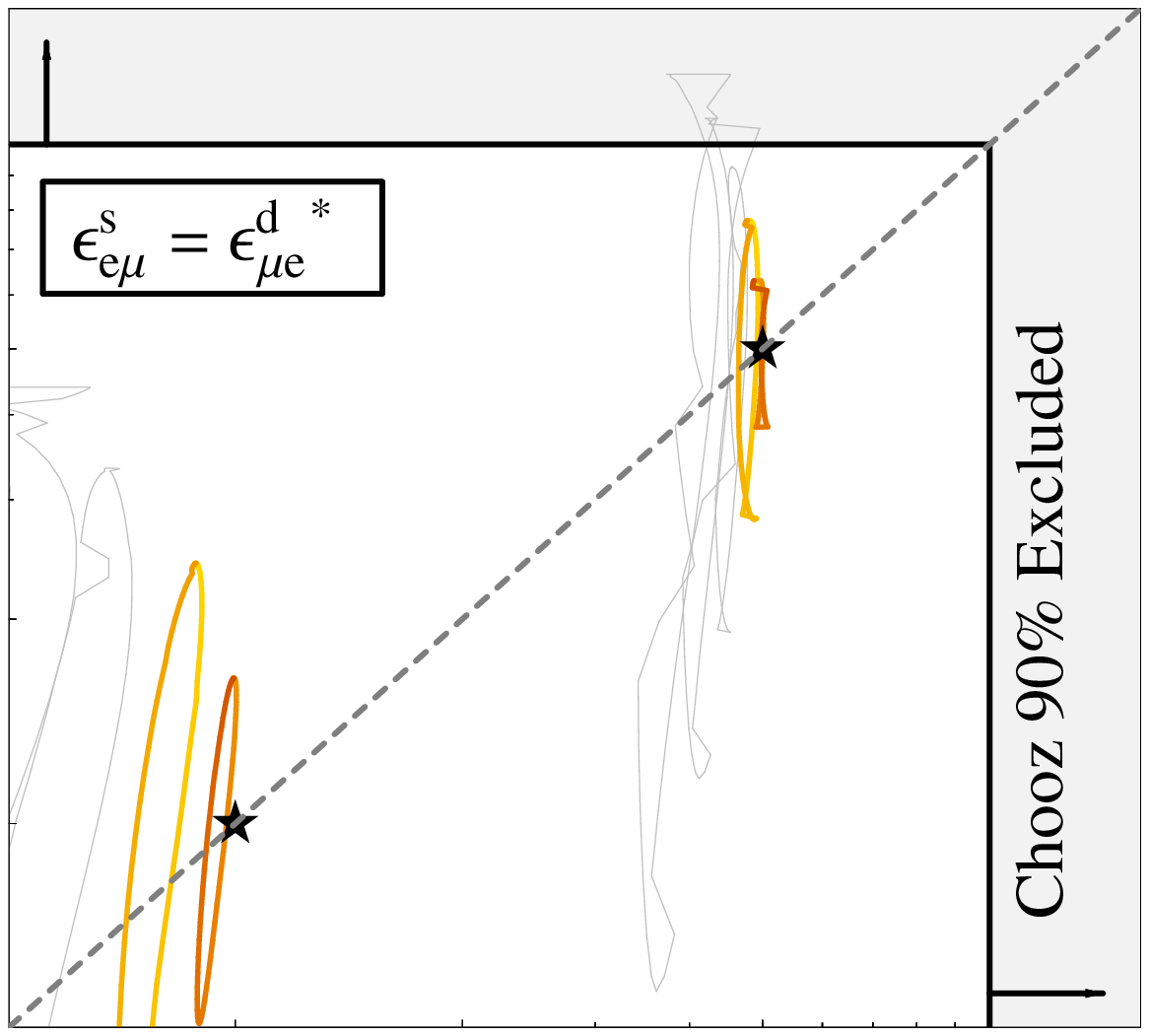}      &
      \includegraphics[width=6.0cm]{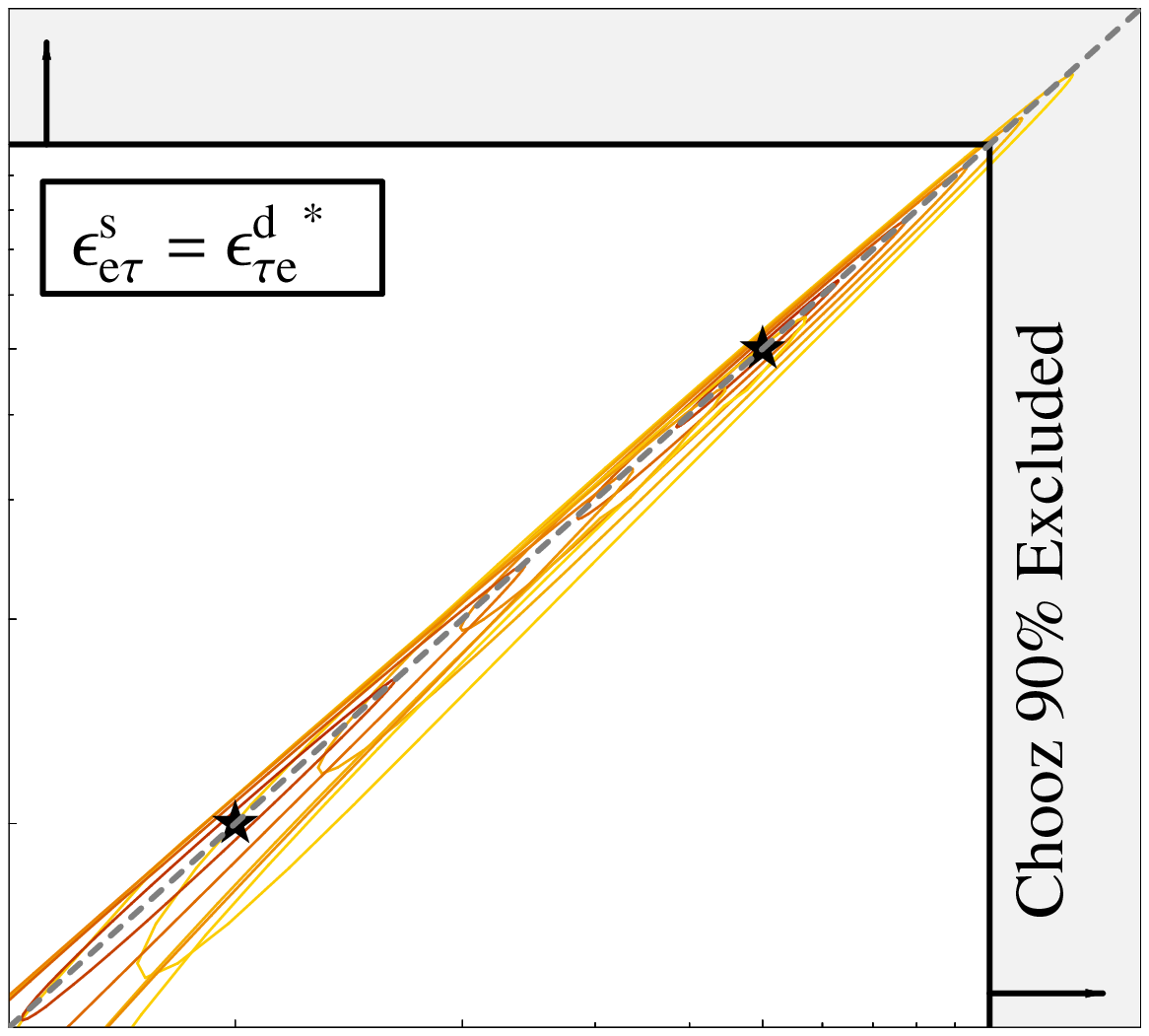}     \\[-0.5cm]
      \includegraphics[width=6.0cm]{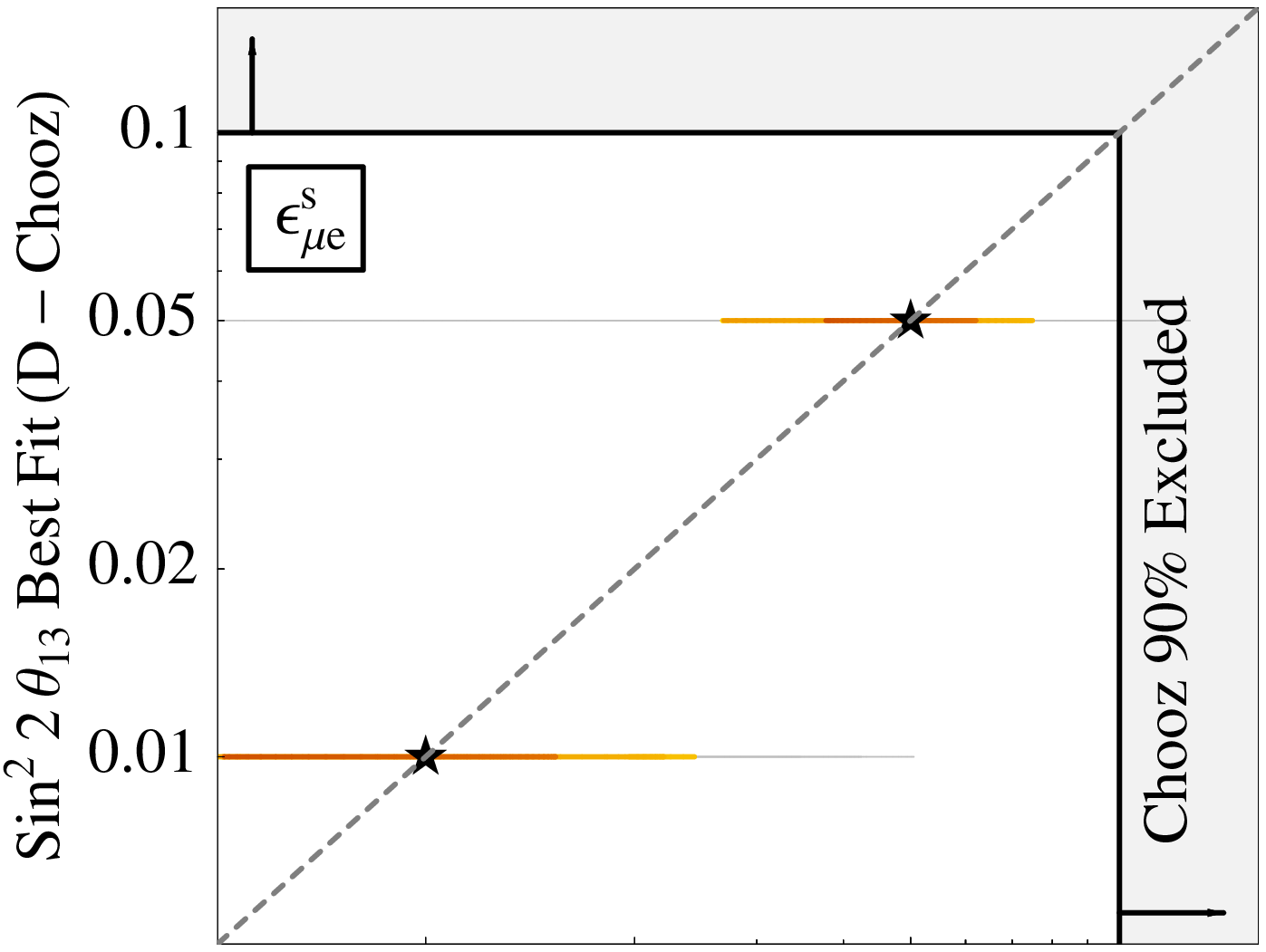}      &
      \includegraphics[width=6.0cm]{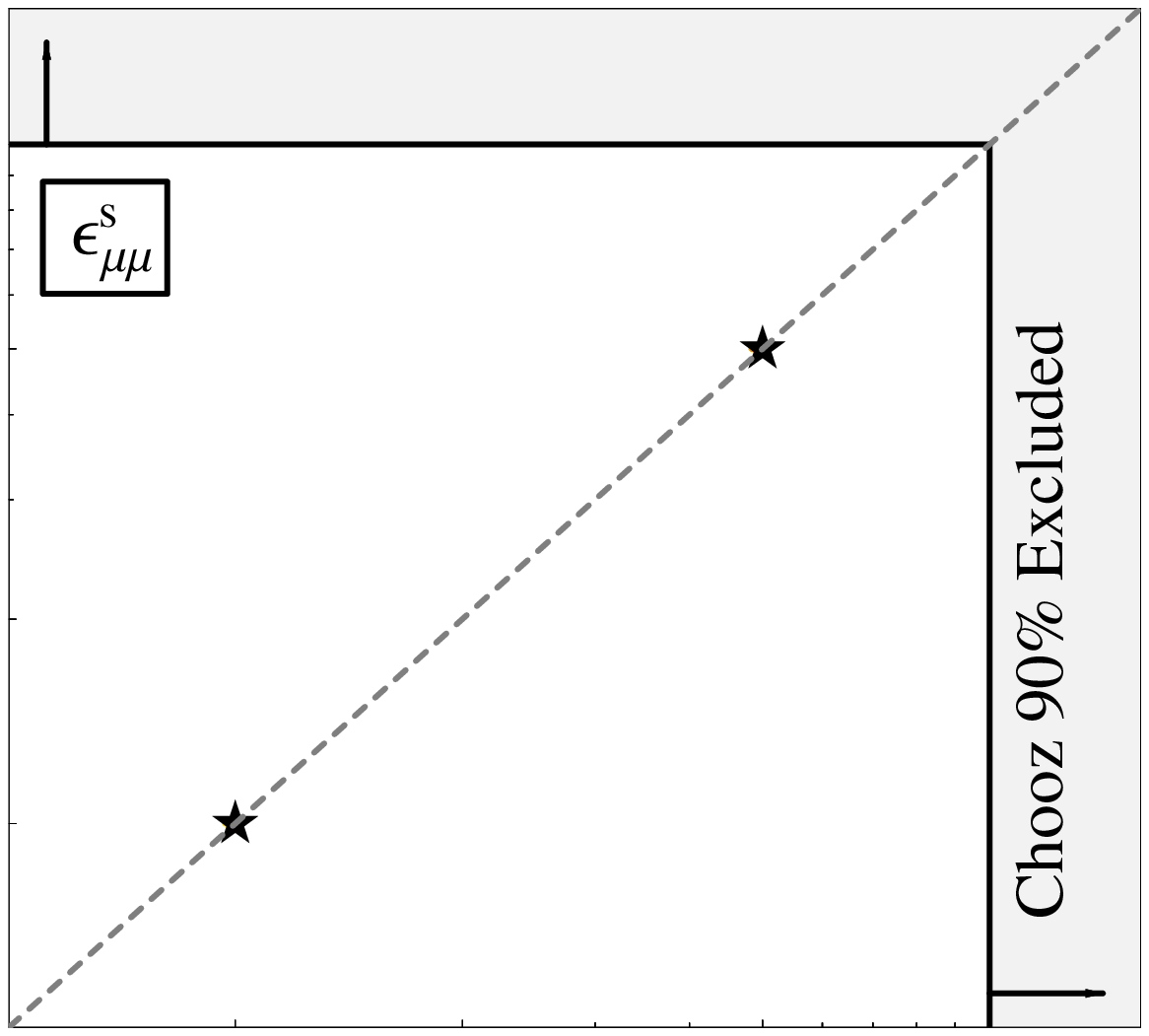}     &
      \includegraphics[width=6.0cm]{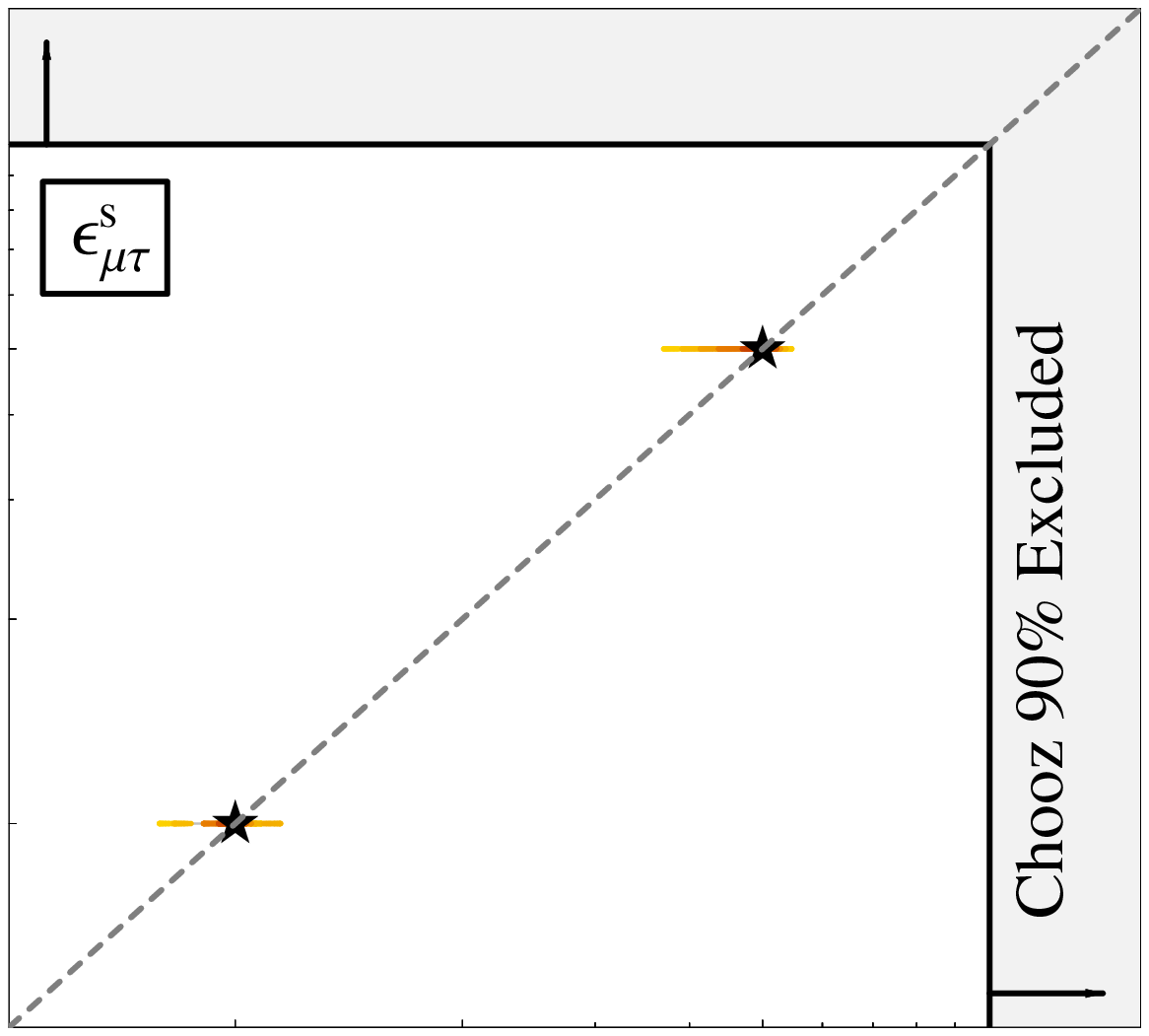}    \\[-0.5cm]
      \includegraphics[width=6.0cm]{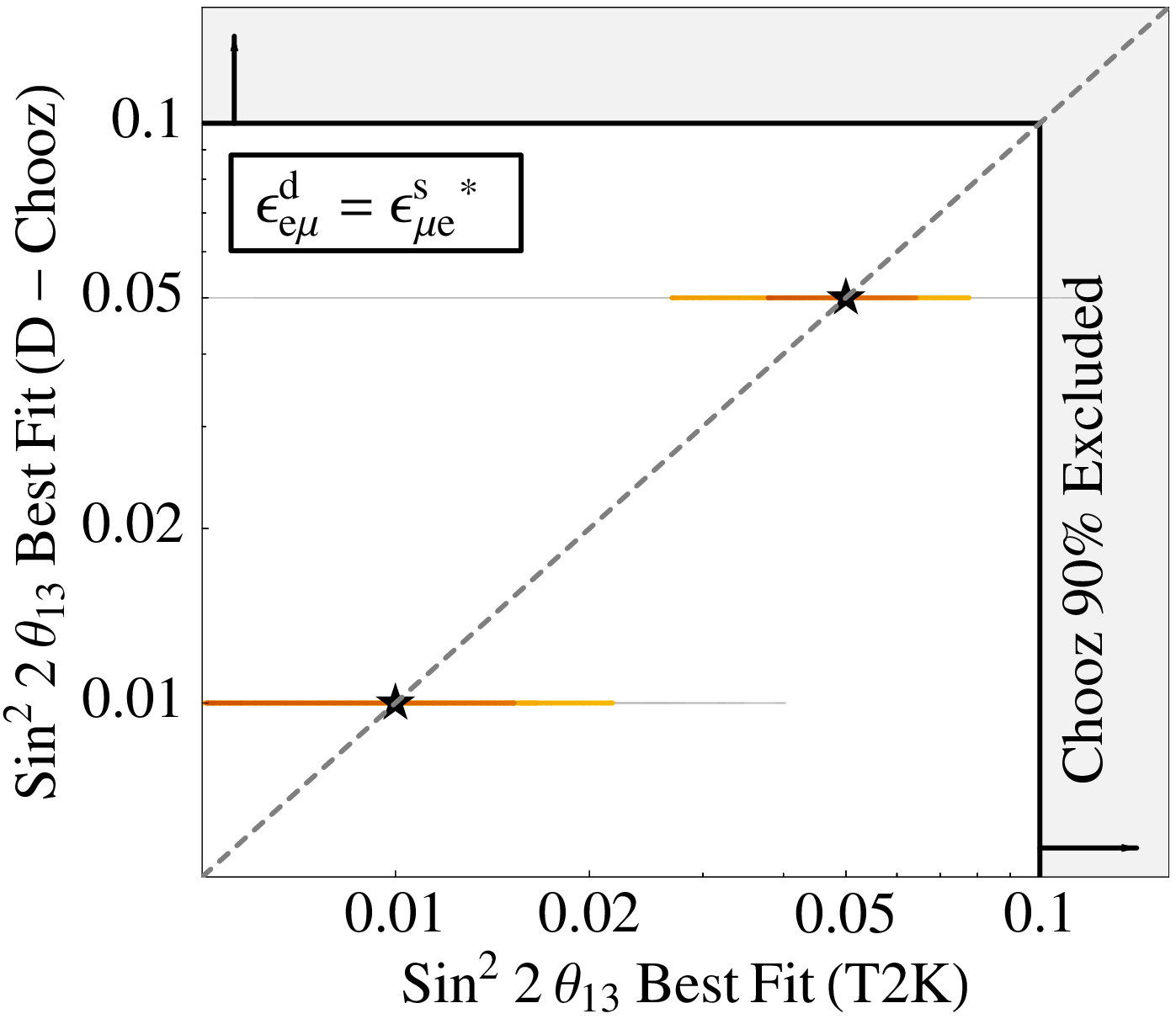}      &
      \includegraphics[width=6.0cm]{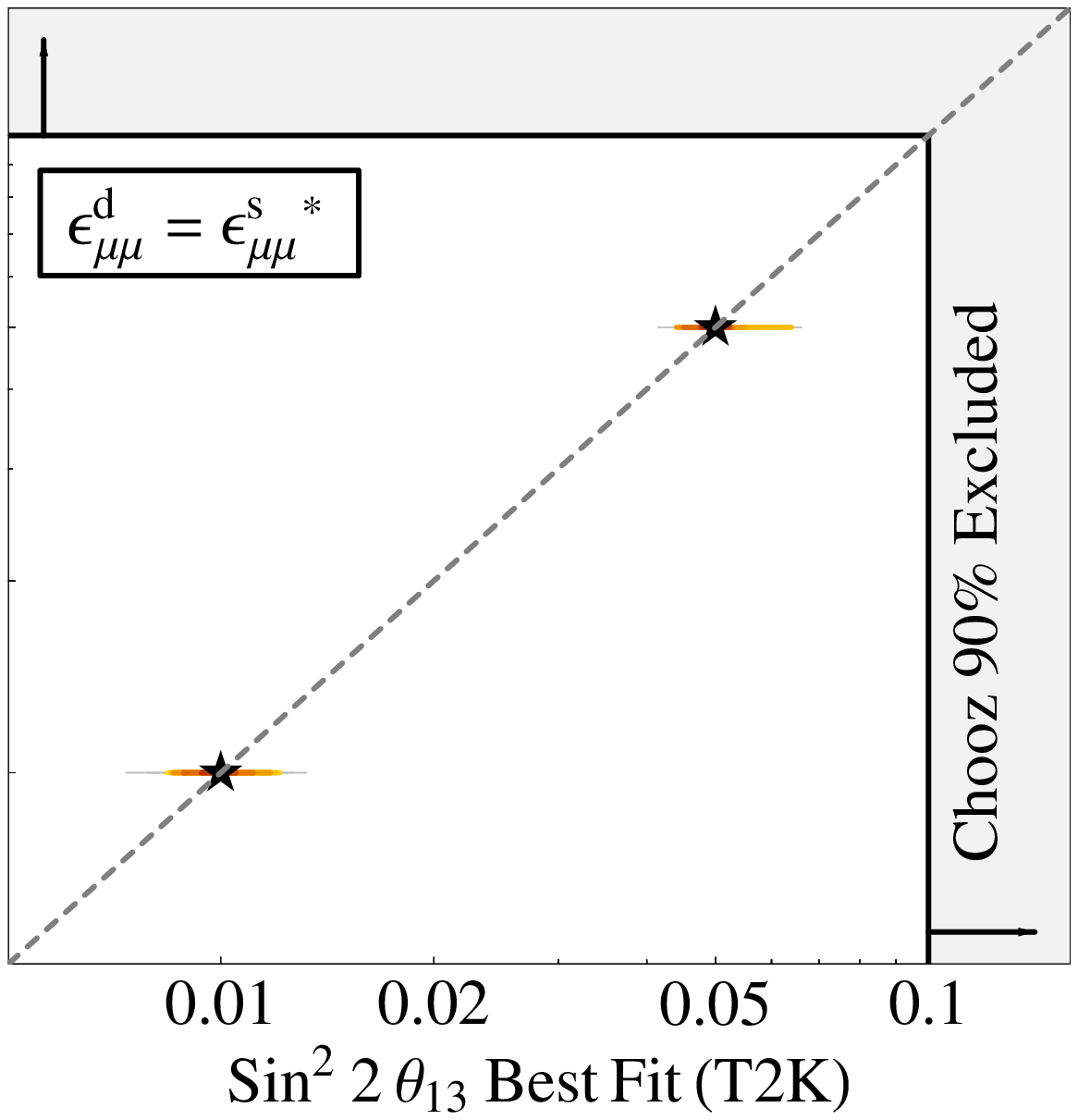}     &
      \includegraphics[width=6.0cm]{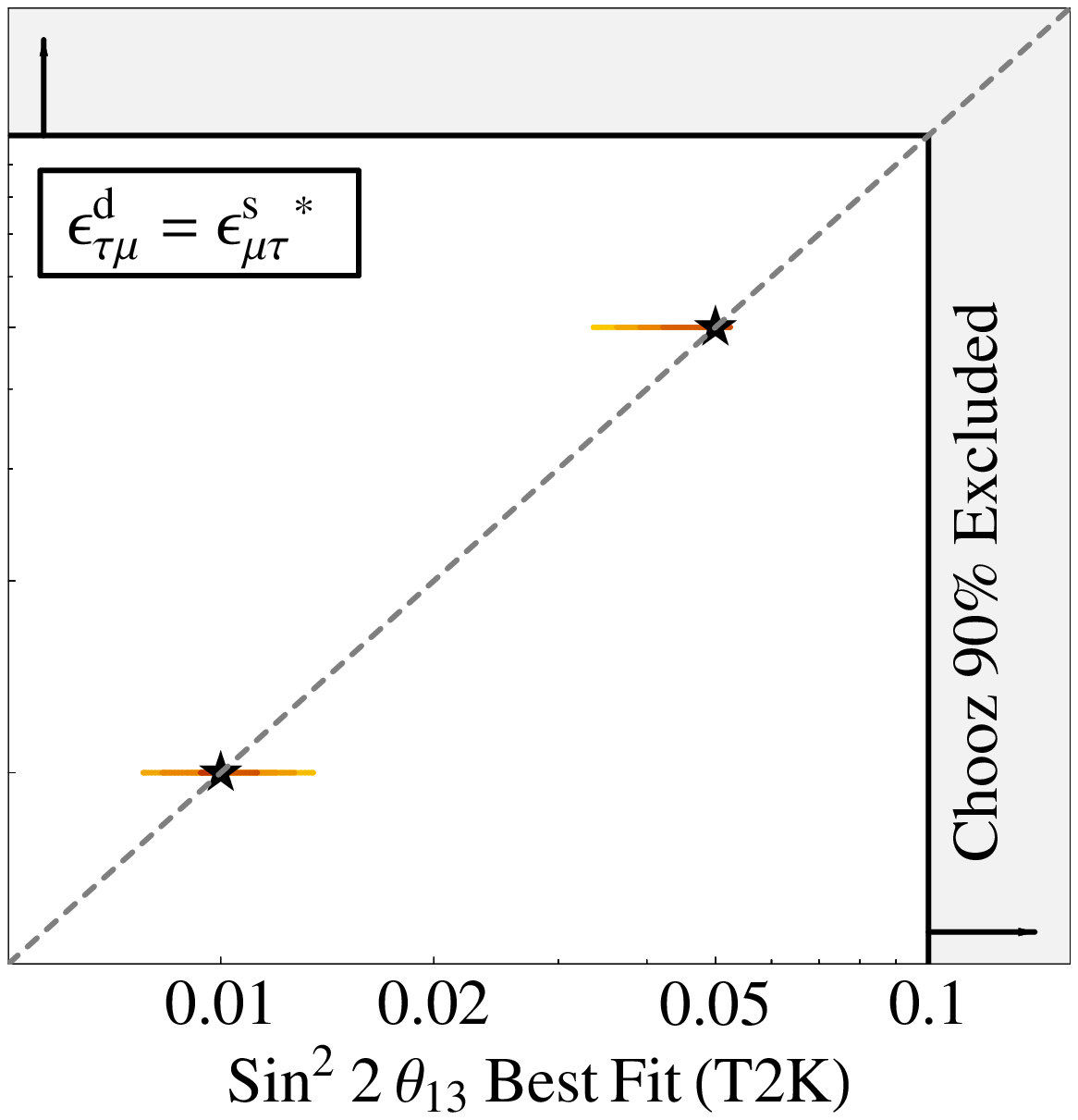}    \\[-0.5cm]
    \end{tabular}
  \end{center}
  \caption{Distortion of the $\theta_{13}$ fits in \TtoK\ and \DoubleChooz\
    in the presence of $\eps^s_{\alpha\beta}$ and $\eps^d_{\beta\alpha}$. For each
    plot, the modulus of the corresponding $\eps$ parameter has been varied from
    $0$ to $0.1$, and its phase from $0$ to $2\pi$. For every such combination,
    we show the result of standard oscillation fits of $\theta_{13}$.
    Connected lines represent contours of equal $|\eps_{\alpha\beta}|$ and varying
    phase. Dark red (dark gray) lines correspond to $|\eps_{\alpha\beta}| = 0$,
    while yellow (medium gray) lines correspond to $|\eps_{\alpha\beta}| = 0.1$.
    Points giving a quality of fit worse than $3\sigma$ in at least one of the two
    experiments are plotted in light gray. The black stars indicate the assumed
    ``true'' $\sthchooz$.}
  \label{fig:th13fits-T2K-epsilon-sd}
\end{figure*}

\begin{figure*}
  \begin{center}
    \begin{tabular}{c@{\hspace{-0.5cm}}c@{\hspace{-0.5cm}}c}
      \includegraphics[width=6.0cm]{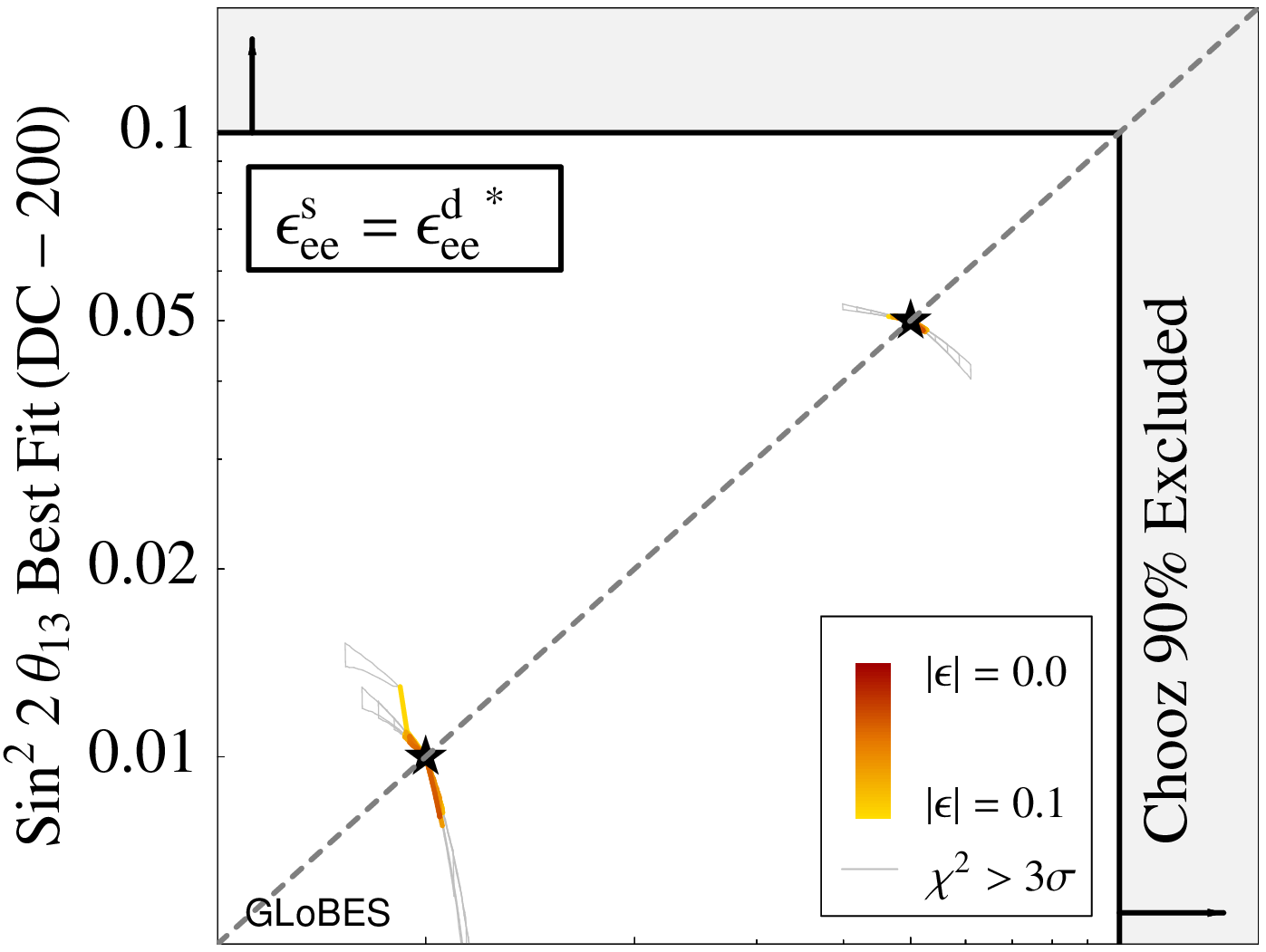}       &
      \includegraphics[width=6.0cm]{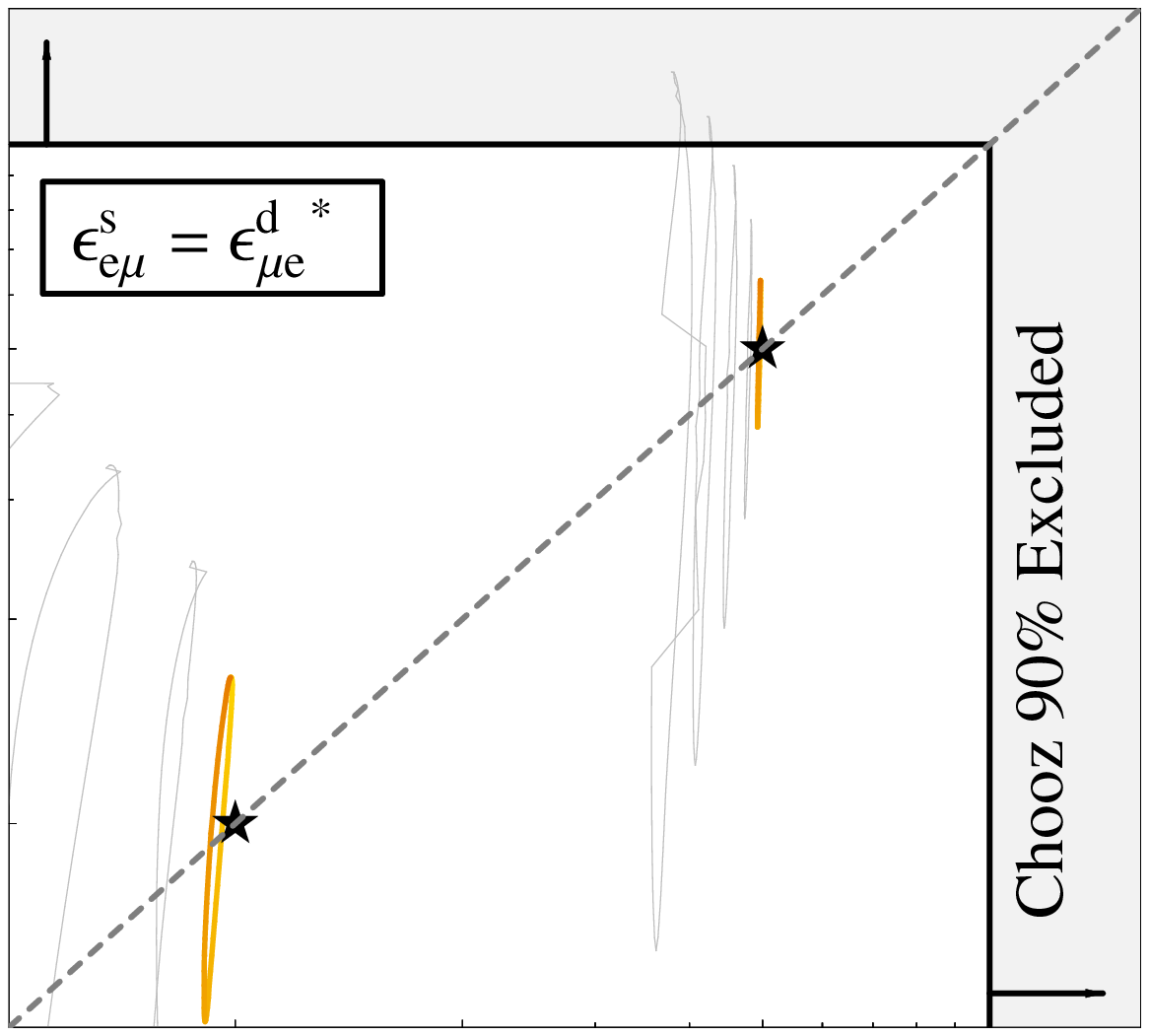}      &
      \includegraphics[width=6.0cm]{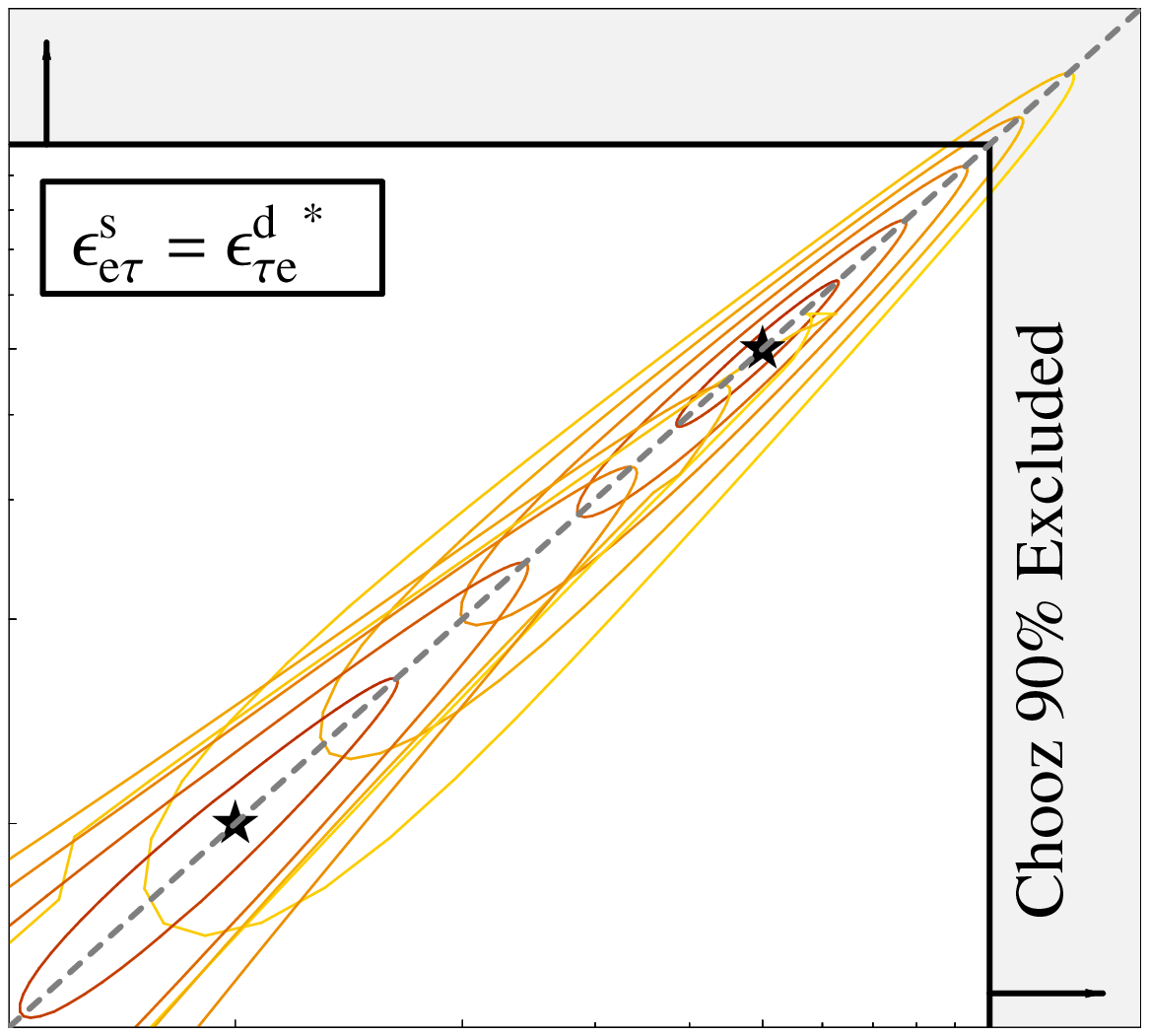}     \\[-0.5cm]
      \includegraphics[width=6.0cm]{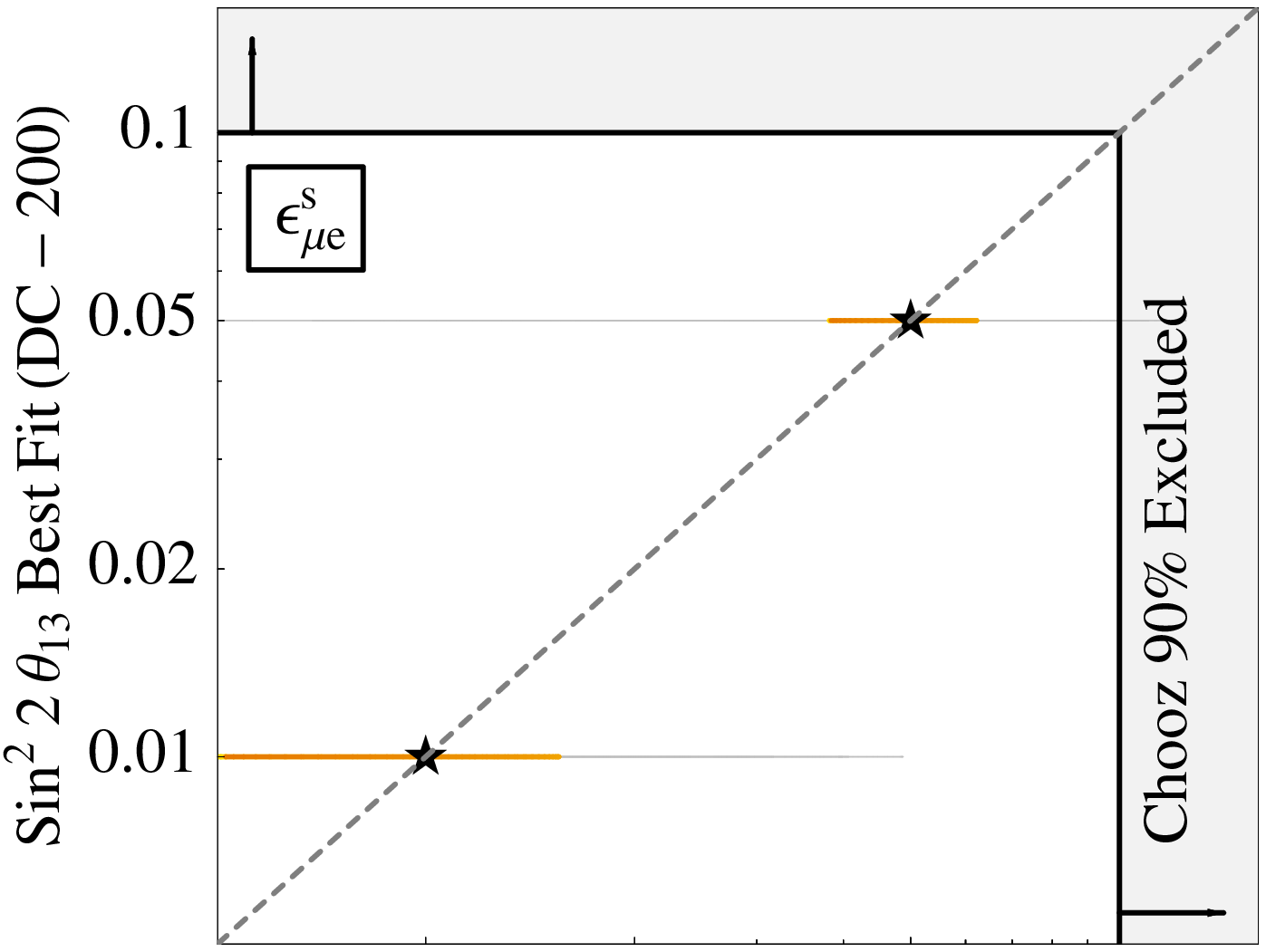}      &
      \includegraphics[width=6.0cm]{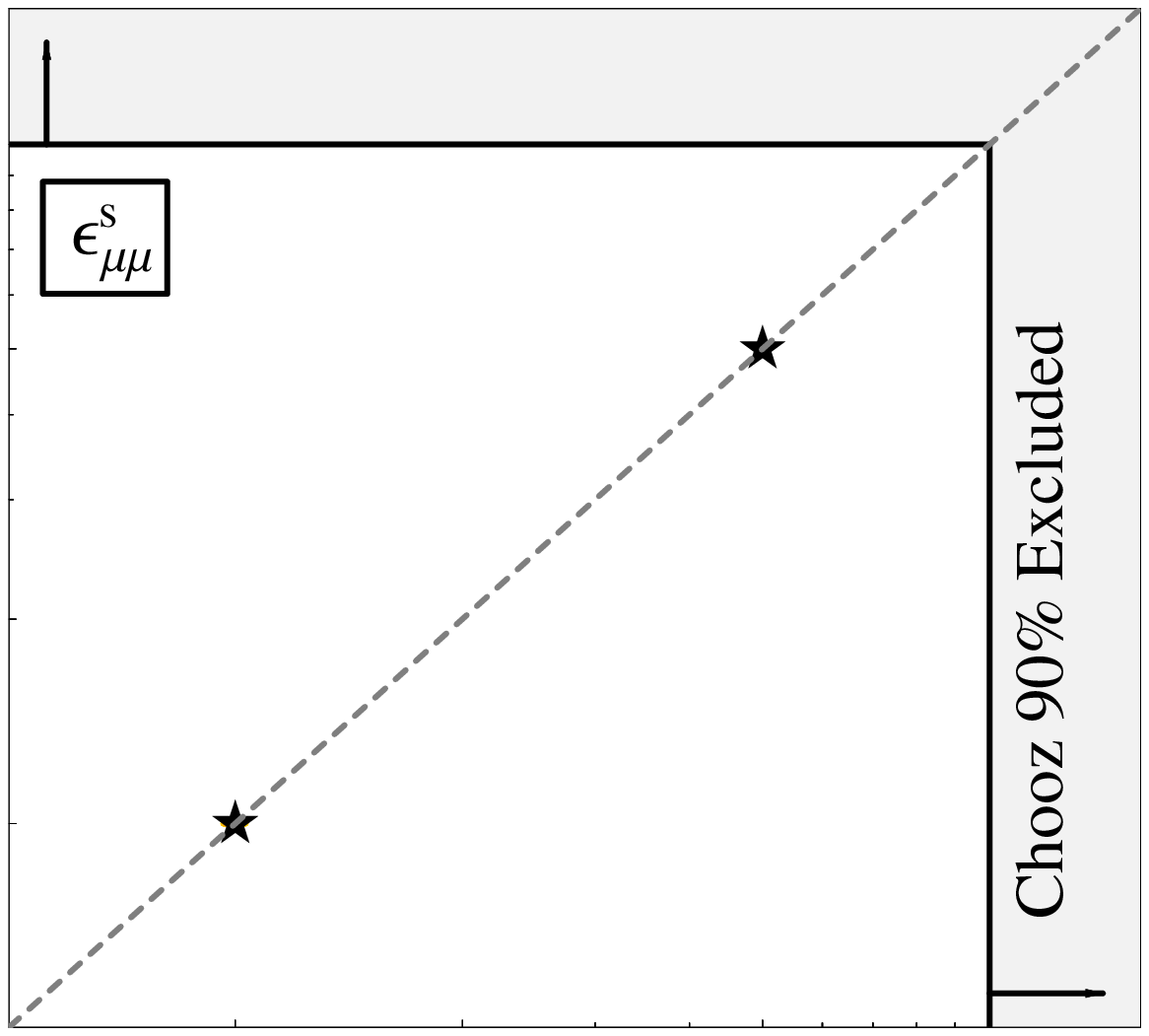}     &
      \includegraphics[width=6.0cm]{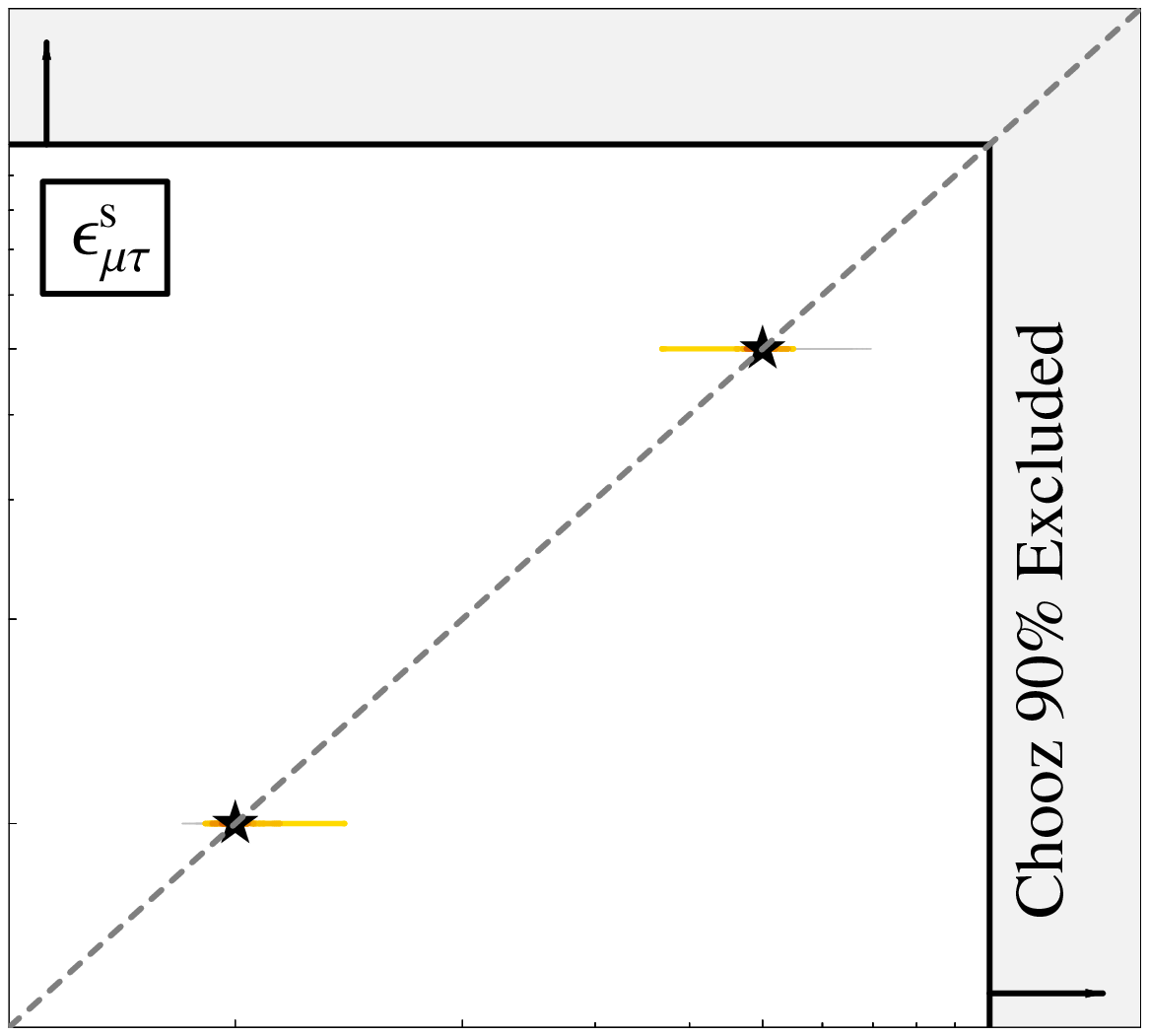}    \\[-0.5cm]
      \includegraphics[width=6.0cm]{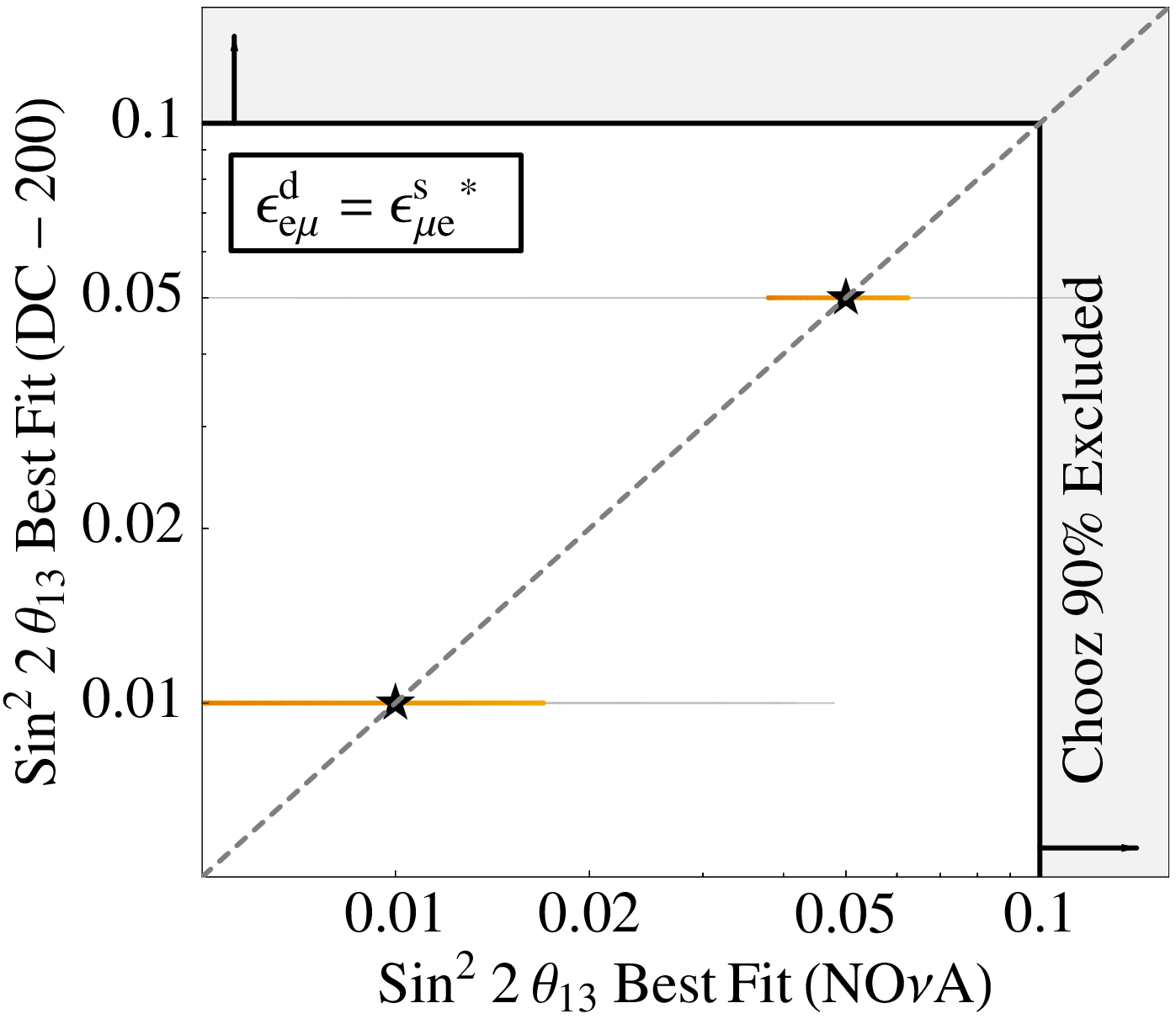}      &
      \includegraphics[width=6.0cm]{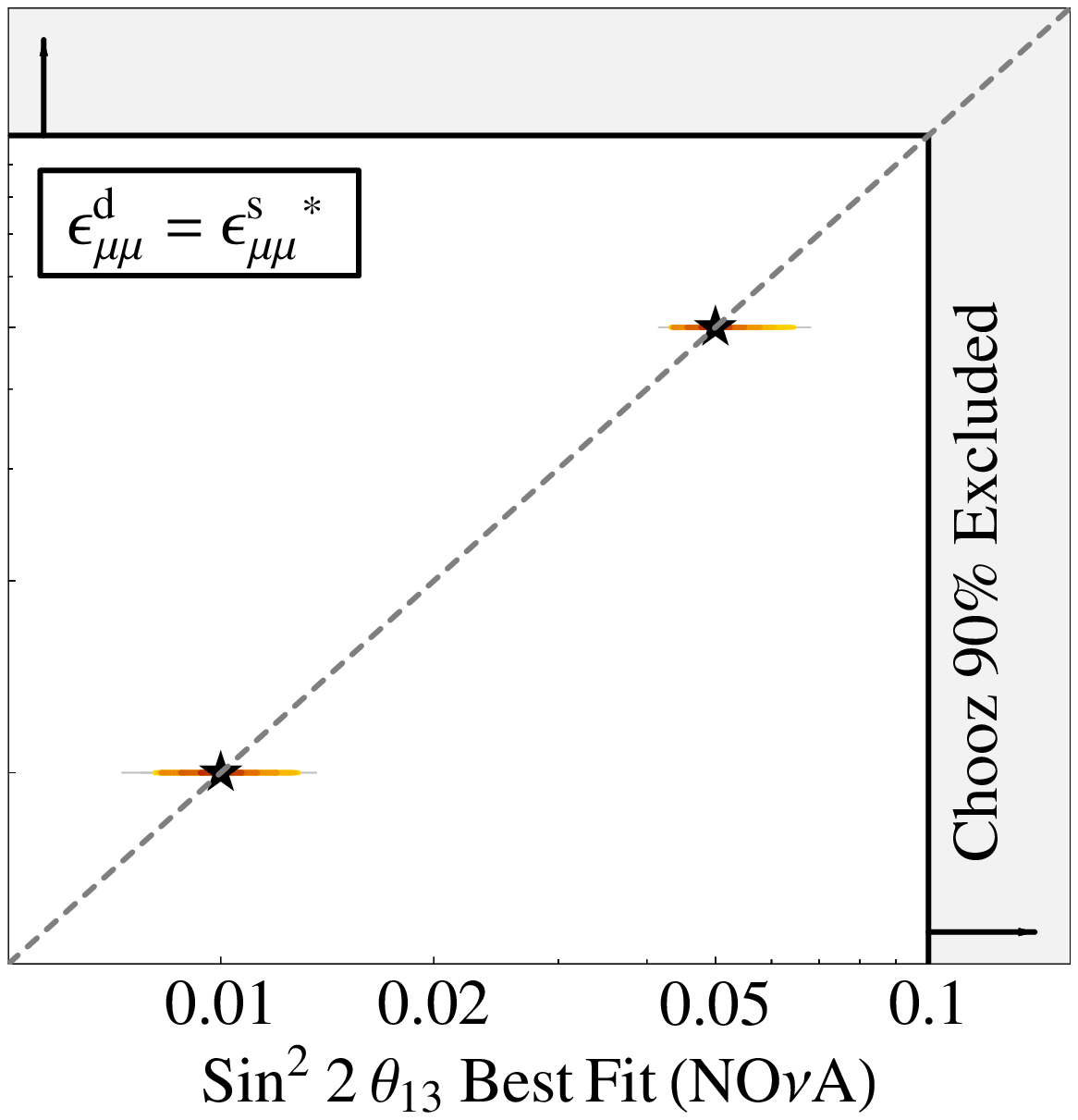}     &
      \includegraphics[width=6.0cm]{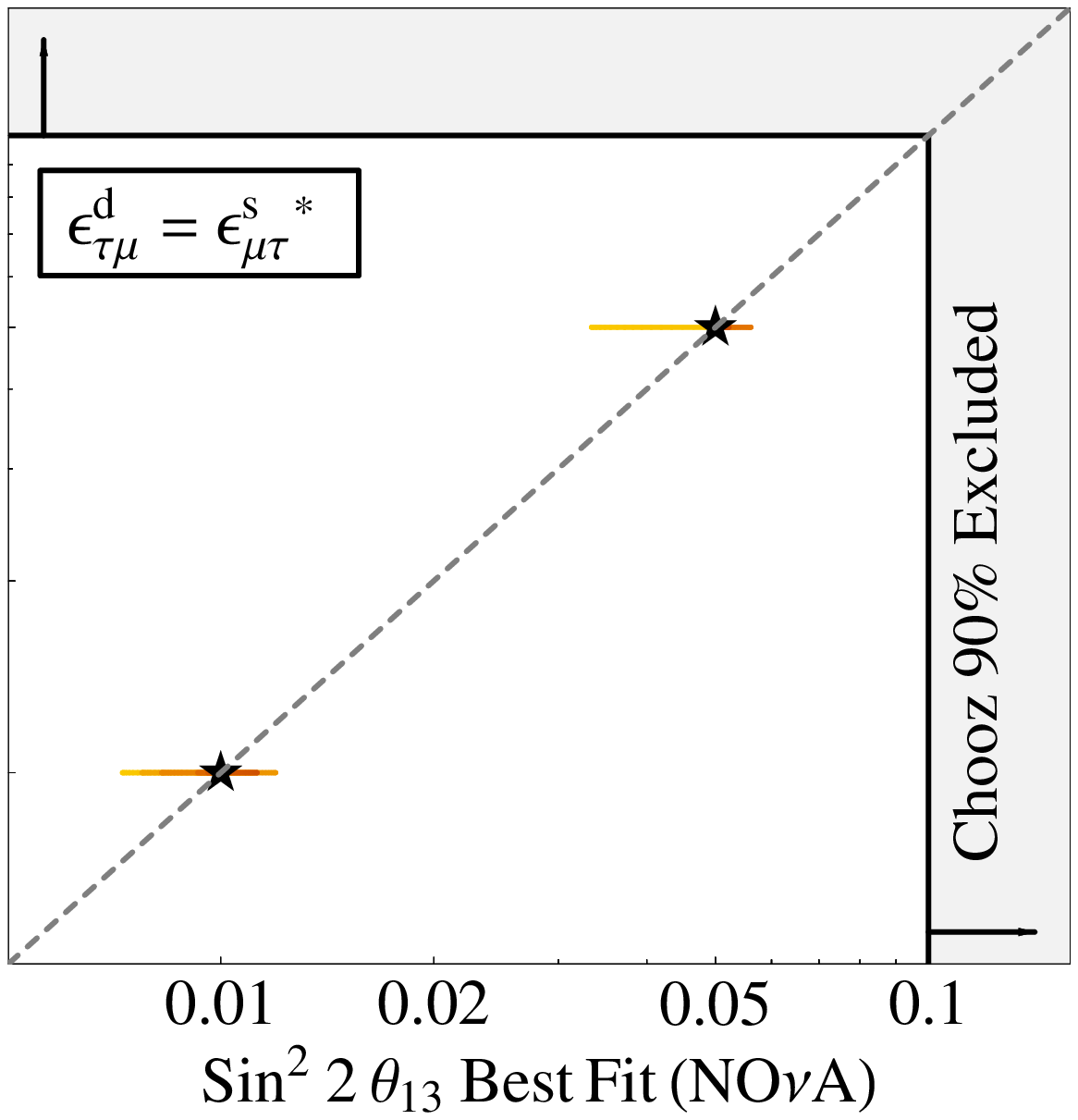}    \\[-0.5cm]
    \end{tabular}
  \end{center}
  \caption{Effect of $\eps^s_{\alpha\beta}$ and $\eps^d_{\beta\alpha}$
    on the $\theta_{13}$ fits in \NOvA\ and \DCext. The color-coding is the
    same as in Fig.~\ref{fig:th13fits-T2K-epsilon-sd}.}
  \label{fig:th13fits-NOvA-epsilon-sd}
\end{figure*}

\begin{figure}
  \begin{center}
    \includegraphics[width=8.0cm]{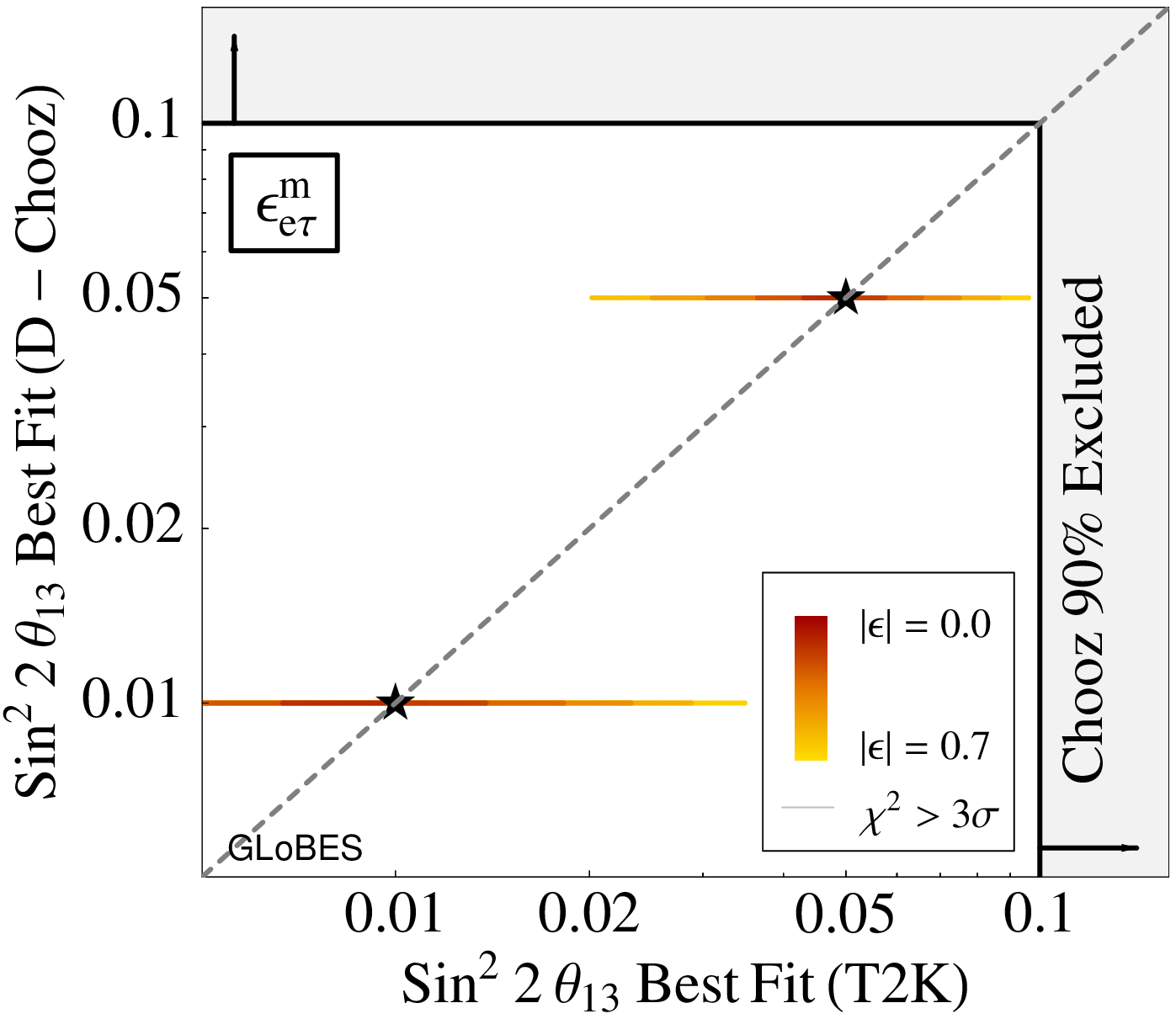} \\[0.5cm]
    \includegraphics[width=8.0cm]{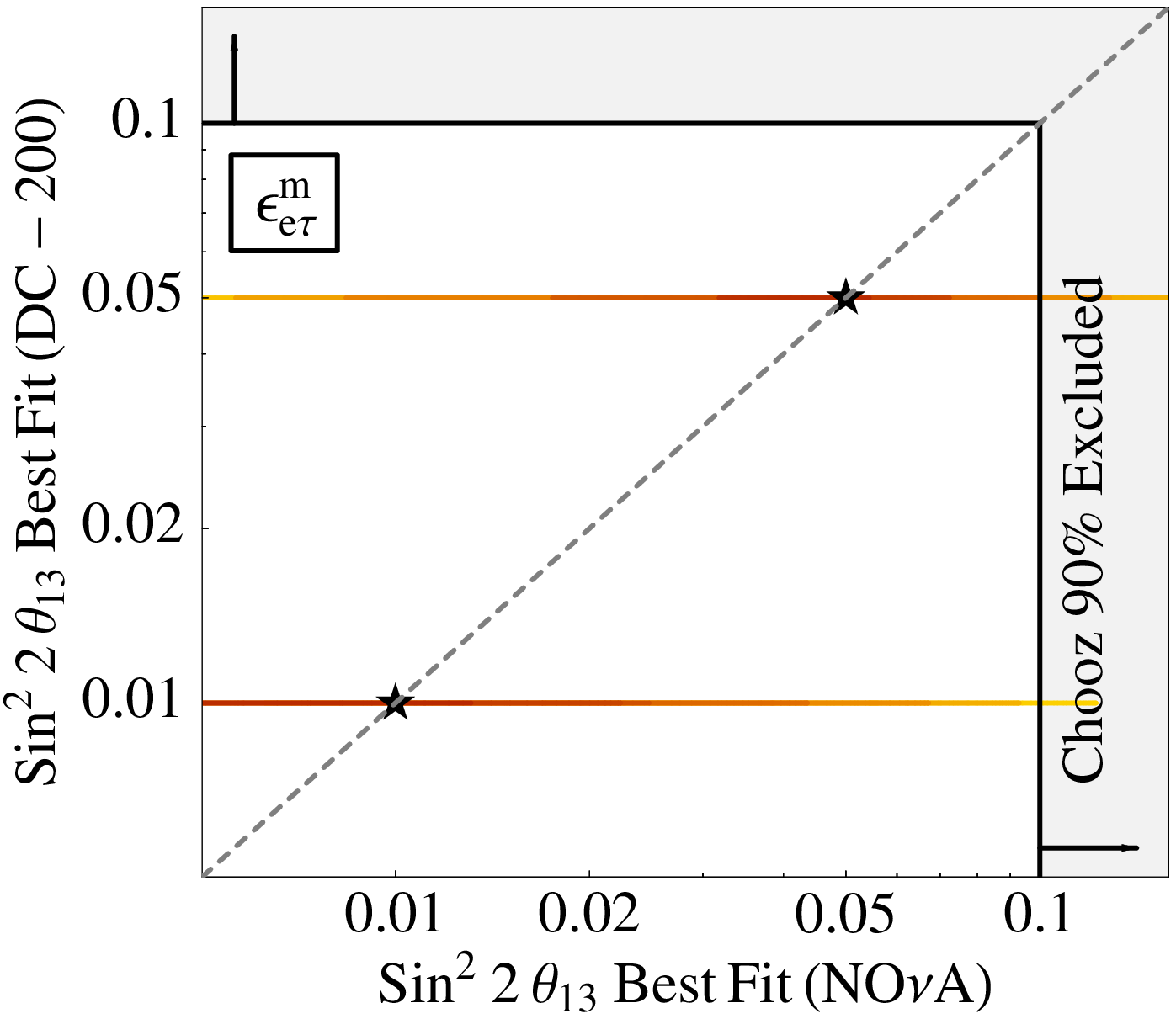}
  \end{center}
  \caption{Effect of $\eps^m_{e\tau}$ on the $\theta_{13}$ fits in
    \TtoK\ and \DoubleChooz, resp.\ in \NOvA\ and \DCext. We do not
    show plots for the other entries of $\eps^m$, since these are
    either strongly constrained already ($\eps^m_{e\mu}$), or do
    not affect the $\theta_{13}$ measurement ($\eps^m_{ee}$,
    $\eps^m_{\mu\mu}$, $\eps^m_{\mu\tau}$, and $\eps^m_{\tau\tau}$).
    The color-coding is the same as in
    Fig.~\ref{fig:th13fits-T2K-epsilon-sd}, but the scale is different
    since the bound on $\eps^m_{e\tau}$ is weaker than that
    for $\eps^{s,d}_{\alpha\beta}$~\cite{Davidson:2003ha,
    Gonzalez-Garcia:2001mp}.}
  \label{fig:th13fits-epsilon-m}
\end{figure}

It is obvious from the plots that the standard oscillation fit to $\theta_{13}$
and $\delta_{\rm CP}$ can be severely wrong if NSI are present. This observation
is similar to the one made in~\cite{Huber:2002bi} for a neutrino factory.
In the case shown in the upper plots of Fig.~\ref{fig:th13delta}, the reactor
experiment gives the correct best fit value, but the superbeam results conflict
with this measurement. In the case of \NOvA, we even obtain a fit value above
the \Chooz\ bound and a fake  hint to the mass hierarchy, indicated by the fact
that, within the resolution of the plot, the 90\% contour reduces to a single
point. In the second case (lower plots), both experiment agree very well, but
they erroneously seem to rule out the ``true'' $\theta_{13}$.

Of course, the NSI scenarios analyzed in Fig.~\ref{fig:th13delta} were only
two examples, and a more systematic analysis of non-standard interactions
in reactor and superbeam experiments is desirable. This is done in
Figs.~\ref{fig:th13fits-T2K-epsilon-sd} -- \ref{fig:th13fits-epsilon-m},
where we show how the (standard oscillation) $\theta_{13}$ fits in
\TtoK/\DoubleChooz\ respectively in \NOvA/\DCext\ may be distorted in the
presence of non-standard interactions. For each diagram,
only one of the independent $\eps$ parameters was assumed to be non-zero, but we
have ensured that Eqs.~\eqref{eq:eps-e-constraint} and~\eqref{eq:eps-mu-constraint}
are fulfilled. In particular, we did not consider the hypothetical case
$\eps^s_{\mu\alpha} = 0$, $\eps^d_{\alpha\mu} \neq 0$, but only
$\eps^s_{\mu\alpha} \neq 0$, $\eps^d_{\alpha\mu} = 0$ and
$\eps^s_{\mu\alpha} = \eps^d_{\alpha\mu} \neq 0$. Moreover we omit
all entries of $\eps^m$ except $\eps^m_{e\tau}$, because they
are either strongly constrained already ($\eps^m_{e\mu}$), or
do not have any impact on the $\theta_{13}$ measurements ($\eps^m_{ee}$,
$\eps^m_{\mu\mu}$, $\eps^m_{\mu\tau}$, and $\eps^m_{\tau\tau}$).
The modulus of each parameter has been varied between 0 and its current upper
bound, which is $0.1$ for $\eps^{s,d}_{\alpha\beta}$ from universality in
charged lepton decays~\cite{Gonzalez-Garcia:2001mp}, and $0.7$ for
$\eps^m_{e\tau}$~\cite{Davidson:2003ha}.%
\footnote{Note that, according to the naive estimate from
Eq.~\eqref{eq:epsilon-estimate}, such large values of $|\eps|$ would correspond
to $M_{\rm NSI} \sim \mathcal{O}(100~{\rm GeV})$. In many models, such
low new physics scales are already ruled out, but in our model-independent
treatment, they are still viable.}
The complex phases were allowed to vary between $0$ and $2\pi$.
For each combination of $|\eps|$ and $\arg(\eps)$, we have then performed a fit
assuming standard oscillations, and the resulting best fit values for
$\theta_{13}$ are shown in the plots. Points giving a good fit (better than
$3\sigma$ in both experiments), are drawn as thick colored lines, with the hue
indicating the respective value of $|\eps|$. Dark red (dark gray) corresponds
to $|\eps| = 0$, while yellow (medium gray) corresponds to the upper bound
of $|\eps|$. Points giving a fit quality worse than $3\sigma$ are shown by
thin light gray lines, and the information on $|\eps|$ is omitted for them.
All computations have been performed for two different ``true'' values,
$\sthchooz = 0.01$ and $\sthchooz = 0.05$, as indicated by the black stars.

By comparing the plots with Tab.~\ref{tab:suppression-factors}, we find
that our expectations for the impact of the different $\eps$ parameters
from the discussion in Sec.~\ref{sec:theory} are confirmed. A particularly
interesting situation arises for $\eps^d_{\tau e}$, because this parameter
has a sizeable effect in both, the reactor experiment and the superbeam setup.
It is especially dangerous because it induces a similar offset in both
experiments, i.e.\ one would find perfectly consistent $\theta_{13}$ fits,
which might, however, be far away from the true value.

Other parameters may lead to fit points far from the diagonal, corresponding
to seemingly conflicting fits. An interesting case is the $\eps^d_{\mu e}$
term, for which the non-standard interaction mimics a significantly
modified $\nu_e$ flux in the near detector. This, in turn, leads
to a miscalibration of the beam-intrinsic backgrounds, so that, at the
far site, many of the actually oscillation-induced $\nu_e$ events will
be mistaken as background. Thus, the fit value for $\theta_{13}$ becomes
too small. However, we can also read off from the plot, that, in this
situation, the quality of the standard oscillation fit becomes so bad
that the NSI effect can actually be detected. Note that the curves for
large $|\eps^d_{\mu e}|$ look slightly untidy, because for some parameter
values, the smallest $\chi^2$ is provided by the normal hierarchy fit,
while for others the inverted hierarchy fit is marginally better. Therefore,
frequent ``jumps'' between these two solutions occur.

When interpreting Figs.~\ref{fig:th13fits-T2K-epsilon-sd} and
\ref{fig:th13fits-NOvA-epsilon-sd}, it is important to keep in mind
that the error bars of the experiments considered here are rather large (cf.\
Fig.~\ref{fig:th13delta}), so that even sizeable deviations from the
diagonals will in most cases only create some tension, but no unambiguous
contradiction between the beam and reactor fits.

Of the non-standard matter effects, we expect from
Tab.~\ref{tab:suppression-factors} that only $\eps^m_{e\mu}$ and
$\eps^m_{e\tau}$ should have any  effect on the $\theta_{13}$ fits.
$\eps^m_{e\mu}$ is already strongly constrained from charged lepton
flavor violation experiments~\cite{Davidson:2003ha}, but
$\eps^m_{e\tau}$ may still have a large impact. In fact, for extreme
values of this parameter, there is even the possibility that
\NOvA\ would erroneously report a $\theta_{13}$ value above the
\Chooz\ bound.

Let us emphasize that, in order to obtain reliable estimates for the impact of
non-standard interactions on reactor and superbeam experiments, it is crucial
to take the information from the near detectors into account. To show this,
we have also studied how Figs.~\ref{fig:th13fits-T2K-epsilon-sd}
and~\ref{fig:th13fits-NOvA-epsilon-sd} get modified if we use a simplified
simulation, in which the near detector does not appear explicitly,
but only through suitably small values for the systematical uncertainties.
In doing so, we have again treated $\eps^s$ and $\eps^d$ as completely
independent matrices, possibly violating Eqs.~\eqref{eq:eps-e-constraint}
and~\eqref{eq:eps-mu-constraint}. Thus, the results are also applicable
to setups where $\eps^s$ and $\eps^d$ are indeed unrelated.
In accordance with our expectations from Tab.~\ref{tab:suppression-factors},
we have found:
\begin{itemize}
  \item For $\eps^s_{ee}$ and $\eps^d_{ee}$, the effect on the reactor
  becomes stronger without the proper treatment of the near
  detector, because these terms no longer cancel then. Moreover, the
  discovery potential becomes worse, i.e.\ there will be fewer gray segments
  in the plot.

  \item For the superbeam, the discovery of $\eps^s_{\mu e}$ and
  $\eps^d_{\mu e}$ becomes also much harder without the near detector, because
  the clear signature of an apparently modified $\nu_e$ flux at the near site
  is no longer available. Moreover, for $\eps^d_{\mu e}$, the strong
  impact of the NSI on the $\theta_{13}$ fit in the superbeam, which
  we have identified as a near detector effect in the above discussion,
  vanishes in the single-detector simulation.

  \item The contours for $\eps^s_{e\tau} = \eps^{d*}_{\tau e}$ are
  deformed without the near detector because in this (unrealistic)
  situation, it is no longer possible to misinterpret the non-standard
  effect as a reactor flux calibration error. Note that this misinterpretation
  is only due to the fact that $\eps^s_{e\tau}$ and $\eps^{d*}_{\tau e}$
  are identical, since the near detector is only affected if both are
  present (cf.\ Tab.~\ref{tab:suppression-factors}). Otherwise, it
  would retain its capability to properly calibrate the reactor
  flux to its true value.
\end{itemize}

\begin{figure}
  \begin{center}
    \includegraphics[width=8cm]{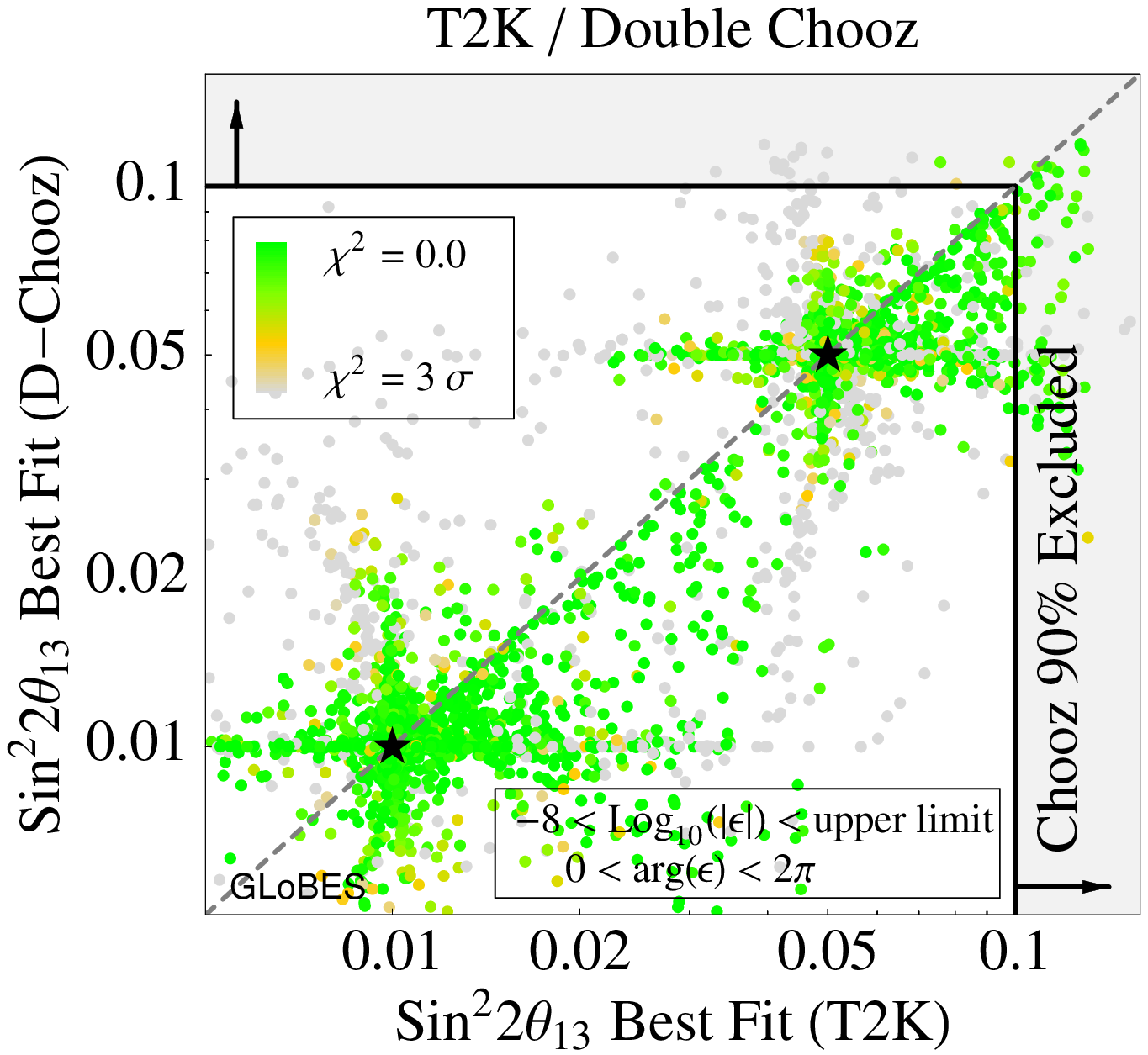} \\[0.5cm]
    \includegraphics[width=8cm]{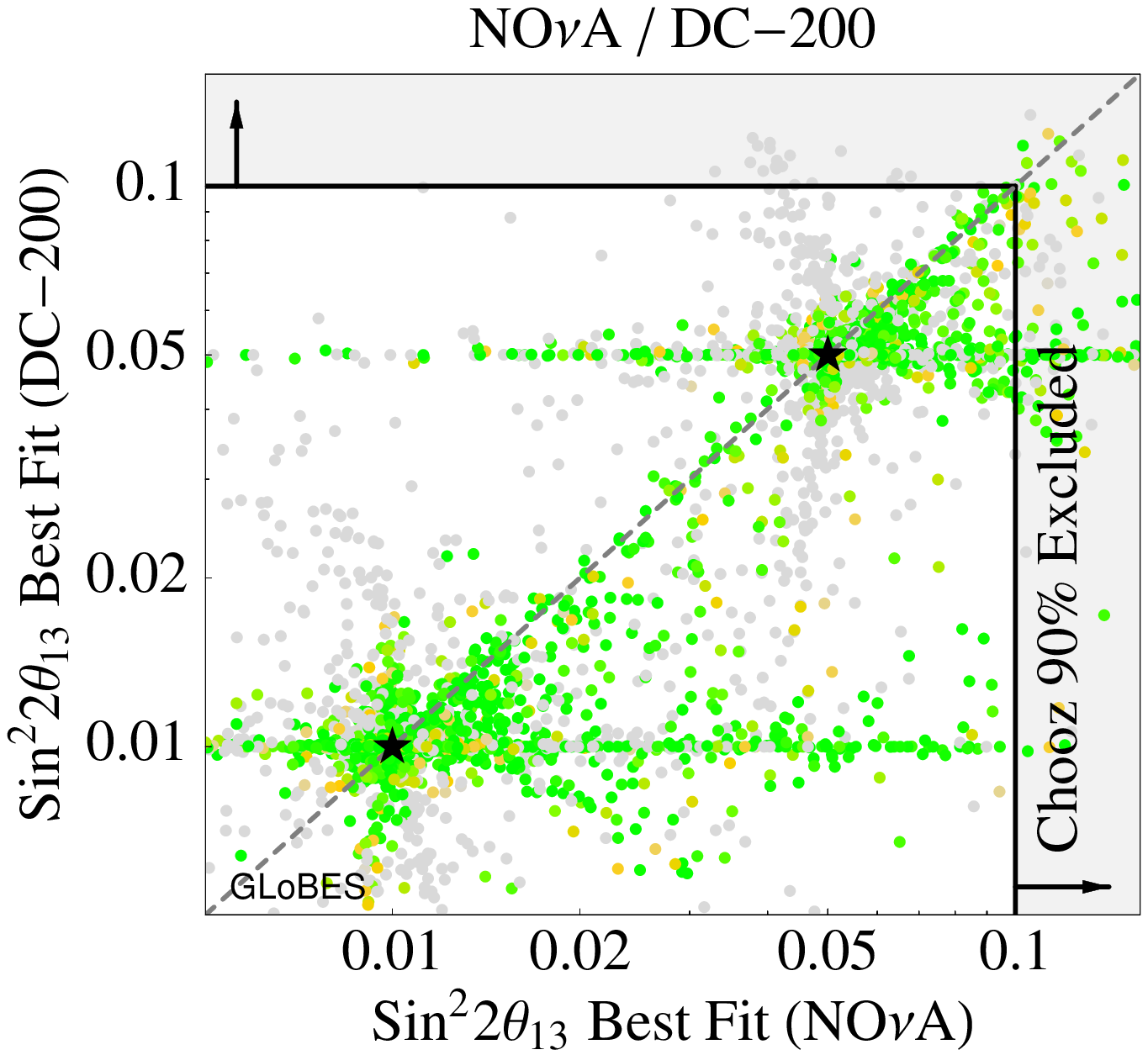}
  \end{center}
  \caption{Possible outcomes of standard three-flavor oscillation fits to
    reactor and superbeam data in the presence of non-standard interactions.
    Each plot contains two datasets, one for $\sthchooz^{\rm true} = 0.01$, and
    one for $\sthchooz^{\rm true} = 0.05$ (indicated by the black stars). For
    each dataset, 5,000 random combinations of $\eps$ parameters where chosen,
    with their moduli being distributed logarithmically between $10^{-8}$ and the
    respective upper bounds~\cite{Gonzalez-Garcia:2001mp,Davidson:2003ha},
    and their phases varying linearly between $0$ and $2\pi$. Each of these
    random non-standard scenarios was then fitted under the assumption
    of standard three-flavor oscillations. Green (dark gray)
    points indicate a very good quality of this fit, while light gray
    points denote a fit quality worse than $3\sigma$ in at least one of the two
    experiments, i.e.\ an effective discovery of the non-standard effect. The
    plots show that non-standard interactions can induce ostensible discrepancies
    between reactor and superbeam data (off-diagonal points), or a common offset
    (close-to-diagonal points), which would lead to consistent, but wrong results.
    Note that some points lie even above the \Chooz\ bound. A reactor fit above
    the \Chooz\ bound indicates that the corresponding combination of
    NSI parameters and $\theta_{13}^{\rm true}$ could already be ruled
    out using existing data.}
  \label{fig:th13discrep-random}
\end{figure}

So far, we have only considered situations in which one non-standard parameter
is dominant, and all others are negligible. In realistic models, however, many
parameters may be of the same order of magnitude. Since it is impossible to
visualize the resulting high-dimensional parameter space, we resort to the
scatter plots shown in Fig.~\ref{fig:th13discrep-random}. These plots were
created by choosing a random value for each NSI parameter, and then performing
a standard oscillation fit to the resulting experimental data. The moduli of
the $\eps$ parameters were logarithmically distributed between $10^{-8}$ and
their current upper limits, where we have assumed the model-independent bound
from universality in charged lepton decays~\cite{Gonzalez-Garcia:2001mp}
for $\eps^s$ and $\eps^d$, and the results of~\cite{Davidson:2003ha} for
$\eps^m$. The phases were distributed linearly between $0$ and $2\pi$.

We can see from Fig.~\ref{fig:th13discrep-random} that there are again
points which yield a clear discrepancy in the $\theta_{13}$ fits of the
reactor and superbeam data, and others which correspond to a common offset
of the fit value. The color coding shows that for a considerable fraction
of the parameter space, the non-standard effect can actually be discovered.
It is interesting to observe that there are some points for which the reactor
fit lies above the \Chooz\ bound. This indicates that already with the present
data, some parts of the parameter space could be ruled out.

\section{Discovery reach for non-standard interactions in a combined analysis
         of reactor and superbeam data}
\label{sec:dis-reach}

\begin{figure*}
  \begin{center}
    \begin{tabular}{c@{\hspace{-0.8cm}}c@{\hspace{-0.8cm}}c}
      \includegraphics[width=6.0cm]{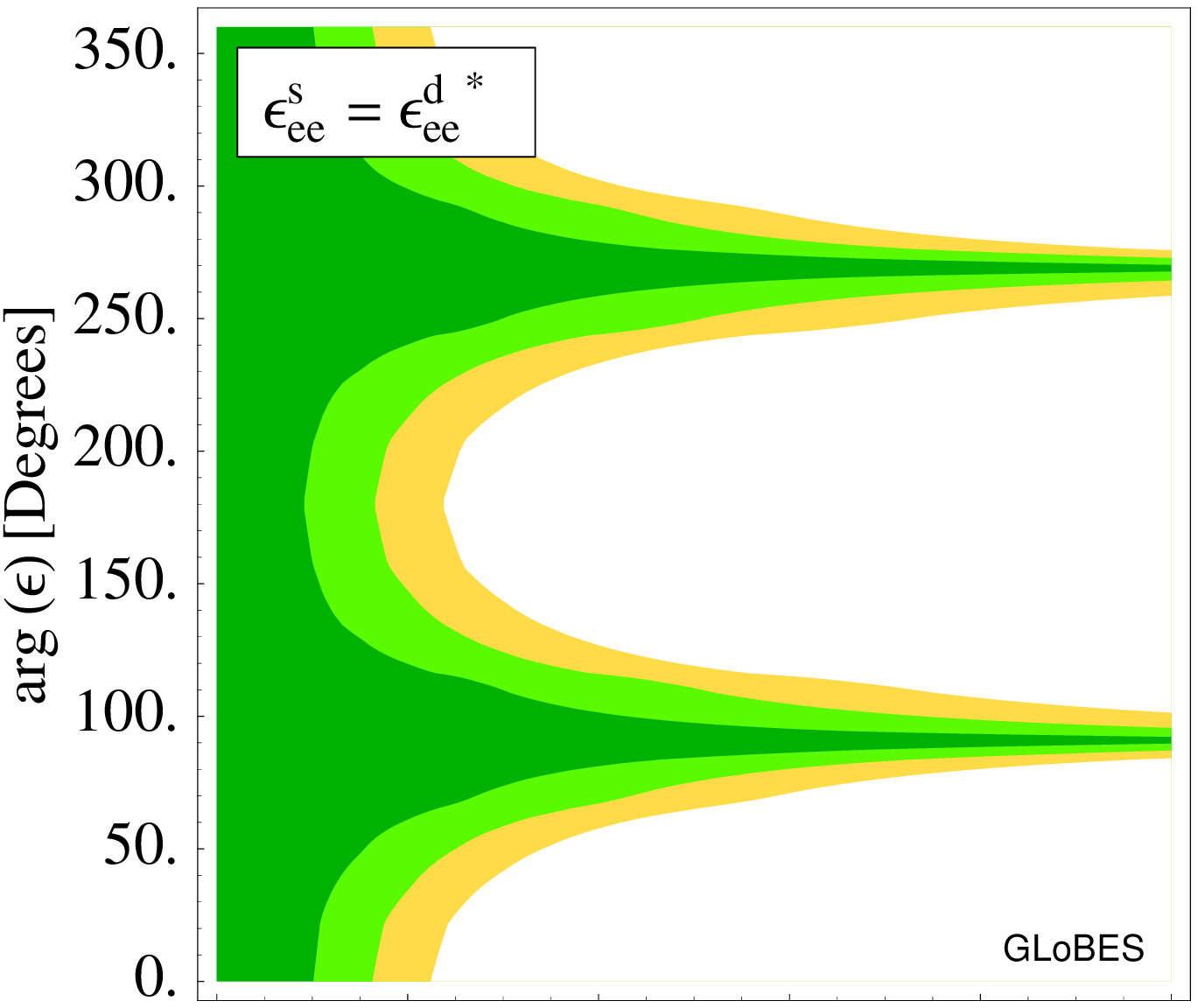}       &
      \includegraphics[width=6.0cm]{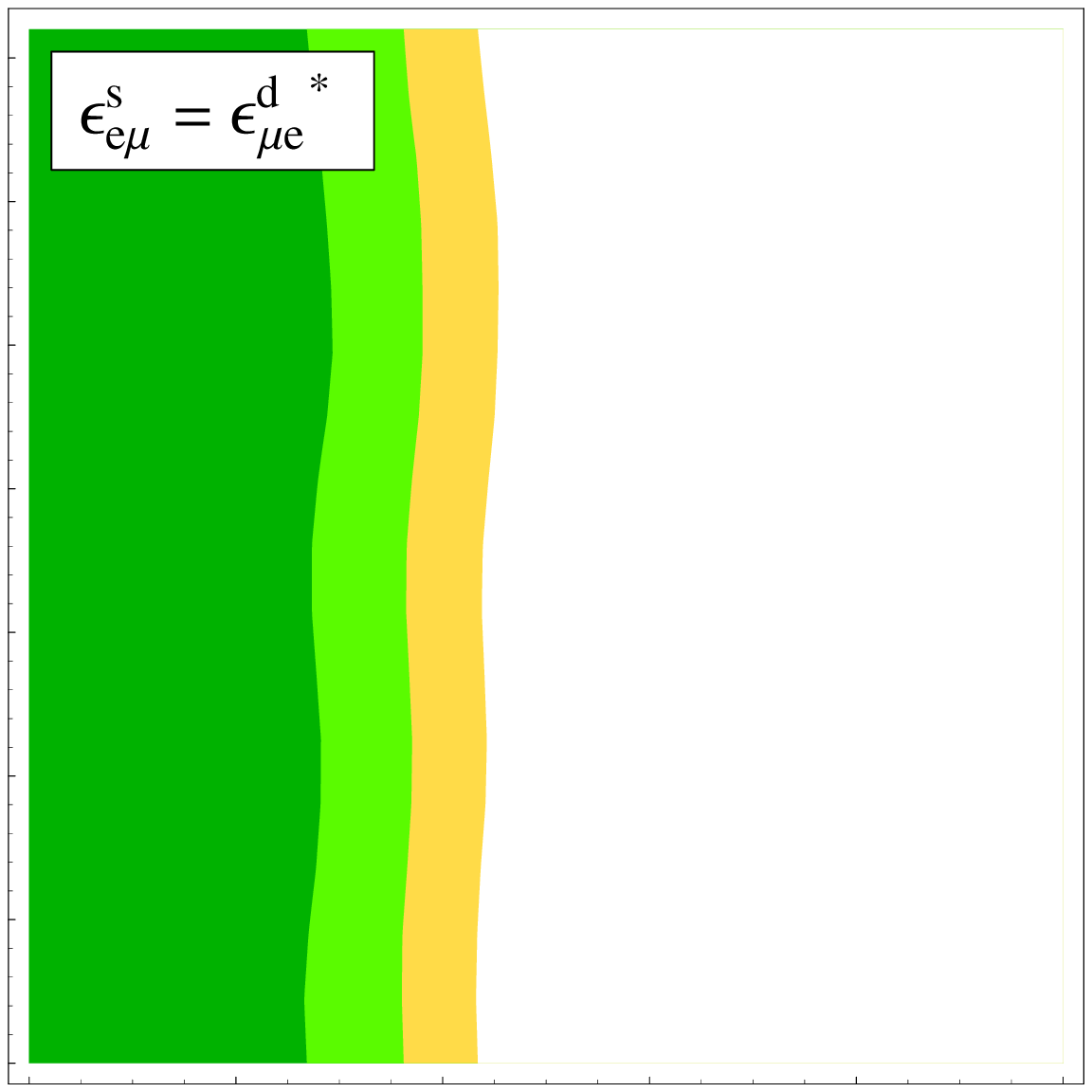}      &
      \includegraphics[width=6.0cm]{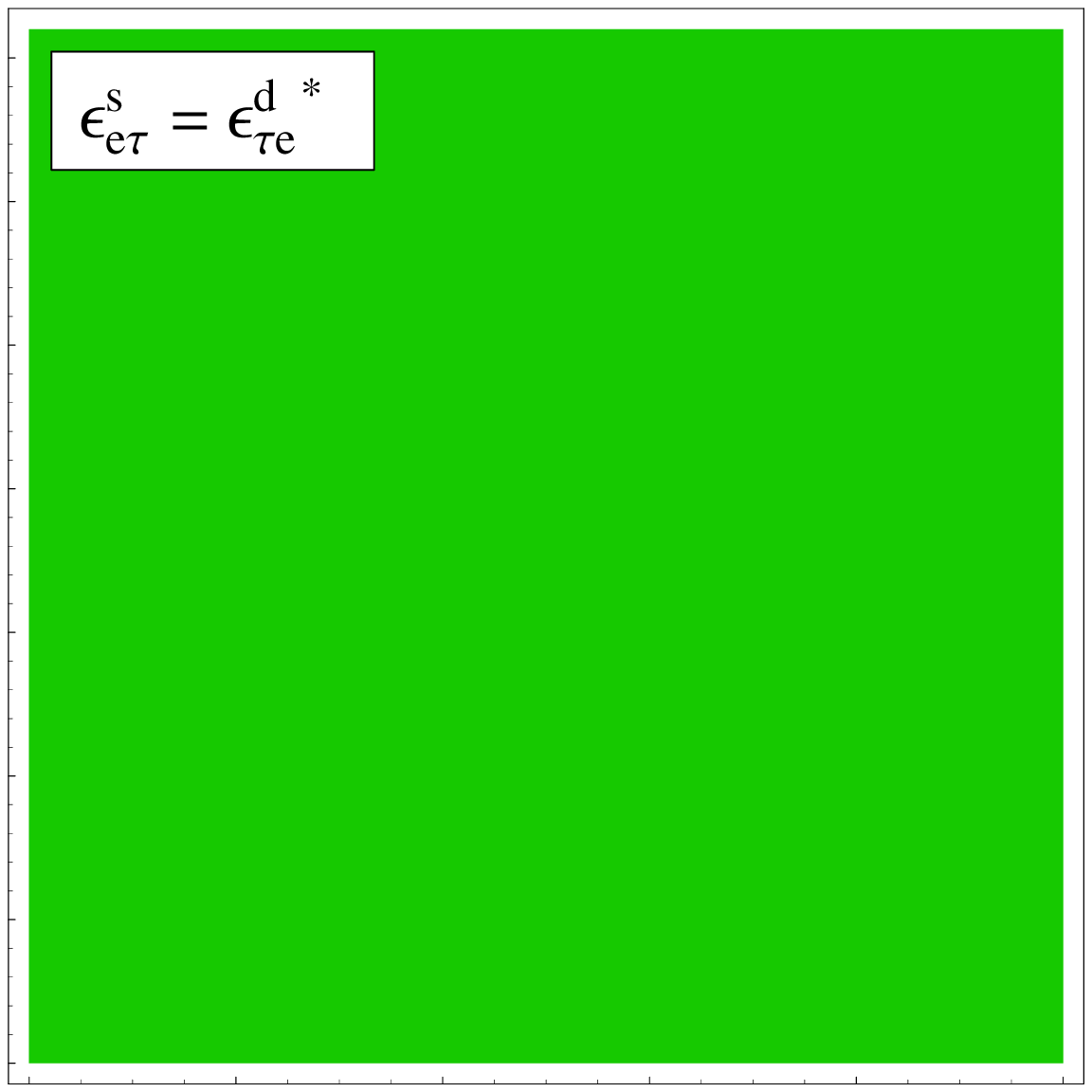}     \\[-0.9cm]
      \includegraphics[width=6.0cm]{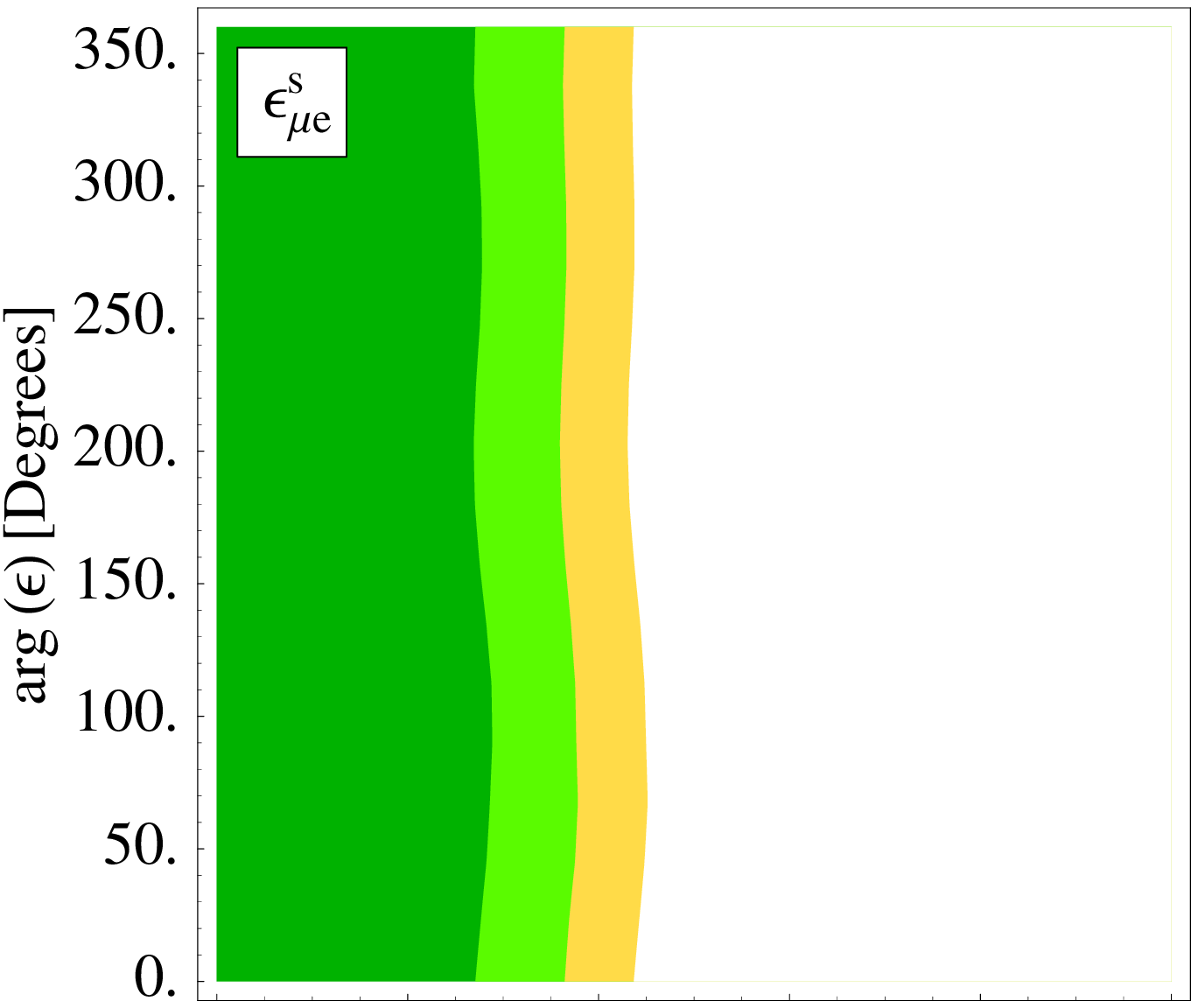}      &
      \includegraphics[width=6.0cm]{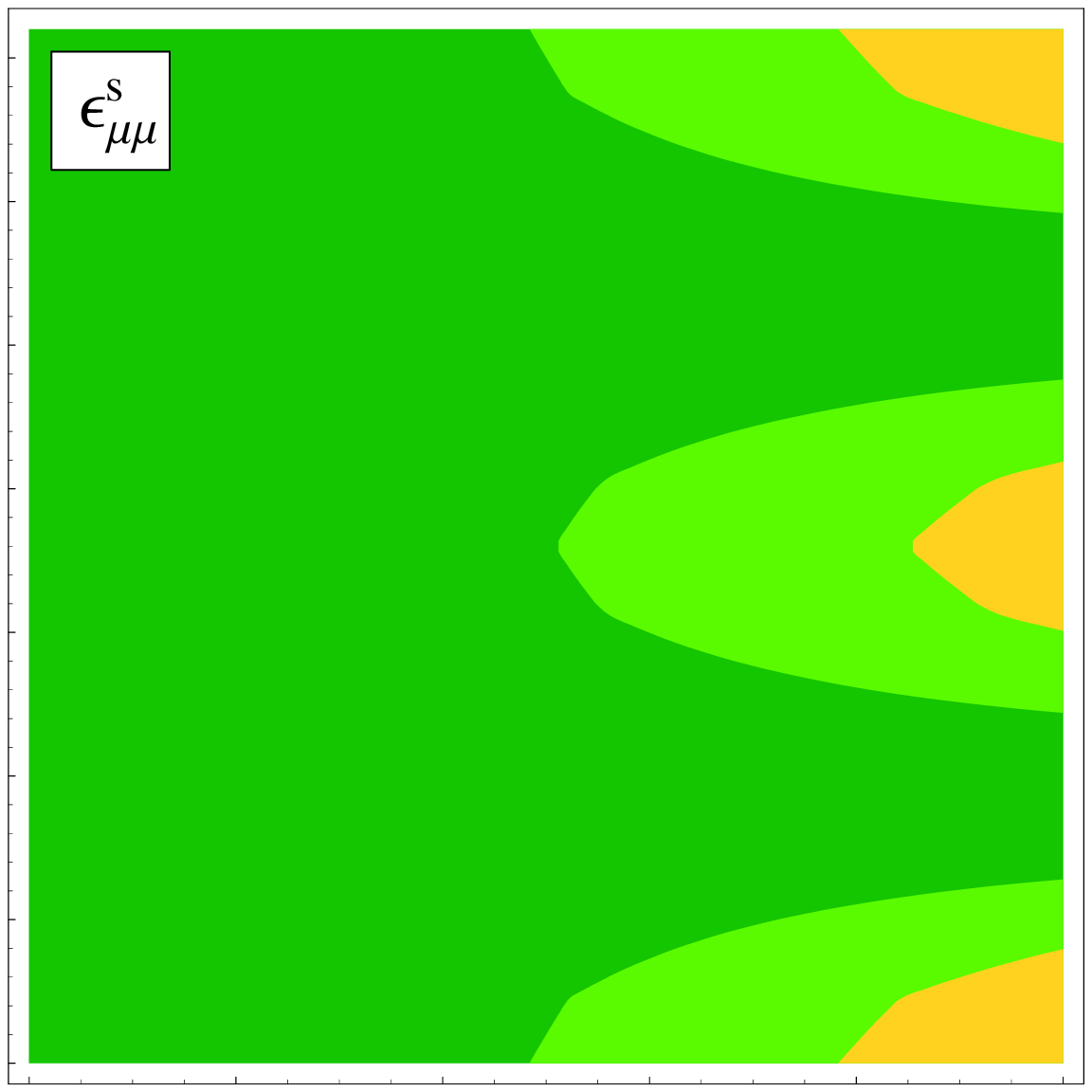}     &
      \includegraphics[width=6.0cm]{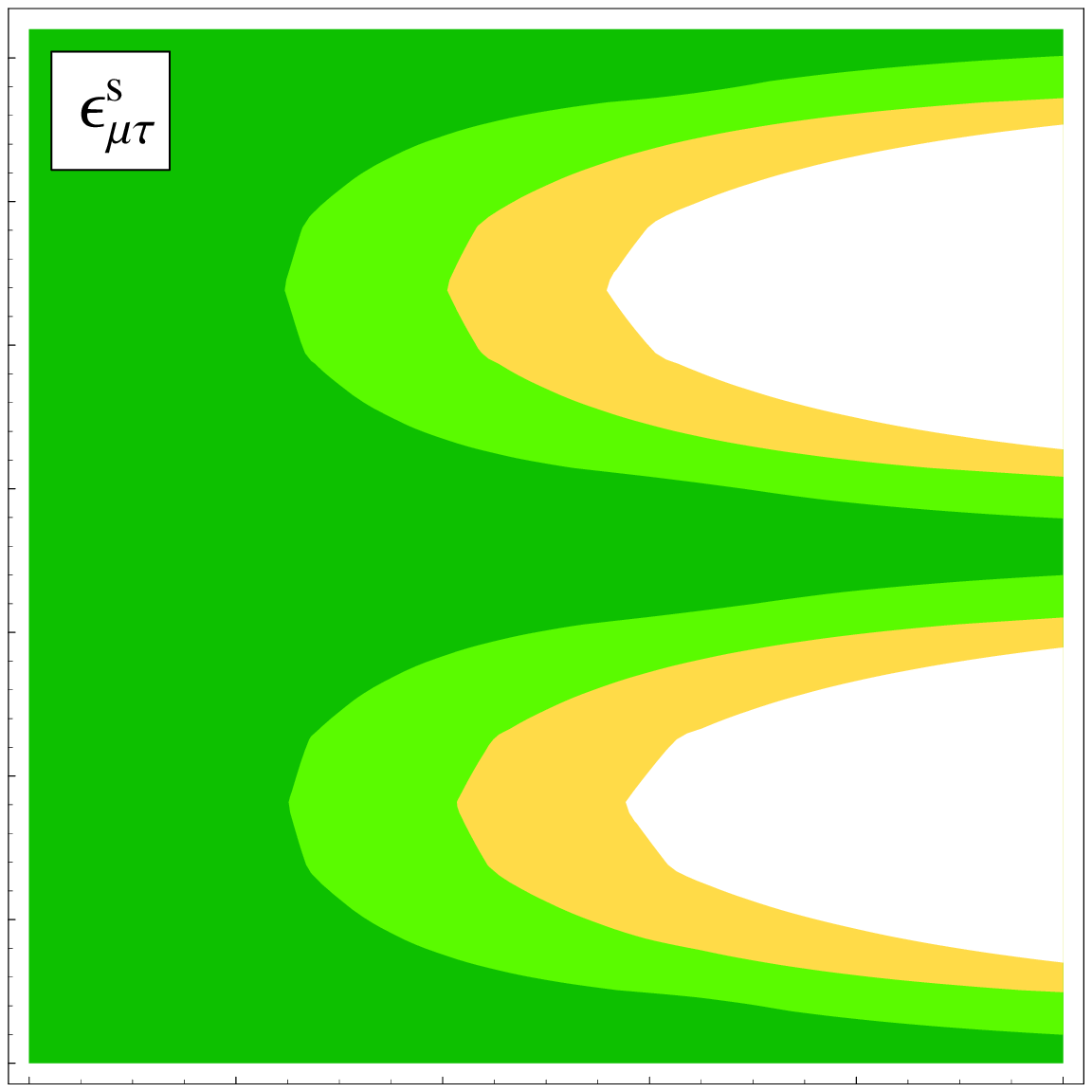}    \\[-0.9cm]
      \includegraphics[width=6.0cm]{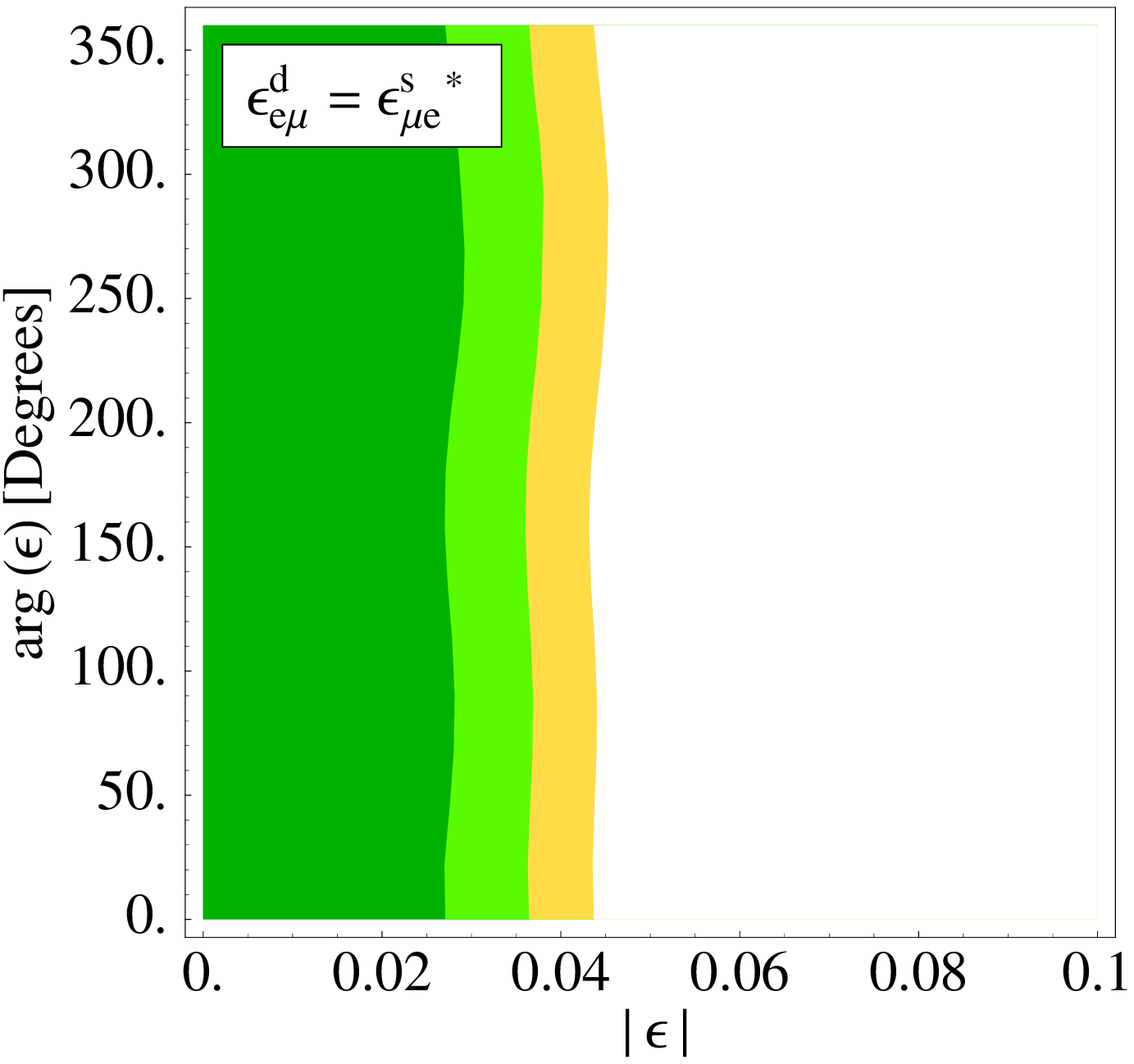}      &
      \includegraphics[width=6.0cm]{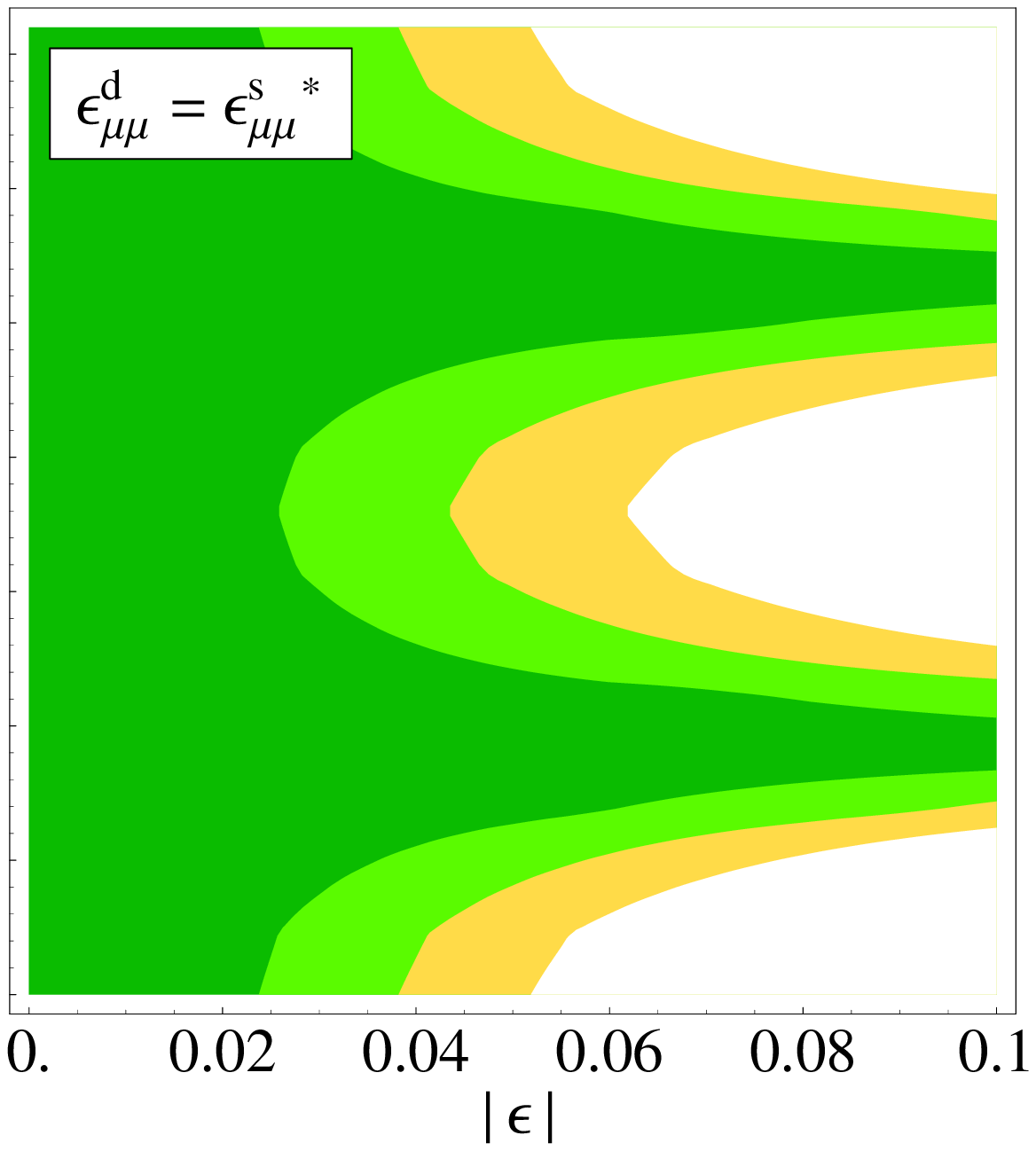}     &
      \includegraphics[width=6.0cm]{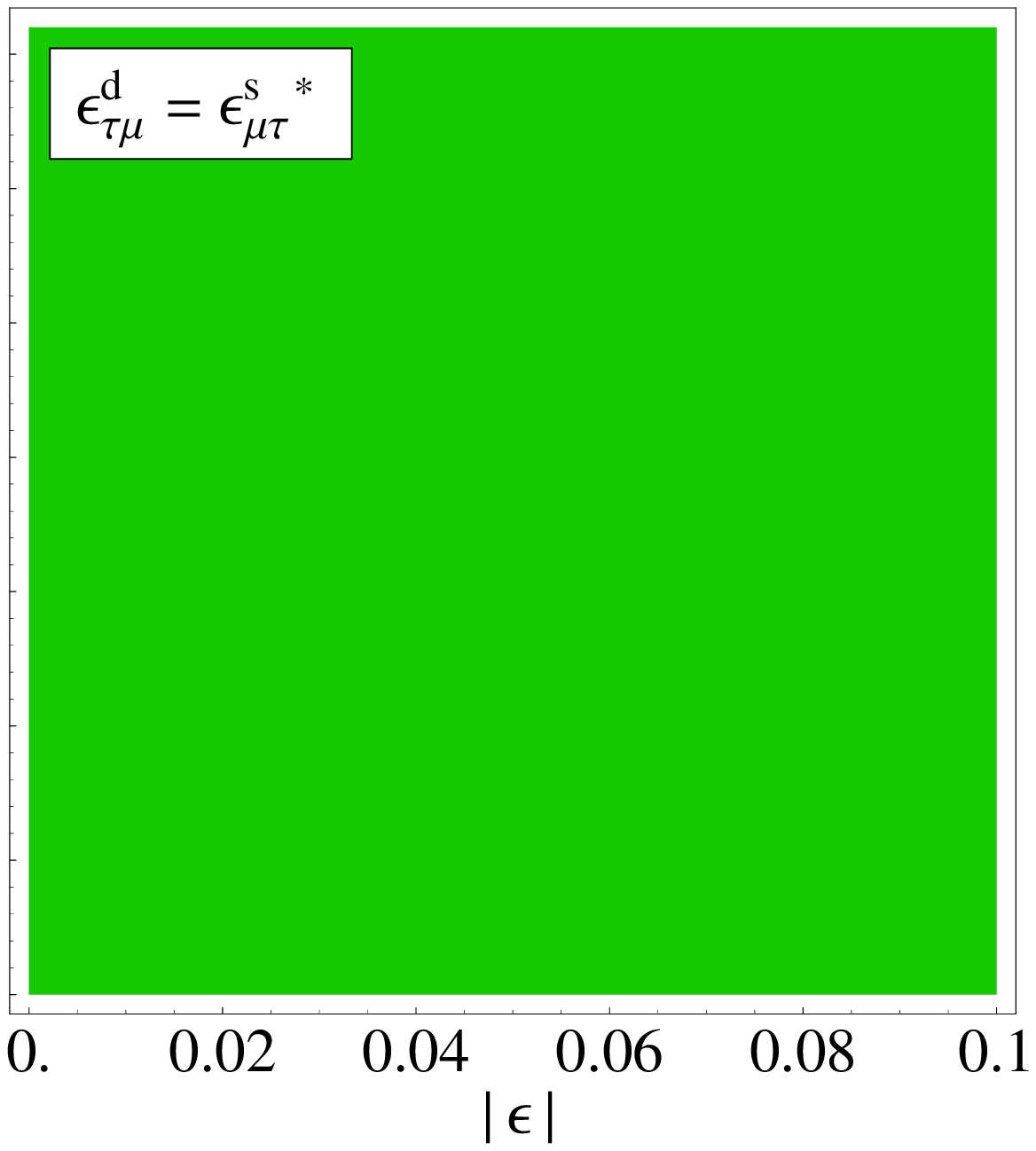}    \\[-0.9cm]
    \end{tabular}
  \end{center}
  \caption{Discovery reach for $\eps^s_{\alpha\beta}$ and $\eps^d_{\beta\alpha}$
    in a combined analysis of \TtoK\ and \DoubleChooz. Contours for $1\sigma$,
    $2\sigma$, and $3\sigma$ are shown. A true value of $\sthchooz^{\rm true}=0.05$
    was assumed, but we have checked that the results do not depend on
    $\sthchooz^{\rm true}$.}
  \label{fig:th13discovery-T2K-epsilon-sd}
\end{figure*}

\begin{figure*}
  \begin{center}
    \begin{tabular}{c@{\hspace{-0.8cm}}c@{\hspace{-0.8cm}}c}
      \includegraphics[width=6.0cm]{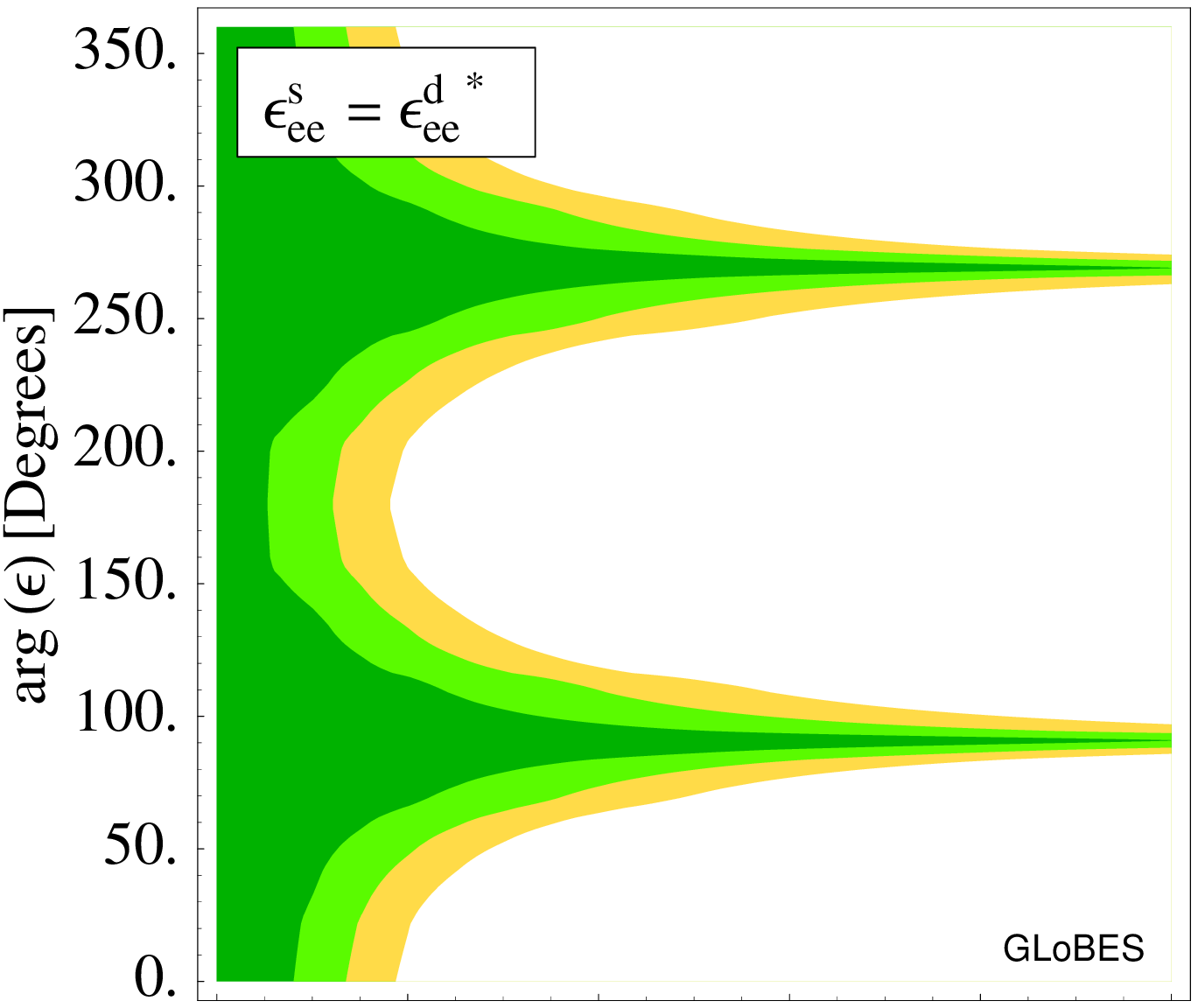}       &
      \includegraphics[width=6.0cm]{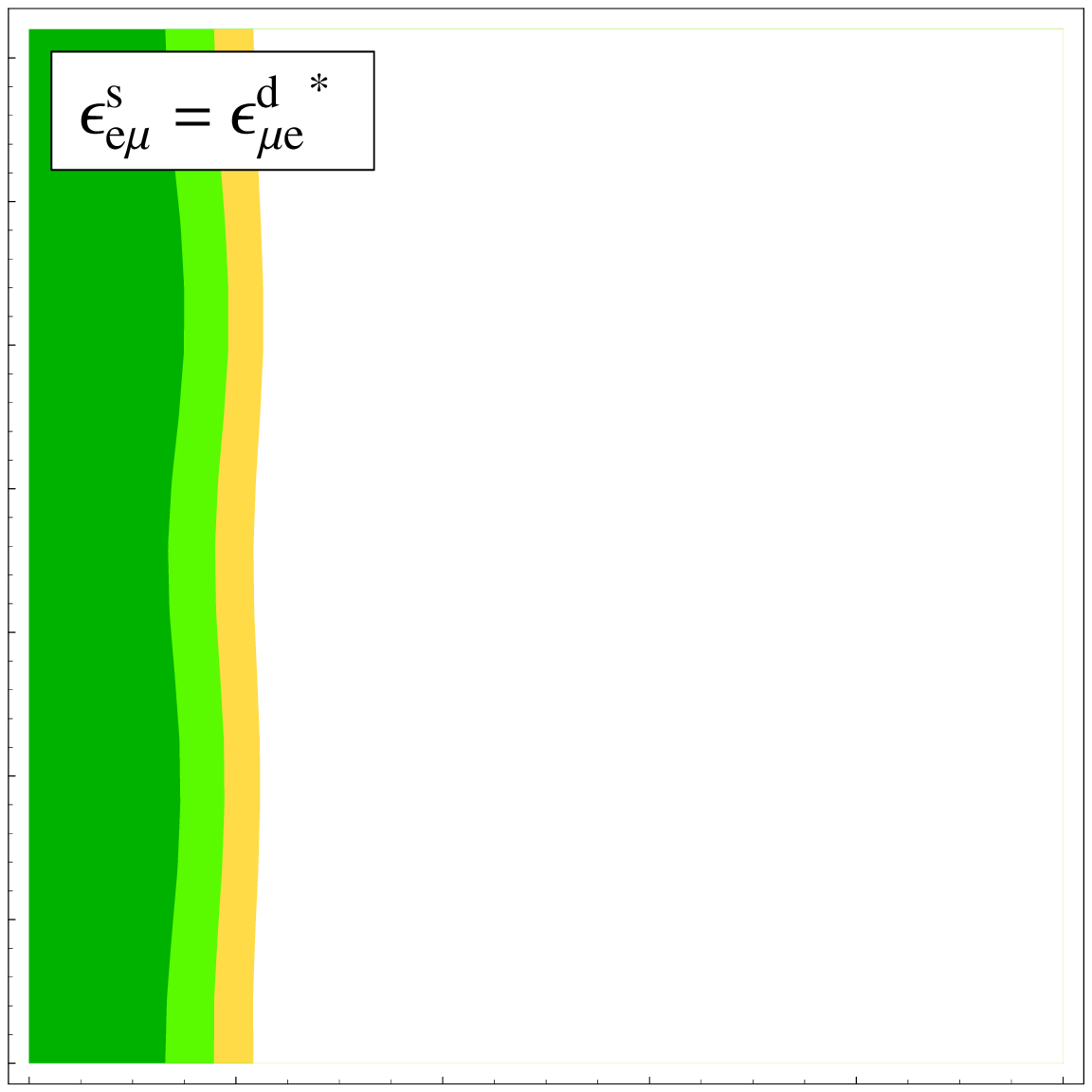}      &
      \includegraphics[width=6.0cm]{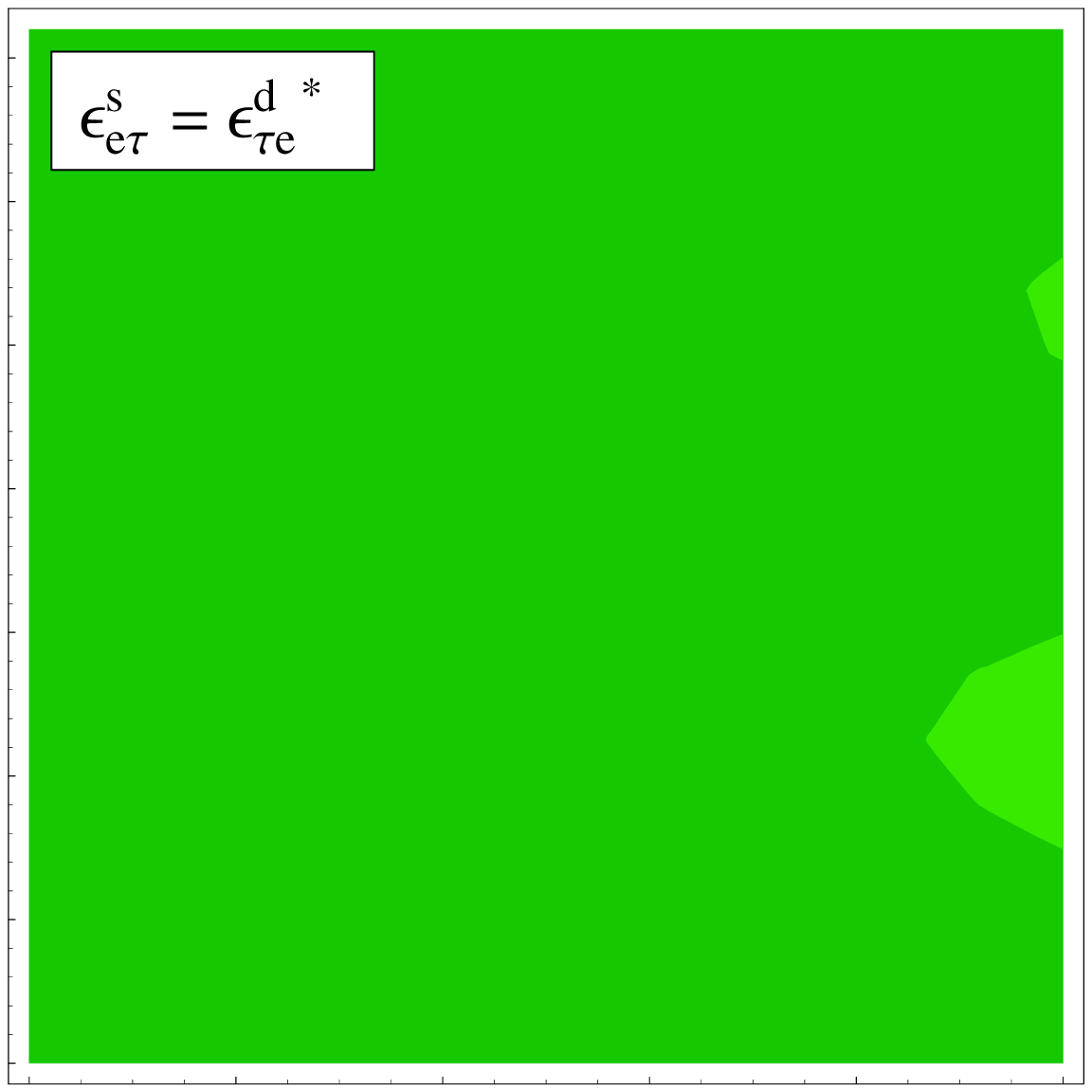}     \\[-0.9cm]
      \includegraphics[width=6.0cm]{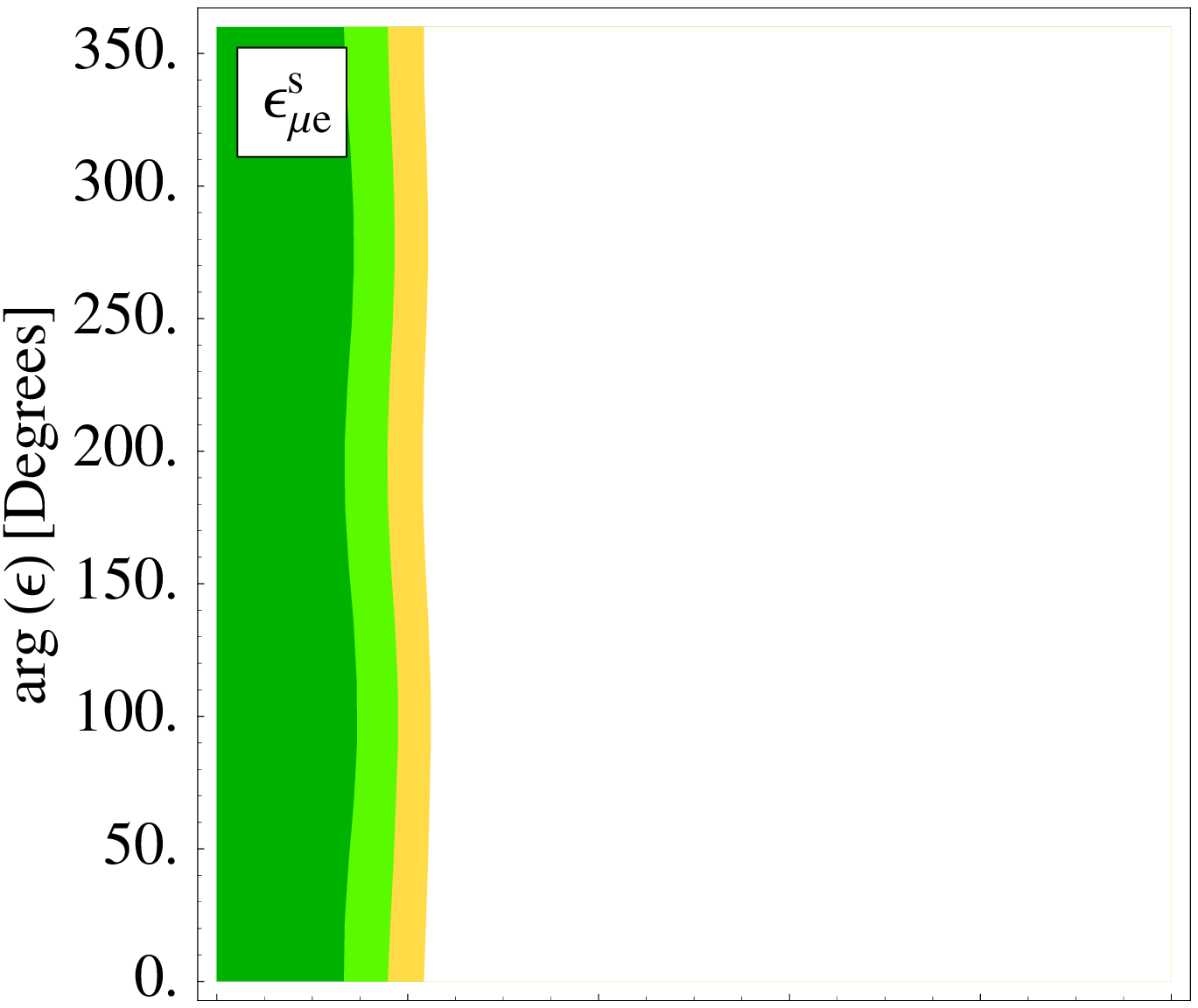}      &
      \includegraphics[width=6.0cm]{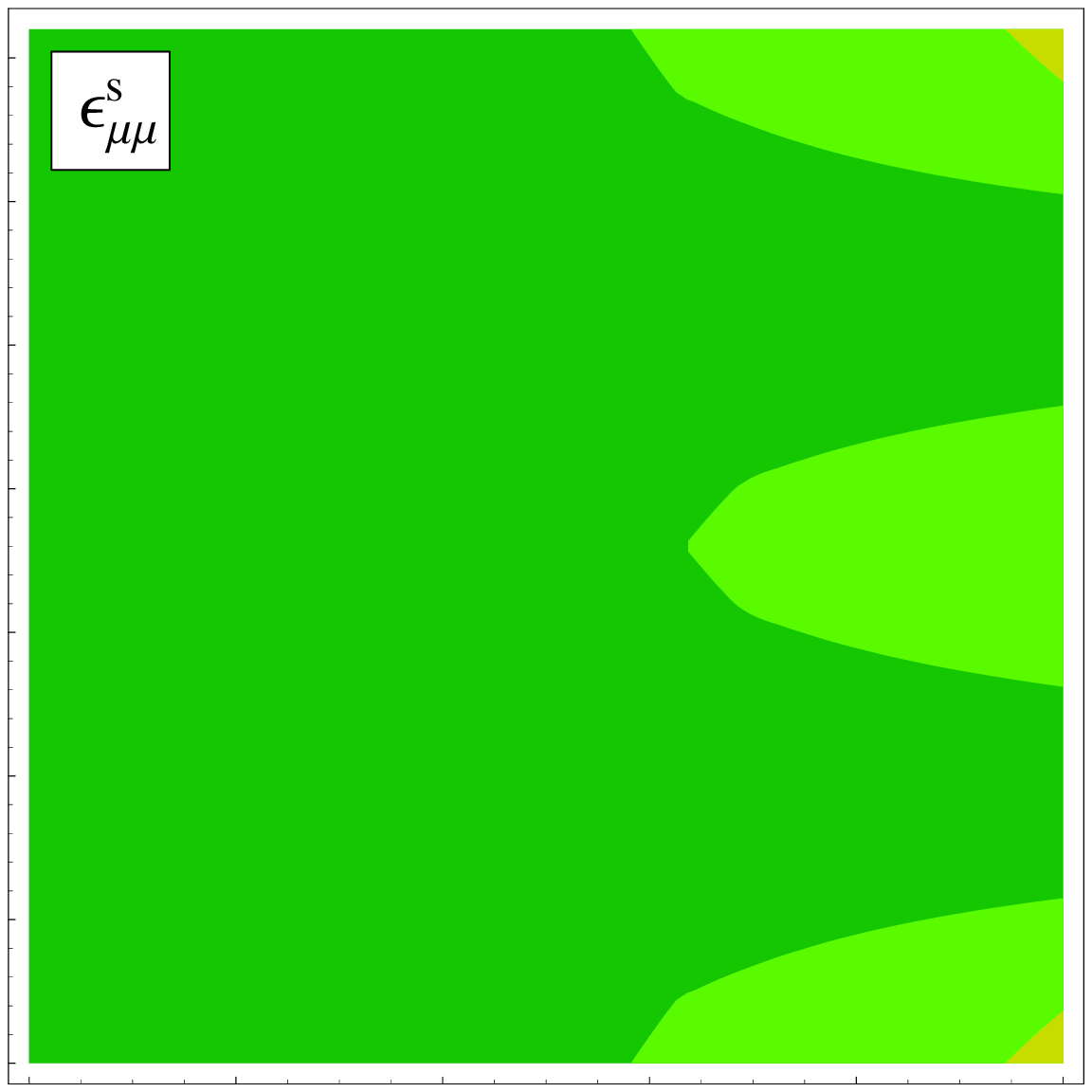}     &
      \includegraphics[width=6.0cm]{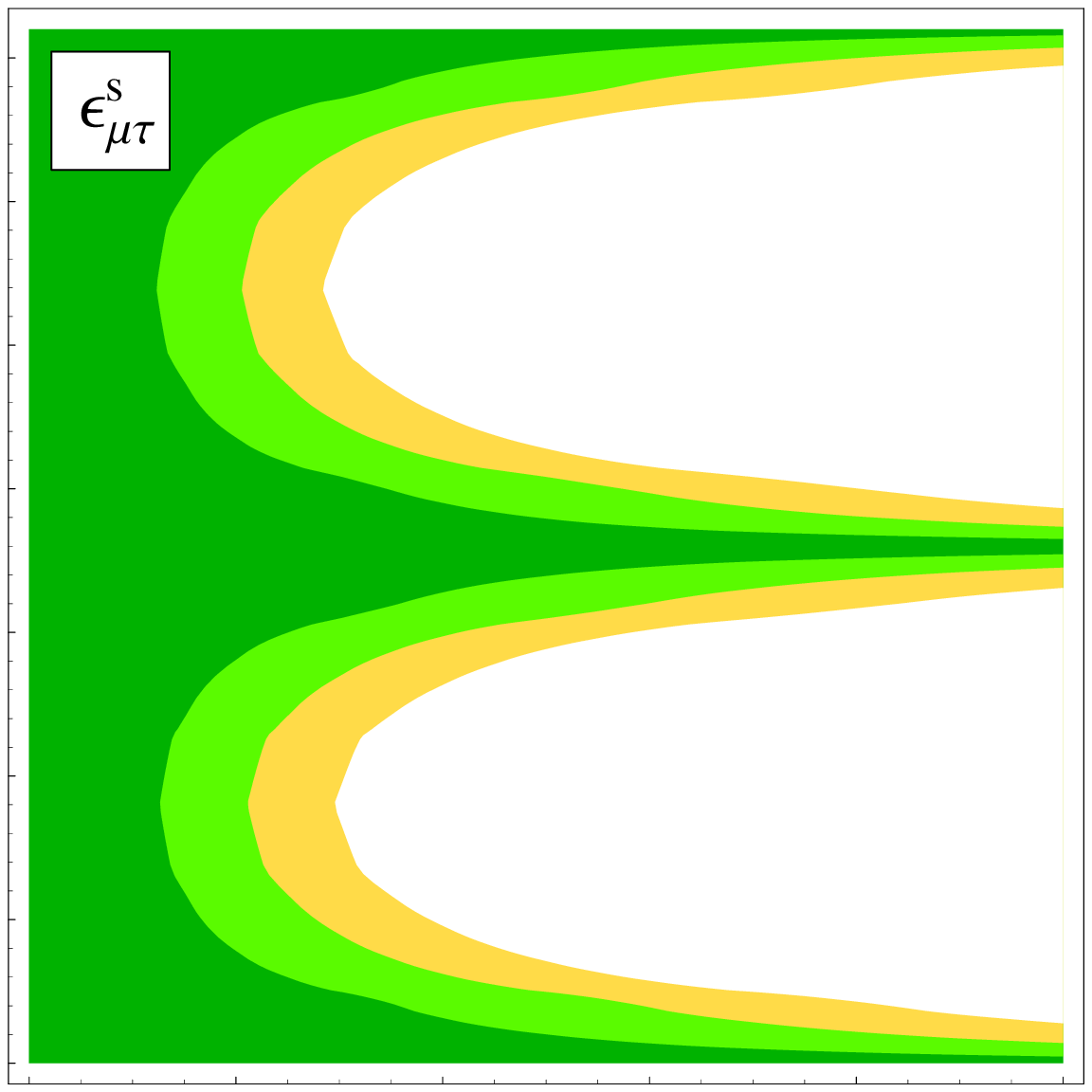}    \\[-0.9cm]
      \includegraphics[width=6.0cm]{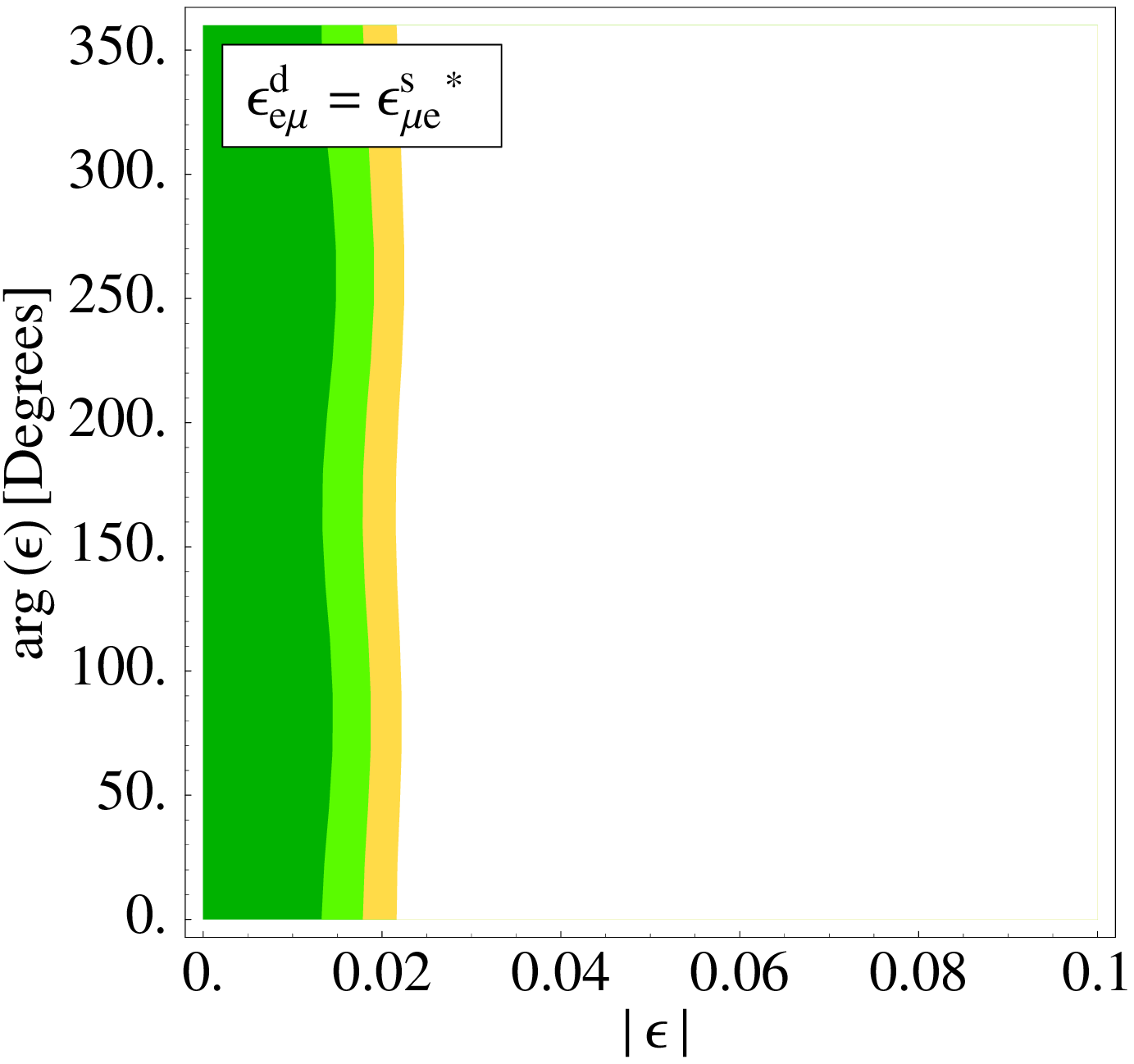}      &
      \includegraphics[width=6.0cm]{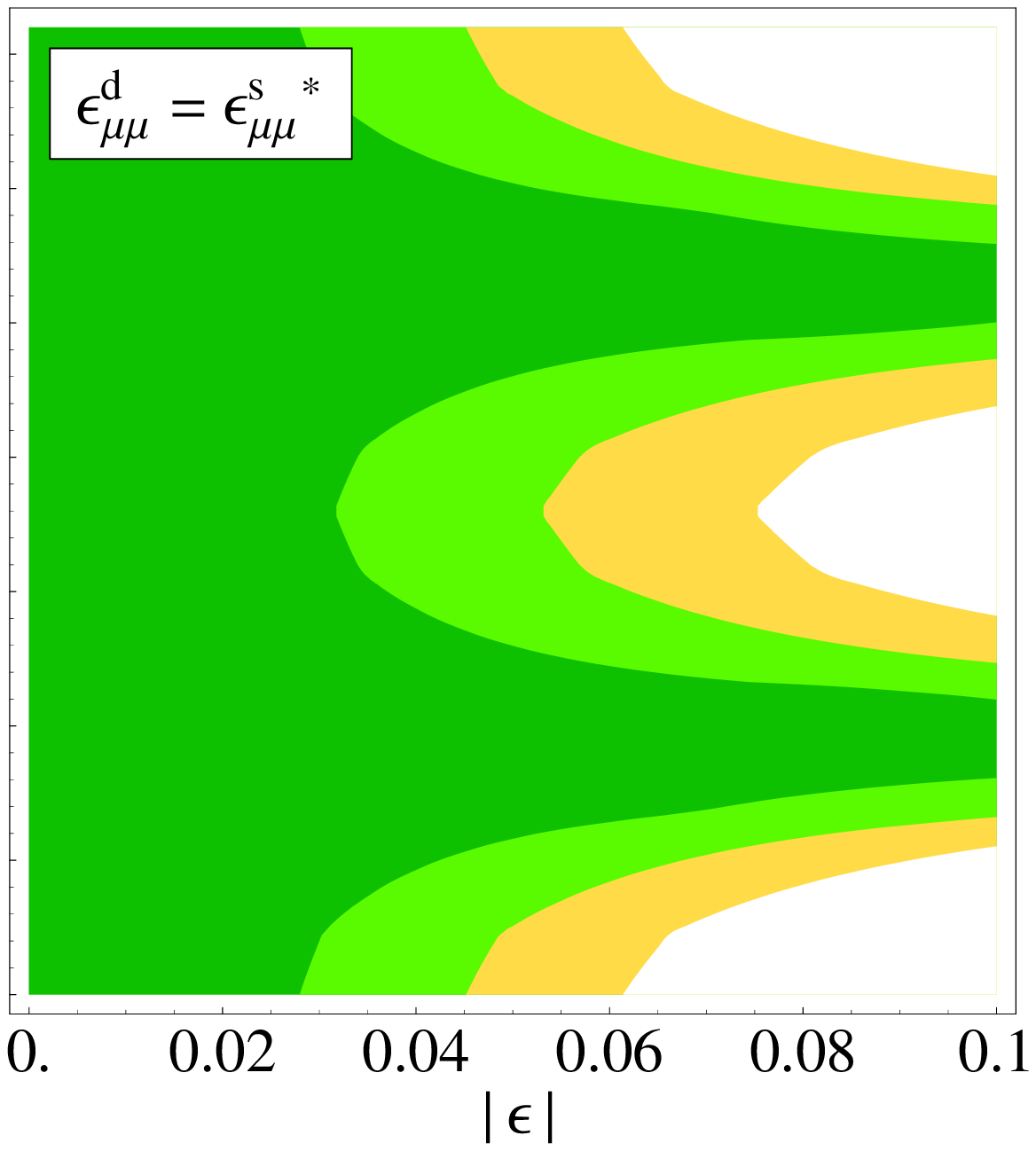}     &
      \includegraphics[width=6.0cm]{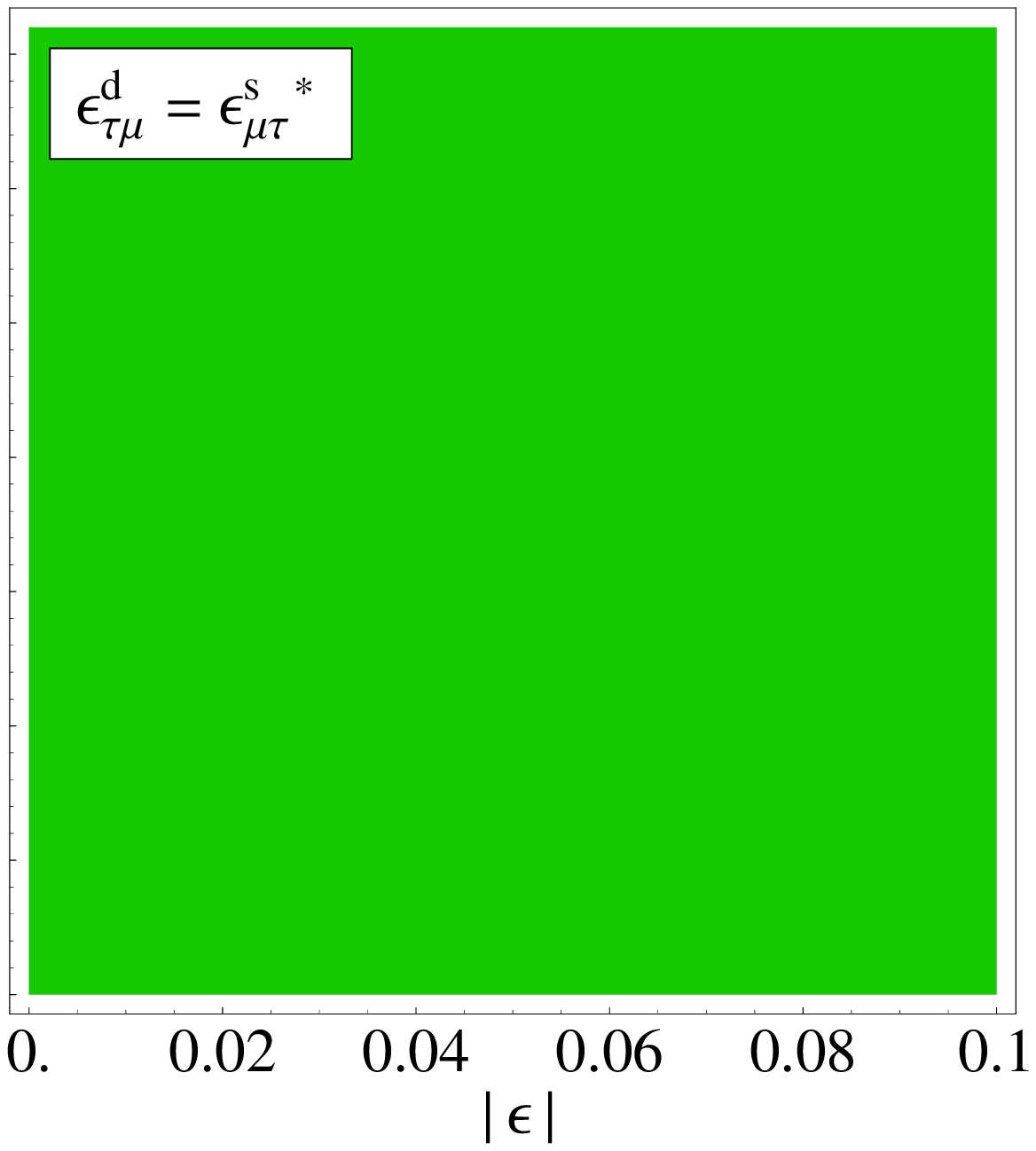}    \\[-0.9cm]
    \end{tabular}
  \end{center}
  \caption{Discovery reach for $\eps^s_{\alpha\beta}$ and $\eps^d_{\beta\alpha}$
    in a combined analysis of \NOvA\ and \DCext. Contours for $1\sigma$,
    $2\sigma$, and $3\sigma$ are shown. A value of $\sthchooz^{\rm true}=0.05$
    was assumed in the simulation, but we have checked that the results remain
    unchanged if we go to a different $\sthchooz^{\rm true}$.}
  \label{fig:th13discovery-NOvA-epsilon-sd}
\end{figure*}

\begin{figure*}
  \begin{center}
    \begin{tabular}{c@{\hspace{-0.5cm}}c@{\hspace{-0.5cm}}c}
      \includegraphics[width=6.0cm]{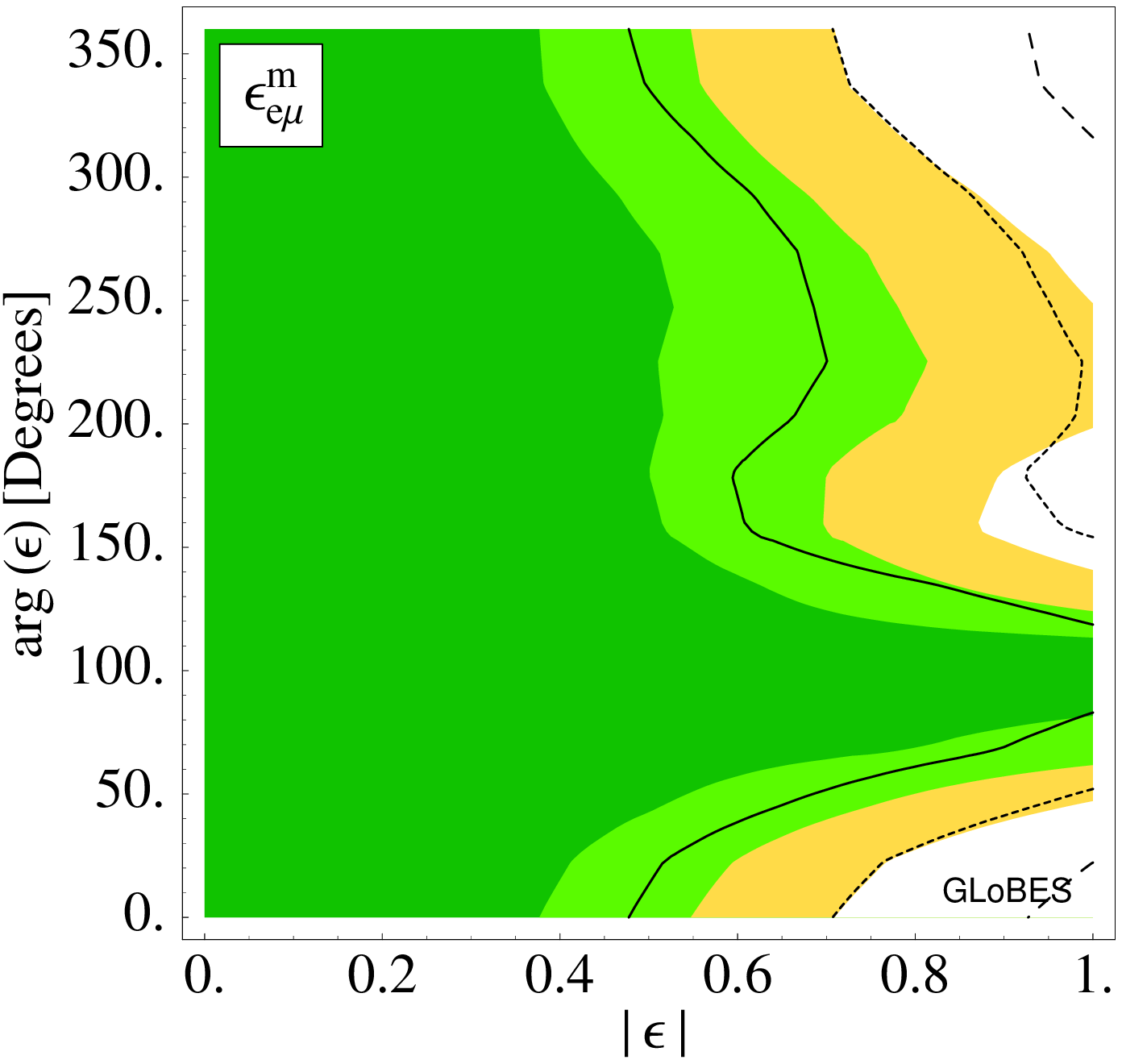}      &
      \includegraphics[width=6.0cm]{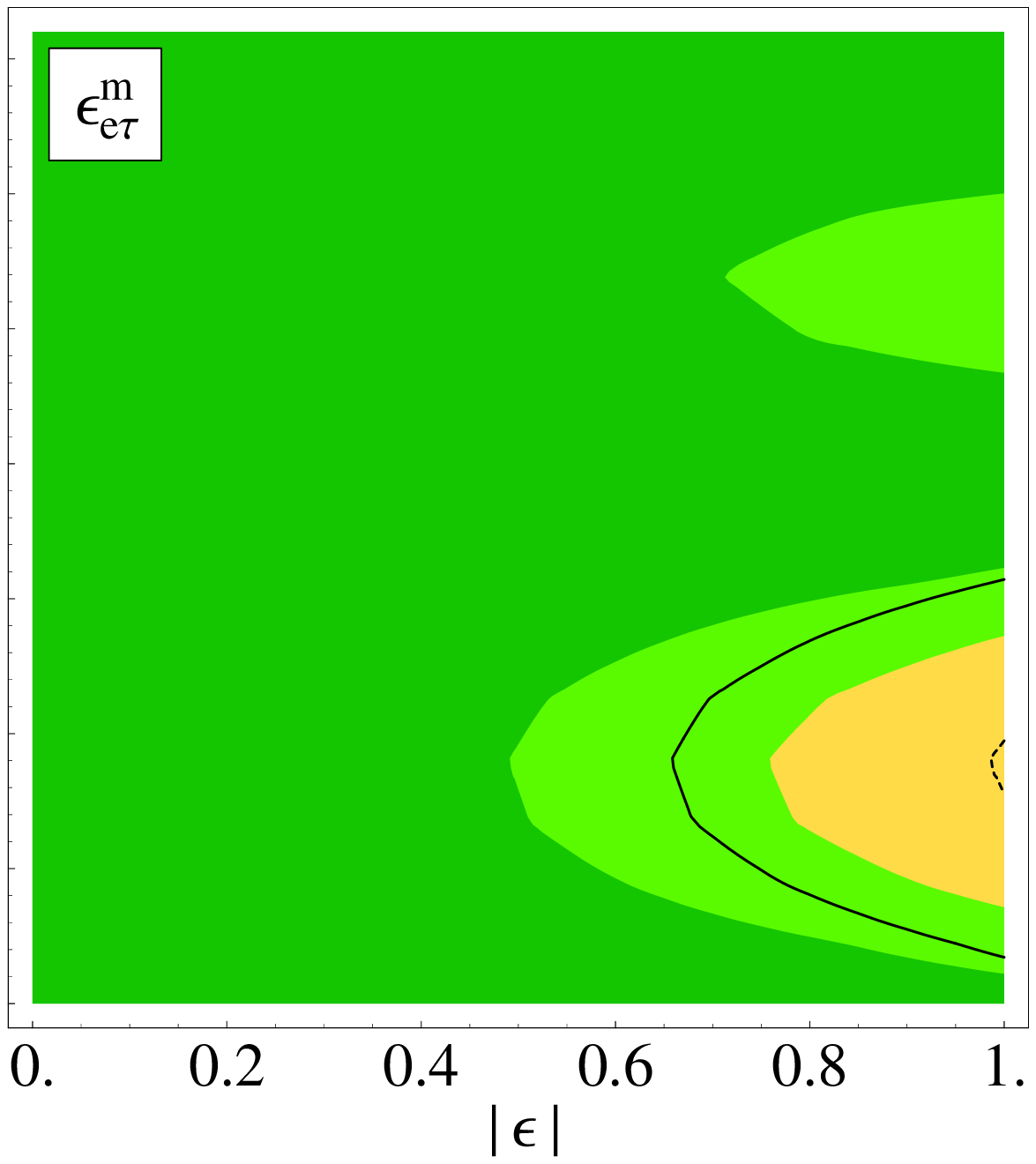}     &
      \includegraphics[width=6.0cm]{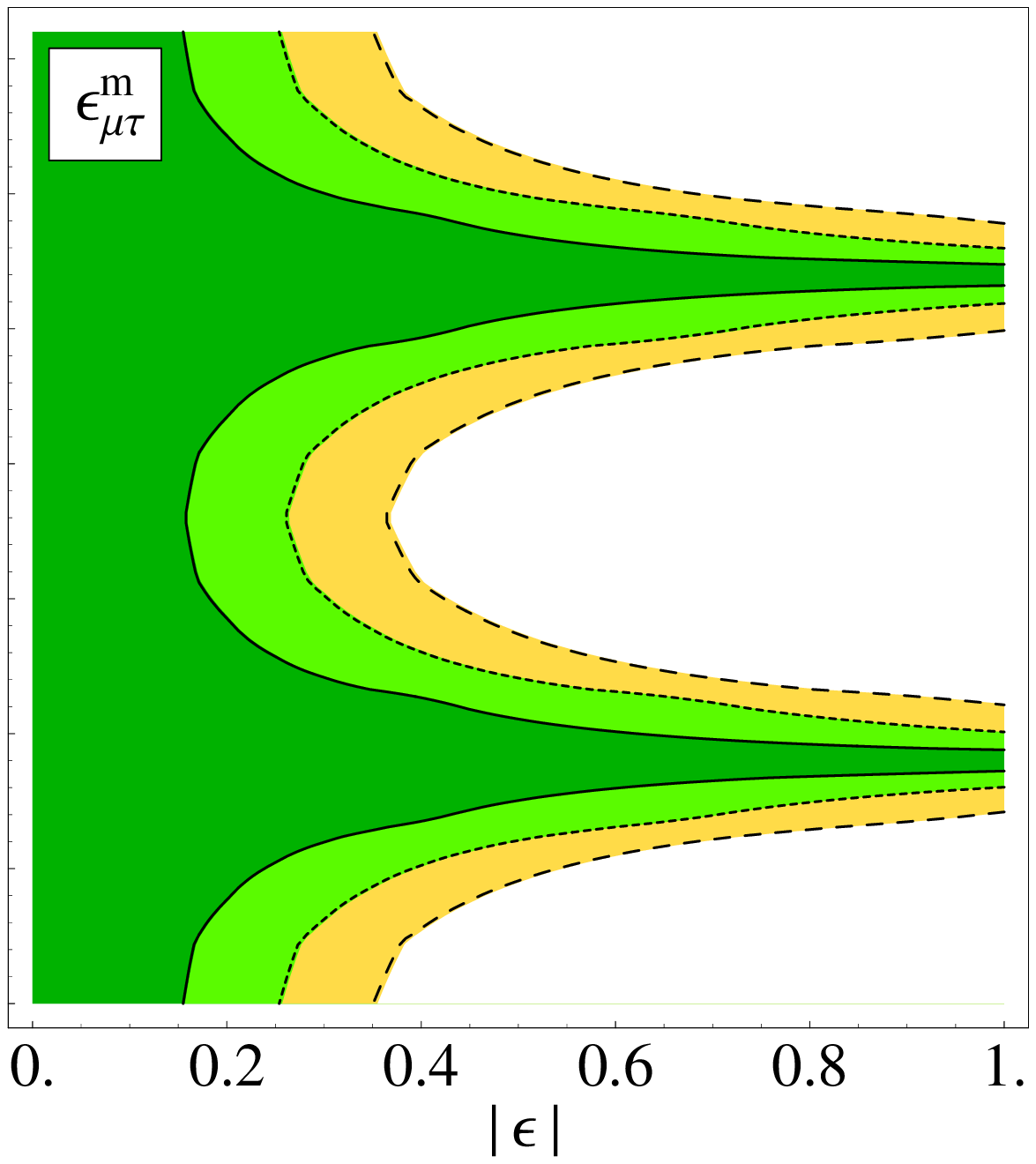}
    \end{tabular}
  \end{center}
  \caption{Discovery reach for $\eps^m_{\alpha\beta}$ in a combined analysis of
    \TtoK\ and \DoubleChooz. Since these experiments have essentially no sensitivity
    to $\eps^m_{ee}$, $\eps^m_{\mu\mu}$, and $\eps^m_{\tau\tau}$, we show
    only the off-diagonal entries of $\eps^m$. The colored areas show the $1\sigma$,
    $2\sigma$, and $3\sigma$ confidence regions for $\sthchooz=0.05$, while the
    black contours are for $\sthchooz=0.01$. Note that the scaling of the
    horizontal axis is different from Figs.~\ref{fig:th13discovery-T2K-epsilon-sd}
    and \ref{fig:th13discovery-NOvA-epsilon-sd}.}
  \label{fig:th13discovery-T2K-epsilon-m}
\end{figure*}

\begin{figure*}
  \begin{center}
    \begin{tabular}{c@{\hspace{-0.5cm}}c@{\hspace{-0.5cm}}c}
      \includegraphics[width=6.0cm]{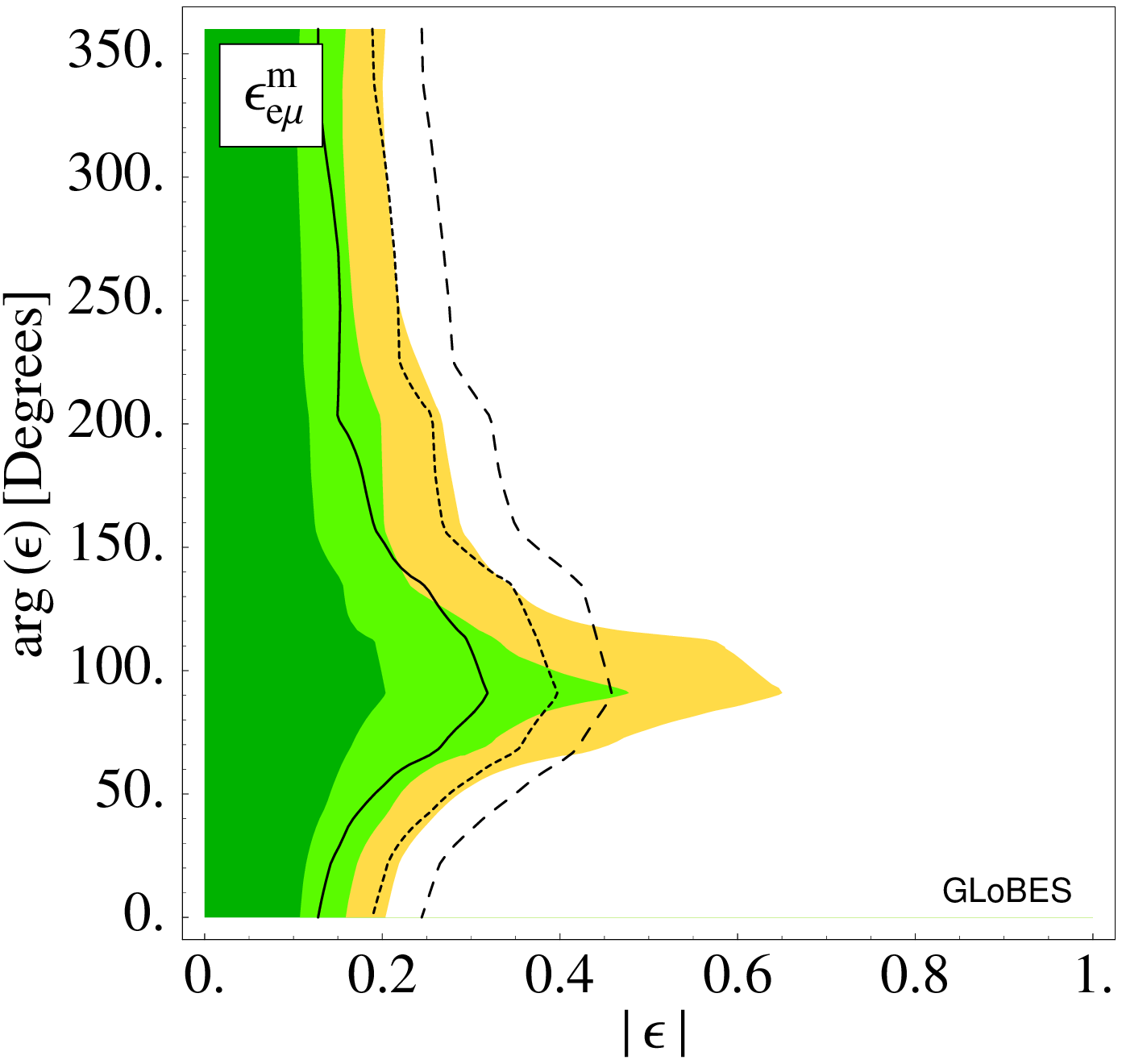}      &
      \includegraphics[width=6.0cm]{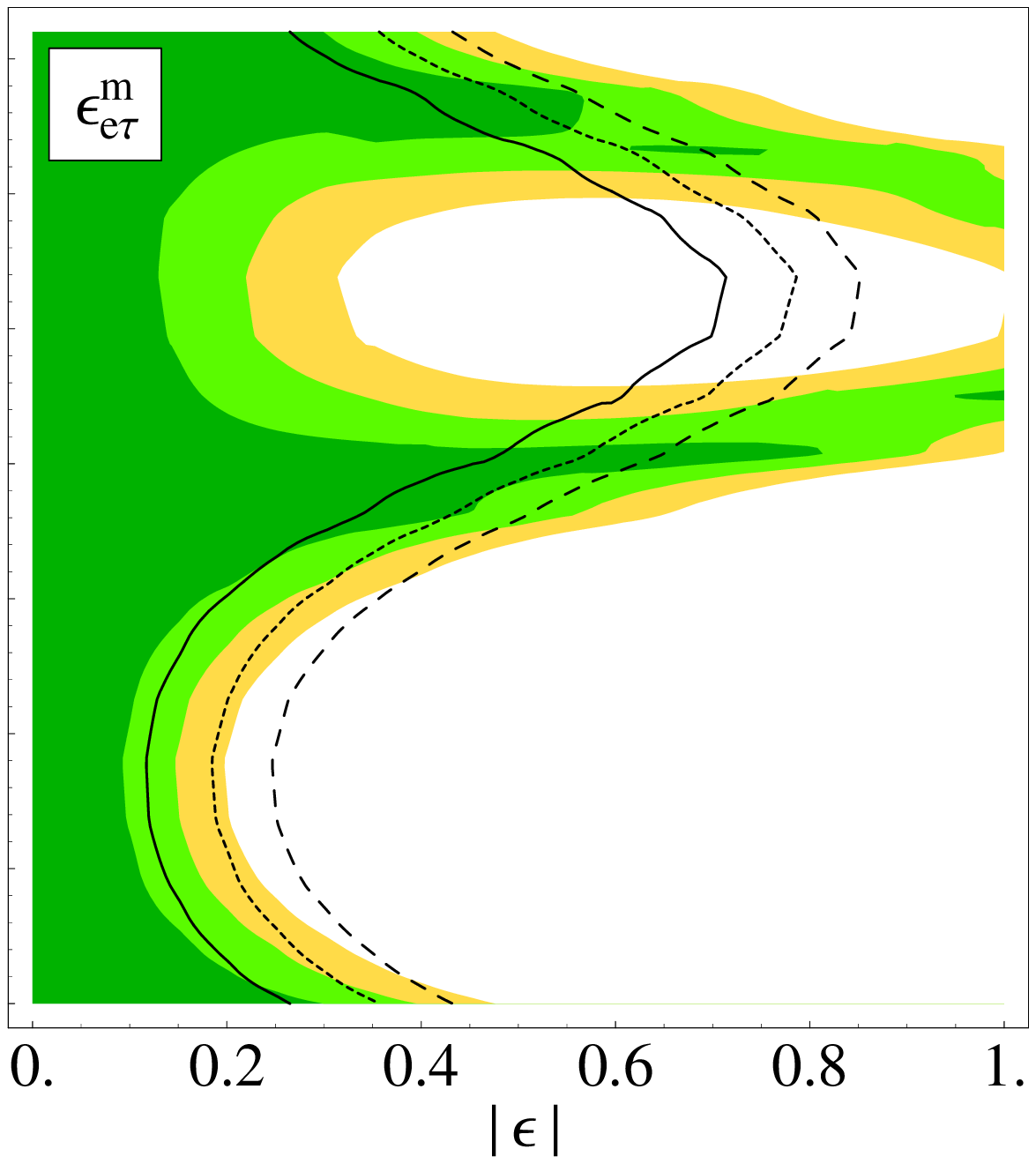}     &
      \includegraphics[width=6.0cm]{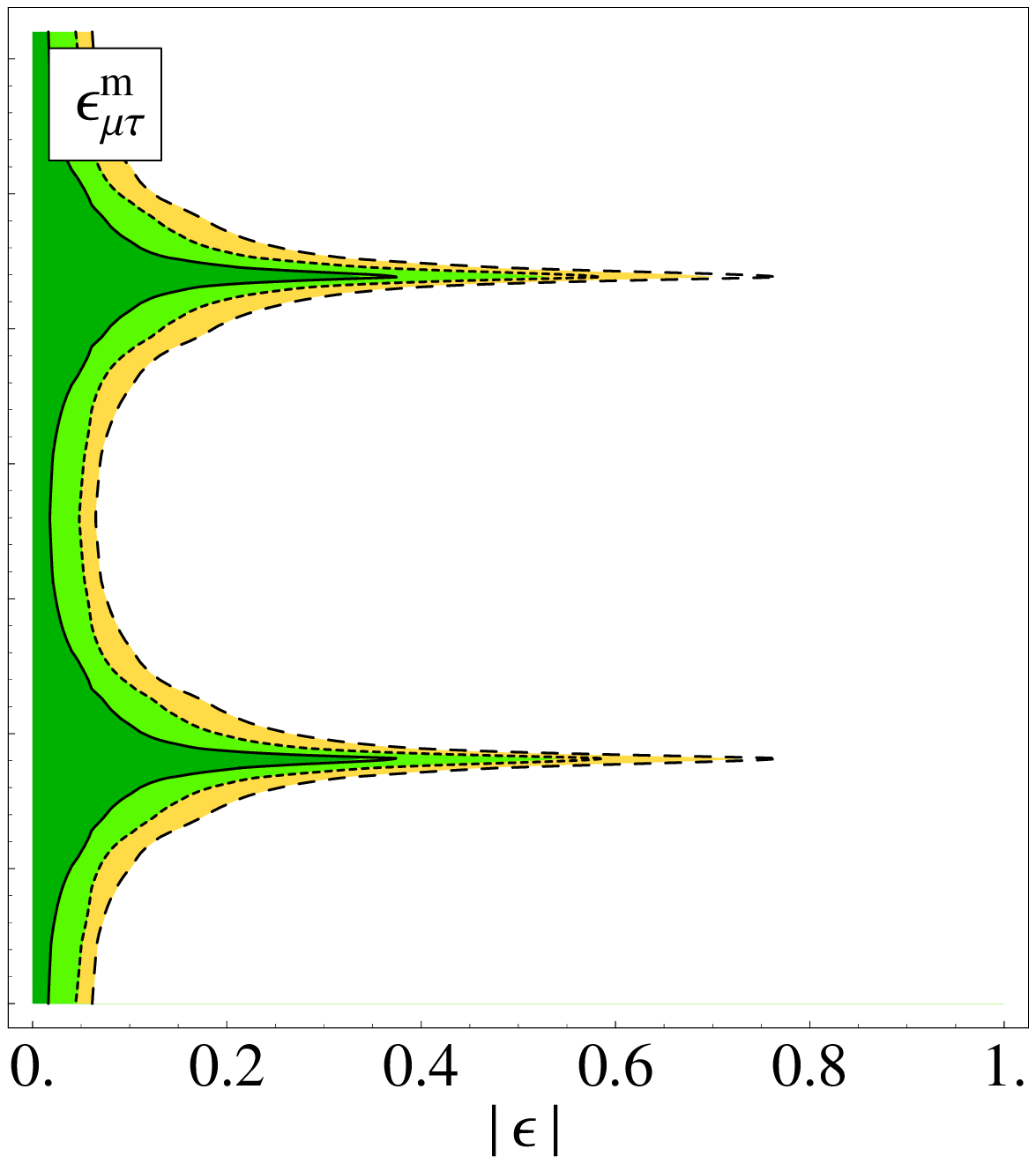}
    \end{tabular}
  \end{center}
  \caption{Discovery reach for $\eps^m_{\alpha\beta}$ in a combined analysis of
    \NOvA\ and \DCext. Since these experiments have essentially no sensitivity
    to $\eps^m_{ee}$, $\eps^m_{\mu\mu}$, and $\eps^m_{\tau\tau}$, we show
    only the off-diagonal entries of $\eps^m$. The colored areas show the $1\sigma$,
    $2\sigma$, and $3\sigma$ confidence regions for $\sthchooz=0.05$, while the
    black contours are for $\sthchooz=0.01$. Note that the scaling of the
    horizontal axis is different from Figs.~\ref{fig:th13discovery-T2K-epsilon-sd}
    and \ref{fig:th13discovery-NOvA-epsilon-sd}.}
  \label{fig:th13discovery-NOvA-epsilon-m}
\end{figure*}

Let us now discuss the prospects of actually detecting the presence of
non-standard interactions in reactor and superbeam experiments. We define
the \emph{discovery reach} as the range of $\eps$ parameters for which the quality
of a standard oscillation fit is below a given confidence level. In
Figs.~\ref{fig:th13discovery-T2K-epsilon-sd} to \ref{fig:th13discovery-NOvA-epsilon-m},
we show numerical results for this quantity, which were obtained by performing
standard oscillation fits to the combined data of \emph{both} experiments.

The results can again be interpreted with the help of Tab.~\ref{tab:suppression-factors}
and of the formulas derived in Sec.~\ref{sec:theory}. We see that for those
non-standard parameters which have a large impact on any of the observed oscillation
channels, there is typically also a good discovery potential. For some parameters,
it comes from the reactor measurement, for others, it is dominated by the superbeam.
It is remarkable, however, that in the case $\eps^s_{e\tau} = \eps^{d*}_{\tau e}$, there
is practically no discovery potential at all, because neither experiment can discover
these parameters on its own, and there is also no significant discrepancy between
them, but only a common offset in their $\theta_{13}$ fits.

It is interesting to observe that the good discovery reach for $\eps^s_{\mu\tau}$
comes from the \emph{disappearance} channel, as can be easily verified from
the corresponding analytical formulas in Sec.~\ref{sec:theory}. Note that in those
plots where $\eps^s_{\mu\tau} = \eps^{d*}_{\tau\mu}$ is assumed, there is no
discovery potential because the corresponding NSI terms in Eq.~\eqref{eq:Pmumu-vac}
cancel.

The discovery reach depends strongly on the phases of the NSI coupling
constants, $\phi^{s,d,m}_{\alpha\beta}$. To first order in $s_{13}$, all
off-diagonal entries of the $\eps$ matrices (except the $\eps^s_{\tau\mu}$
and $\eps^d_{\mu\tau}$ contributions in the $\nu_\mu \rightarrow \nu_\mu$
disappearance channel) are accompanied by a combination of  $\phi^{s,d,m}_{\alpha\beta}$
and $\delta_{\rm CP}$. To first order in $\sdm/\ldm$, they typically appear
together with factors of $\cos\phi^{s,d}_{\alpha\beta}$ or $\sin\phi^{s,d}_{\alpha\beta}$.
The diagonal components of the $\eps$ matrices usually have prefactors of
$\cos\phi^{s,d}_{\alpha\alpha}$.

The plots do not exhibit any phase dependence in the discovery reach for $\eps^s_{\mu e}$
and $\eps^d_{\mu e}$ because the sensitivity to these parameters comes
mainly from the modified $\nu_e$ flux in the near detector of the superbeam experiment.
We have checked, that, in accordance with Eq.~\eqref{eq:Pmue-vac}, the phase
dependence would reappear if the near detector were omitted in the simulation.

Turning to non-standard matter effects described by $\eps^m$, it is clear that the
discovery potential will be very limited, since already standard
matter effects are small in \TtoK\ and \NOvA, and completely negligible
in \DoubleChooz\ and \DCext. Therefore, we use a different scale
for the horizontal axis in Figs.~\ref{fig:th13discovery-T2K-epsilon-m}
and~\ref{fig:th13discovery-NOvA-epsilon-m}. However, for some entries of
$\eps^m$, the present bounds are very weak. In particular we have
$\eps^m_{ee} \lesssim 1.0$, $\eps^m_{e\tau} \lesssim 0.7$, and
$\eps^m_{\tau\tau} \lesssim 1.4$~\cite{Davidson:2003ha}.
Figs.~\ref{fig:th13discovery-T2K-epsilon-m} and~\ref{fig:th13discovery-NOvA-epsilon-m}
thus show that the bound on $\eps^m_{e\tau}$  could be improved
by \NOvA, but not by \TtoK. We should, however, keep in mind that,
according to Eq.~\eqref{eq:epsilon-estimate}, $|\eps^m_{e\tau}| \sim 0.7$
corresponds to $M_{\rm NSI} \sim 100$~GeV, and it is hard to imagine
a model, that could yield such a low NSI scale without violating
present electroweak precision data. Both \TtoK\ and \NOvA\ have some
sensitivity also to $\eps^m_{e\mu}$ and $\eps^m_{\mu\tau}$, but they cannot
compete with the current bounds $\eps^m_{e\mu} \lesssim 5 \cdot 10^{-4}$
and $\eps^m_{\mu\tau} \lesssim 0.1$.

Note that, according to Eqs.~\eqref{eq:Pmue-mat} and~\eqref{eq:Pmumu-mat},
the sensitivity to $\eps^m_{e\mu}$ and $\eps^m_{e\tau}$ comes from the\
$\nu_e$ appearance channel, while the sensitivity to $\eps^m_{\mu\tau}$
has its origin in the disappearance channel.

Let us dwell for a moment on the interesting shape of the sensitivity contours
for $\eps^m_{e\tau}$ and $\eps^m_{e\mu}$, which can only be understood by taking
into account terms proportional to $|\eps|^2$. Let us consider, for example,
$\eps^m_{e\mu}$. According to Eq.~\eqref{eq:Pmue-mat}, the NSI contribution
to the oscillation probability is, to first order in $|\eps|$ and neglecting
$\sdm$,
\begin{align}
  & - 4 |\epsilon^{m}_{e\mu}| s_{23} c_{23}^{2} \tilde{s}_{13} 
      \cos(\phi^{m}_{e\mu} + \delta_{\rm CP})                                      \nonumber\\
  &\hspace{0.5cm} \cdot
       \left[  \sin^{2} \frac{a_{\rm CC} L}{4E}
             - \sin^{2} \frac{\ldm L}{4E}
             + \sin^{2} \frac{(\ldm - a_{\rm CC})L}{4E}
      \right]                                                                      \nonumber\\
  & - 2 |\epsilon^{m}_{e\mu}| s_{23} c_{23}^{2} \tilde{s}_{13}
      \sin(\phi^{m}_{e\mu} + \delta_{\rm CP})                                      \nonumber\\
  &\hspace{0.5cm} \cdot
      \left[   \sin \frac{a_{\rm CC} L}{2E} 
             - \sin \frac{\ldm L}{2E}
             + \sin \frac{(\ldm - a_{\rm CC})L}{2E}
      \right]                                                                      \nonumber\\
  & + 8 |\epsilon^{m}_{e\mu}| s_{23}^{3} \tilde{s}_{13}
      \cos(\phi^{m}_{e\mu} + \delta_{\rm CP}) \frac{a_{\rm CC}}{\ldm - a_{\rm CC}} \nonumber\\
  &\hspace{0.5cm} \cdot
      \sin^{2} \frac{(\ldm - a_{\rm CC})L}{4E}. \label{eq:Pemu-eps-m}
\end{align}
By carefully studying this expression, one finds that the energy dependence in
the first and third terms of Eq.~\eqref{eq:Pemu-eps-m} is quite different from
that of standard oscillations, which is proportional to
$\sin^2 (\ldm - a_{\rm CC})L/4E$. Therefore, these terms will be easy to detect,
while the second term, which modulates the spectrum in the same way as standard
oscillations, can be absorbed into a modified $\theta_{13}$ fit, and will therefore
be hard to detect. From the phase dependence of these terms, we expect that, for
our choice of $\delta_{\rm CP}^{\rm true} = 0$, the discovery reach should be good
for $\phi^m_{e\mu} \sim 0, \pi$, and poor for $\phi^m_{e\mu} \sim
\tfrac{1}{2} \pi, \tfrac{3}{2} \pi$. The plots in
Figs.~\ref{fig:th13discovery-T2K-epsilon-m} and \ref{fig:th13discovery-NOvA-epsilon-m}
reveal that the discovery reach indeed shows this behavior, except for an
unexpectedly good sensitivity at $\phi^m_{e\mu} = \tfrac{3}{2} \pi$. To
understand this, we have to take into account the second order terms, which
we have found to be
\begin{align}
  & \phantom{+\ } 4 |\epsilon^{m}_{e\mu}|^2 c_{23}^4
        \sin^{2} \frac{a_{\rm CC} L}{4E}                   \nonumber\\
  & + 4 |\epsilon^{m}_{e\mu}|^2 s_{23}^4
        \Big( \frac{a_{\rm CC}}{\ldm - a_{\rm CC}} \Big)^2
        \sin^{2} \frac{(\ldm - a_{\rm CC})L}{4E}           \nonumber\\
  & + 2 |\epsilon^{m}_{e\mu}|^2 s_{2 \times 23}^2
        \frac{a_{\rm CC}}{\ldm - a_{\rm CC}}               \nonumber\\
  &\hspace{1cm} \cdot
        \cos\frac{\ldm L}{4E} \sin\frac{a_{\rm CC} L}{4E}
        \sin\frac{(\ldm - a_{\rm CC})L}{4E}.                           
\end{align}
The important observation is that the net effect of the second order terms
is always positive, while for the first order terms, it is positive at
$\phi^m_{e\mu} \simeq \tfrac{3}{2} \pi$, and negative at $\phi^m_{e\mu} \simeq
\tfrac{1}{2} \pi$. In the first case, we would therefore need a much stronger deviation
of the fitted $\theta_{13}$ from its true value in order to absorb the non-standard
term. This, however, is disfavored by the reactor measurement, so that the combined
fit improves the discovery reach by a considerable amount at $\phi^m_{e\mu} \simeq
\tfrac{3}{2} \pi$. We are here in the interesting situation that the combination of seemingly
redundant data sets can be beneficial if there are deviations from standard three-flavor
oscillations. For most other non-standard parameters, the discovery reach is dominated
by either the reactor or the superbeam.

\section{Conclusions}
\label{sec:conclusions}

In this paper, we have studied the impact of non-standard neutrino interactions
on upcoming reactor and accelerator neutrino experiments. We have first classified
the allowed NSI terms in the Lagrangian according to their Lorentz structure, and
have found that many of them are irrelevant to reactor and superbeam setups.
Those which can have an impact are mostly of the $(V-A)(V \pm A)$ type, but in
superbeam experiments, also $(S+P)(S \pm P)$ type effects can be important. Since
reactor and superbeam experiments are not able to distinguish different Lorentz
structures, we have reparameterized the NSI coupling constants in order to
greatly reduce the number of free parameters in the problem.

Using this reparameterization, we have then derived approximate analytic
expressions for the non-standard neutrino oscillation probabilities, both in
vacuum and in matter of constant density. We have developed an intuitive
understanding of the terms relevant to specific oscillation channels, and
have classified them accordingly.

In the second part of our work, we have performed detailed numerical simulations
using \GLoBES. We have considered two scenarios: \TtoK\ combined with \DoubleChooz,
and \NOvA\ combined with a 200~t reactor experiment, dubbed \DCext. Our
simulations take into account parameter correlations, degeneracies, and
systematical errors, and in particular, we employ a realistic treatment of
the near detectors. We have found that non-standard interactions can have a
sizeable impact on future reactor and superbeam experiments, if the coupling
constants are close to their current upper limits, and if complex phases do not
conspire to cancel them. The biggest impact is on the $\theta_{13}$ measurement:
If NSI are not properly taken into account in the fit, the results may be
significantly wrong. There are scenarios in which a clear discrepancy between
reactor and superbeam experiments shows up, but we can also have the situation
that both fits sets seem to agree very well, but the derived $\theta_{13}$ value
has a significant offset from the true value. It is even possible that the
true  $\theta_{13}$ is erroneously ``ruled out'' at $3\sigma$. To detect this kind
of problems, a third experiment, complementary to the other two, would be
required. Thus, we see that the possibility of non-standard effects should
always be kept in mind when planning or analyzing upcoming experiments. 

We have also studied the discovery potential for NSI in reactor and
superbeam experiments, i.e.\ the range of non-standard parameters, which
can actually be detected by these experiments because the quality of a standard
oscillation fit becomes poor. We have found that, depending on the complex phases,
some NSI may be discovered if their coupling constants are not more than a factor
of 5 smaller than the current upper bounds. The best discovery reach is
obtained only if both, reactor and superbeam experiments, and also the
respective near detectors are considered in the analysis. In most cases, one
of the experimental channels dominates the discovery reach, but there are
also situations where only the discrepancy between the single-experiment
fits indicates the presence of NSI. Our discussion thus shows that
reactor and superbeam measurements, which might seem to be redundant in the
standard three-flavor framework, turn out to be highly complementary once
non-standard effects are considered.

\begin{acknowledgments}
This work was in part supported by the Transregio Sonderforschungsbereich TR27
``Neutrinos and Beyond'' der Deutschen Forschungsgemeinschaft. The work of JS
is supported in part by the Grant-in-Aid for the Ministry of Education, Culture,
Sports, Science, and Technology, Government of Japan (No. 17740131 and 18034001).
JK would like to acknowledge support from the Studienstiftung des Deutschen Volkes.
\end{acknowledgments}

\bibliography{./nsi}
\bibliographystyle{apsrev}

\end{document}